\documentclass[aps,pra,reprint,twocolumn,superscriptaddress,nofootinbib]{revtex4-1}
\pdfoutput=1 

\usepackage{graphicx}
\usepackage{bm}
\usepackage{amsmath}
\usepackage{amssymb}
\usepackage{amsfonts}
\usepackage{amsthm}
\usepackage{mathtools}
\usepackage{hyperref}
\usepackage{braket}
\usepackage[normalem]{ulem}
\usepackage{wrapfig}
\usepackage{tikz}
\usepackage{dsfont}
\usepackage{comment}
\usepackage{thmtools,thm-restate}
\usepackage[export]{adjustbox}

\usepackage[resetlabels]{multibib}
\newcites{app}{Appendix References}

\hypersetup{colorlinks=true,urlcolor=[rgb]{0,0,0.5},citecolor=[rgb]{0.5,0,0},linkcolor=[rgb]{0,0,0.4}}

\usetikzlibrary{decorations.pathmorphing}

\newtheorem*{theorem*}{Theorem}

\newtheorem*{task*}{Task}
\newtheorem*{proposition*}{Proposition}

\def\autorefapp#1{\hyperref[#1]{Appendix~\ref{#1}}}

\def\phib{\boldsymbol{\phi}}
\def\pib{\boldsymbol{\pi}}
\def\chib{\boldsymbol{\chi}}

\def\and{\quad {\rm and} \quad}

\newcommand{\calD}{\mathcal{D}}

\newcommand{\calS}{\mathcal{S}}

\DeclarePairedDelimiter{\brk}{[}{]}

\makeatletter
\def\Pr{\@ifnextchar[{\@witha}{\@withouta}}
\def\@witha[#1]{\mathop{\operator@font Pr}_{#1}\brk}
\def\@withouta{\mathop{\operator@font Pr}\brk}
\makeatother

\makeatletter
\def\E{\@ifnextchar[{\@withb}{\@withoutb}}
\def\@withb[#1]{\mathop{\mathbb{E}}_{#1}\brk}
\def\@withoutb{\mathop{\mathbb{E}}\brk}
\makeatother

\makeatletter
\def\Var{\@ifnextchar[{\@withc}{\@withoutc}}
\def\@withc[#1]{\mathop{\mathbb{V}}_{#1}\brk}
\def\@withoutc{\mathop{\mathbb{V}}\brk}
\makeatother

\renewcommand{\vec}[1]{\mathbf{#1}}

\begin{document}
\title{Quantum Scars in Quantum Field Theory}

\author{Jordan Cotler}
\email{jcotler@fas.harvard.edu}
\affiliation{Society of Fellows, Harvard University, Cambridge, MA, USA}
\affiliation{Black Hole Initiative, Harvard University, Cambridge, MA, USA}

\author{Annie Y. Wei}
\email{anniewei@mit.edu}
\affiliation{Center for Theoretical Physics, Massachusetts Institute of Technology, Cambridge, MA 02139}

\begin{abstract}
We develop the theory of quantum scars for quantum fields.  By generalizing the formalisms of Heller and Bogomolny from few-body quantum mechanics to quantum fields, we find that unstable periodic classical solutions of the field equations imprint themselves in a precise manner on bands of energy eigenfunctions.  This indicates a breakdown of thermalization at certain energy scales, in a manner that can be characterized via semiclassics.  As an explicit example, we consider time-periodic non-topological solitons in complex scalar field theories.  We find that an unstable variant of Q-balls, called Q-clouds, induce quantum scars.  Some technical contributions of our work include methods for characterizing moduli spaces of periodic orbits in field theories, which are essential for formulating our quantum scar formula.  We further discuss potential connections with quantum many-body scars in Rydberg atom arrays.
\\
\end{abstract}

\maketitle

\section{Introduction}


Most systems in nature are chaotic, and the interplay between chaos and quantum mechanics has long been a topic of intense study.  While the subject of quantum chaos is not yet in its final form, it hosts a diverse set of phenomena~\cite{haake2010q, ChaosBook} which have been essential to our understanding of mesoscopic systems~\cite{simons2002mesoscopic, heller2018semiclassical}, quantum thermalization~\cite{DAlessio:2015qtq}, and even quantum gravity~\cite{Maldacena:2015waa, Cotler:2016fpe}. 

In classical Hamiltonian mechanics, chaos provides a mechanism for systems to exhibit ergodic behavior.  For instance, a chaotic Hamiltonian system at fixed energy is generically expected to equitably explore all configurations in phase space at that energy.  This forms the theoretical basis for the classical microcanonical ensemble.  The quantum analog is to consider a Hamiltonian operator $H$, and a projector $P_E$ onto a band of eigenstates concentrated around $E$.  Then the Wigner phase space distribution corresponding to $P_E$ is expected to be approximately uniform over all configurations in the classical phase space with energy $\approx E$.  This would justify the usual quantum version of the microcanonical ensemble.

Remarkably, a large variety of chaotic, few-body quantum systems fail to conform to the microcanonical ensemble, even approximately.  What happens is that the Wigner phase space distribution of $P_E$ is enhanced in the vicinity of
unstable classical periodic trajectories~\cite{heller1984bound, bogomolny1988smoothed, berry1989quantum, kaplan1998linear}.

The seminal analysis of Heller leveraged semiclassical techniques and time-smeared correlators to establish that the position-space density of eigenfunctions is enhanced around unstable, classical periodic orbits in few-body systems~\cite{heller1984bound, kaplan1998linear}.  Building off of this, Bogomolny showed that these imprints on the wavefunction could be made visible as oscillatory fringes after smoothing over an energy and position window~\cite{bogomolny1988smoothed}, while Berry showed that these fringes are also visible in phase space using Wigner functions~\cite{berry1989quantum}. See~\cite{kaplan1999scars} for a comprehensive review.  The question we wish to address is whether quantum scars occur in quantum field theories by an analogous mechanism.

There are related phenomena~\cite{turner2018weak,turner2018quantum,choi2019emergent,ho2019periodic} that have received attention following experiments on Rydberg atom arrays~\cite{bernien2017probing} where periodic revivals were observed after a quantum quench. This is a situation where one might normally expect the system to thermalize, and be composed of energy eigenstates obeying the eigenstate thermalization hypothesis (ETH)~\cite{Srednicki:1994mfb, deutsch1991quantum, srednicki1999approach, DAlessio:2015qtq}. Ref.~\cite{turner2018weak} showed that a small number of non-thermal eigenstates are responsible for the periodic revivals, while Ref.~\cite{ho2019periodic} showed that these states could be approximately described by matrix product states that support unstable periodic orbits.  These thermalization-breaking eigenstates have been suggestively called ``quantum many-body scars''~\cite{turner2018weak,turner2018quantum,choi2019emergent,ho2019periodic}, although their precise relation to the few-body scars of Heller and others has not been understood.  An important question is thus: are quantum many-body scars in the sense of~\cite{turner2018weak,turner2018quantum,choi2019emergent,ho2019periodic} due to the same mechanism as the few-body scars of Heller?  To begin to address this, we must first develop the quantum field-theoretic version of Heller's quantum scars.  We will put particular emphasis on the approach of Bogomolny since it admits a rather direct generalization to field theory.  This is the goal of the present paper.

The rest of the paper is organized as follows.  In Section~\ref{sec:scarformula} we generalize Bogomolny's scar formula to quantum field theory. In Section~\ref{sec:qballs} we pursue a particular example of quantum field-theoretic scars, furnished by unstable periodic orbits of non-topological solitons called Q-clouds. In Section~\ref{sec:moduli} we characterize the moduli space of time-periodic Q-cloud solutions in a fixed energy window, and show that it satisfies the assumptions of our scar formula.  Finally, in Section~\ref{sec:discuss} we conclude with a discussion.

\section{Scar formula for QFT}
\label{sec:scarformula}

We adopt the convention that capital Greek letters $\Phi(x,t), \Pi(x,t),...$ denotes fields as a function of spacetime coordinates, whereas lowercase Greek letters $\phi(x), \pi(x),...$ denote fields at a fixed moment of time.  We will set the speed of light to be $c = 1$.  Consider a relativistic scalar quantum field theory with action $S[\Phi]$, Hamiltonian $H[\Phi, \Pi]$, and associated eigenfunctions $\{\Psi_n[\phi]\}_n$.  These eigenfunctions may in general come in a continuous family, in which case $n$ is replaced by a continuous index.  The field theory in question is assumed to be nonintegrable, for instance an interacting scalar field theory in $d+1$ spacetime dimensions where the potential $U(\Phi)$ is not fine-tuned.  It will be convenient for us to consider a complex scalar field, in particular with a potential of the form $U(|\Phi|^2)$.

We are interested in rendering a semiclassical description of the probability functional $|\Psi_n[\phi]|^2$ associated to a single eigenstate $\Psi_n[\phi]$.  However, in general there is no semiclassical description of \textit{individual} eigenstates of a chaotic system~\cite{bogomolny1988smoothed, berry1989quantum}, but there is such a semiclassical description of smeared \textit{bands} of eigenstates.  Accordingly, in the spirit of~\cite{bogomolny1988smoothed}, we make the following simplifications: (i) We work with an average over probability functionals $|\Psi_n[\phi]|^2$ with energy in the range $[E-\varepsilon/2, E + \varepsilon/2]$, corresponding to an energy band around $E$ of size $\varepsilon$.  We assume that $\varepsilon$ is much larger than the mean level spacing, and also much less than $E$.  We denote the energy band average by $\langle |\Psi[\phi]|^2\rangle_E$. (ii) Additionally, we smear slightly in field space using a Gaussian functional.  The smeared version of $\langle |\Psi[\phi]|^2\rangle_E$ is denoted by $\langle |\Psi[\phi]|^2\rangle_{E,\Delta}$, and is given as
\vspace{-.1cm}
\begin{align*}
&\langle |\Psi[\phi]|^2\rangle_{E,\Delta} \!:= \mathcal{N} \!\!\int \![d\chi]\langle |\Psi[\chi]|^2\rangle_{E} \, e^{- \frac{1}{2 \Delta^2}\! \int\! d^{d}\! x \, |\phi(x) - \chi(x)|^2}
\end{align*}
where $\mathcal{N}$ is the normalization.  Note that we have
$\langle |\Psi[\phi]|^2\rangle_{E,\Delta} \to \langle |\Psi[\phi]|^2\rangle_{E}$ as $\Delta \to 0$.

We can derive a semiclassical formula for $\langle |\Psi[\phi]|^2\rangle_{E,\Delta}$ by generalizing the analysis of Bogomolny which applies to the few-body setting~\cite{bogomolny1988smoothed}.  Bogomolny's scar formula and its derivation are reviewed in great detail in Appendix~\ref{sec:bogomolny}; our treatment of the derivation is organized to most readily generalize to quantum field theory.  The quantum field theory scar formula is derived in Appendix~\ref{app:derivation}, and has numerous ingredients which are unique to the field theory setting.  To state our result, let us develop some further notation.

Let $\mathcal{F}$ be the infinite-dimensional space of field configurations $\phi : \mathbb{R}^d \to \mathbb{C}$. We equip the tangent bundle of $\mathcal{F}$ with the standard $L^2$ inner product so that $\mathcal{F}$ becomes a Riemannian manifold.  Then a classical periodic orbit $\Phi_c(x,t)$ can be regarded as a map $\Phi_c(x,\,\cdot\,) : \mathbb{R} \to \mathcal{F}$, whose image specifies a 1-dimensional submanifold of $\mathcal{F}$.  Consider the collection of periodic solutions to the classical equations of motion such that their energies lie in the range $[E - \varepsilon/2, E + \varepsilon/2]$, and their periods $T$ are less than $\sim \hbar/\varepsilon$. In a chaotic system, these orbits are generically unstable.  Let $\mathcal{O}$ denote the union of the images of these periodic orbits; momentarily assuming that any two distinct orbits (which are not time-reversals of one another) do not intersect, $\mathcal{O}$ is a union of connected submanifolds of $\mathcal{F}$, possibly with different dimensions.  We denote the individual connected components by $\mathcal{O}_i$.  Each $\mathcal{O}_i$ is a moduli space of periodic orbits (satisfying our above desiderata) which are continuously deformable into one another.

With these notations at hand, we can formulate our result.  If $\langle |\Psi[\phi]|^2 \rangle_{E, \Delta}$ conformed to our expectations for the microcanonical ensemble, then for small $\varepsilon$ we would expect it to approximately equal
\begin{equation}
\label{E:micro1}
P_{\text{micro}}[\phi] = \frac{\int [\frac{d\pi}{2\pi\hbar}]\,\delta_\varepsilon(E-H(\phi, \pi))}{\int [d\chi]\, [\frac{d\pi}{2\pi\hbar}]\,\delta_\varepsilon(E-H(\chi, \pi))}\,,
\end{equation}
where $[\frac{d\pi}{2\pi \hbar}]$ denotes the formal path integral measure $\prod_{x \in \mathbb{R}^{d}} \!\frac{d\pi(x)}{2\pi\hbar}$ and similarly for $[d\chi]$.  Here $\delta_\varepsilon$ is an $\varepsilon$-smearing of the $\delta$ function.  However, our periodic orbits will provide deviations from this microcanonical answer.

In particular, suppose we evaluate $\langle |\Psi[\phi]|^2\rangle_{E,\Delta}$ at a field configuration that lies on a periodic orbit, or is only slightly off of that periodic orbit.  More precisely, let $\phi = \phi_c + \delta\phi_\perp$, where $\phi_c$ is a point in $\mathcal{O}$, and $\delta\phi_\perp$ is in the normal bundle of $\mathcal{O}$ at $\phi_c$, notated as $\text{N}_{\phi_c}\mathcal{O}$.  The normal bundle $\text{N}\mathcal{O}$ can be thought of as field fluctuations which move us off of $\mathcal{O}$.  Letting $T_{\text{max}}$ be the period of the longest orbit in $\mathcal{O}$, we choose $\Delta$ to be
\begin{equation}
\Delta = \sqrt{\hbar \,T_{\text{max}}} \left(\frac{\hbar}{E\,T_{\text{max}}}\right)^\gamma \,\,\,\,\text{for any}\,\,\,\,\frac{1}{4} < \gamma < \frac{1}{2}\,,
\end{equation}
and consider fluctuations $\delta\phi_\perp$ satisfying
\begin{equation}
\| \delta \phi_\perp \|_{L^2} \lesssim \sqrt{\hbar \, T_{\text{max}}} \left(\frac{\hbar}{E\,T_{\text{max}}}\right)^{1/4}\,.
\end{equation}
\begin{figure*}[t!]
\includegraphics[width=15cm]{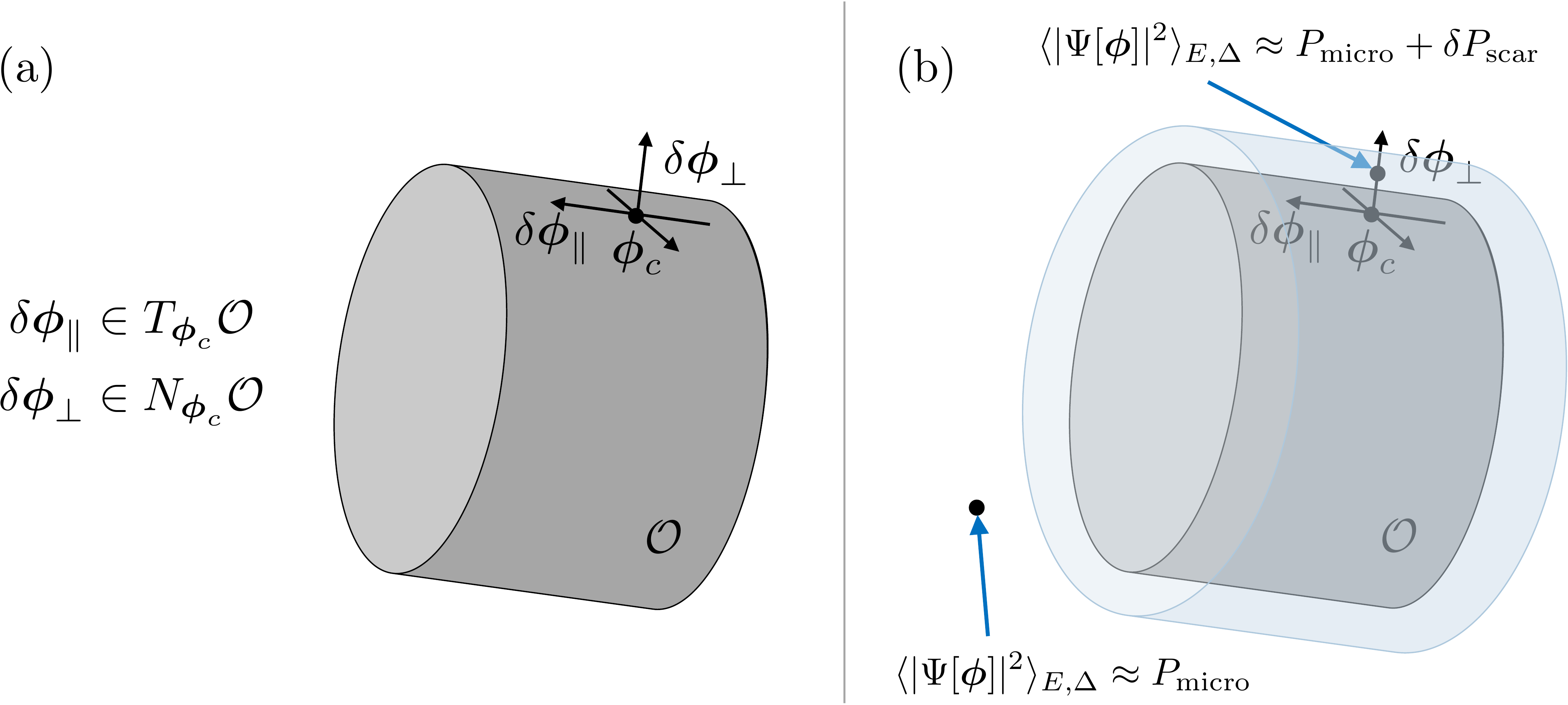}
\caption{(a) A schematic of $\mathcal{O}$, the union of the images of our collection of periodic orbits.  We are interested in the vicinity of a particular orbit $\phi_c$, and construct coordinates $\delta \phi_\parallel$ which lie in the tangent space $T_{\phi_c} \mathcal{O}$ (and include a direction parallel to the orbit $\phi_c$), and coordinates $\delta \phi_\perp$ orthogonal to the orbit which lie in the normal space $\text{N}_{\phi_c} \mathcal{O}$.  (b)  If we evaluate $\langle |\Psi[\phi]|^2\rangle_{E,\Delta}$ at a point $\phi = \phi_c + \delta \phi_\perp$ in the vicinity of $\mathcal{O}$, then we have $\langle |\Psi[\phi]|^2\rangle_{E,\Delta} \approx P_{\text{micro}} + \delta P_{\text{scar}}$ which has a scar contribution.  The neighborhood of $\mathcal{O}$ for which the scar contribution $\delta P_{\text{scar}}$ is sizable is shown in light blue.  If we instead evaluate $\langle |\Psi[\phi]|^2\rangle_{E,\Delta}$ at a point $\phi$ sufficiently far away from $\mathcal{O}$, then we only have $\langle |\Psi[\phi]|^2\rangle_{E,\Delta} \approx P_{\text{micro}}$\,.}
\label{fig:sketch}
\end{figure*}
In this regime, $\langle |\Psi[\phi]|^2 \rangle_{E, \Delta} = P_{\text{micro}}[\phi] + \delta P_{\text{scar}}[\phi_c, \delta\phi_\perp]$, where there is an additional scar contribution.  This is depicted in Figure~\ref{fig:sketch}.   Since $\phi_c$ belongs to some periodic orbit $\Phi_c(x,t)$, say with $\Phi_c(x,0) = \phi_c$, then let us denote $\dot{\phi}_c = \partial_t \Phi_c(x,t)\big|_{t=0}$.
Further writing $\delta \phi_{1,\perp}, \delta \phi_{2,\perp}$ as the real and imaginary parts of $\delta \phi_\perp$, we have the scar formula
\begin{widetext}
\begin{align}
\label{E:mainresult1}
&\delta P_{\text{scar}}[\phi_c, \delta\phi_\perp] \approx -\frac{2}{\pi\hbar\int [d\chi] \left[\frac{d\pi}{2\pi\hbar}\right]\delta_\varepsilon(E-H(\chi, \pi))} \, \text{Im}\Bigg\{  \frac{1}{i} \frac{1}{\|\dot{\phi}_{c}\|_{L^2}} \Bigg|\det\!\left(\frac{1}{2\pi i \hbar}\frac{\delta^2 S(\phi^A, \phi^B, E)}{\delta \phi_{\perp}^A \delta \phi_{\perp}^B}\right)_{\phi^A = \phi^B = \phi_c}\Bigg|^{1/2} \nonumber \\
& \times \exp\!\Bigg[- \frac{\varepsilon}{\hbar}\, T(\phi_c, \phi_c, E)-i \nu(\phi_c, \phi_c, E) \,\frac{\pi}{2}+  \frac{i}{\hbar}\Bigg(\!S(\phi_c, \phi_c, E)   \\
& \qquad \quad + \frac{1}{2}\int \! d^d x \, d^d y \!\!\sum_{\ell,m=1}^2\!\! \delta\phi_{\ell, \perp}(x)\! \left(\frac{\delta^2 S(\phi^A, \phi^B, E)}{\delta\phi_\ell^A(x)\delta\phi_m^A(y)}\!+2\, \frac{\delta^2 S(\phi^A, \phi^B, E)}{\delta\phi_\ell^A(x)\delta\phi_m^B(y)}\!+\!\frac{\delta^2 S(\phi^A, \phi^B, E)}{\delta\phi_\ell^B(x)\delta\phi_m^B(y)}\right)_{\phi^A=\phi^B=\phi_c} \delta\phi_{m,\perp}(y)\Bigg)\Bigg]  \Bigg\}\,.\nonumber 
\end{align}
\end{widetext}
This equation requires some unpacking.  Above, $S(\phi^A, \phi^B, E)$ is the Legendre transform of Hamilton's principal function, so that $S(\phi^A, \phi^B, E) = \int_0^T dt \int d^{d}x \sum_{i=1}^2 \Pi_i(x,t) \,\partial_t \Phi_i(x,t)$ is the `abbreviated action' of a classical solution with energy $E$ starting at $\phi^A$ at time zero and ending at $\phi^B$ at time $T = T(\phi^A, \phi^B, E)$.  Moreover, in~\eqref{E:mainresult1} the term $T(\phi_c, \phi_c, E)$ is the period of the periodic orbit passing through $\phi_c$ with energy $E$.  The term $\nu(\phi_c, \phi_c, E)$ contains phase factors from the square root of the determinant which organize into the Maslov index.  Finally, the $\approx$ means that the equation should be understood as including multiplicative corrections $\big( 1 + O\big(\frac{\varepsilon}{E}\,,(\!\frac{\hbar}{E \,T_{\text{max}}}\!)^\gamma\big)\,\big)$.

Notice in~\eqref{E:mainresult1} that periodic orbits with $T \gg \hbar/\varepsilon$ are exponentially suppressed on account of the $\exp(- \varepsilon\,T(\phi_c, \phi_c, E)/\hbar)$ term.  This is why we are justified in building $\mathcal{O}$ out of orbits with period less than $\sim \hbar/\varepsilon$.  Interestingly, quantum scarring occurs for orbits which are classically unstable; this is a signature feature of quantum scars~\cite{heller1984bound}, and is fortuitous since generic periodic orbits of Hamiltonian systems are unstable.  We note that our formula for corrections to the microcanonical ensemble also hold for stable periodic orbits, although these are terminologically not called quantum scars.

It is clarifying to write out our expression for $\langle |\Psi[\phi]|^2 \rangle_{E, \Delta}$ in the following way:
\begin{widetext}
\begin{equation}
\label{E:piecewise1}
\langle |\Psi[\phi]|^2\rangle_{E,\Delta} \approx \begin{cases}  P_{\text{micro}}[\phi] + \delta P_{\text{scar}}[\phi_c, \delta \phi_\perp] & \text{if  }\phi = \phi_c + \delta \phi_\perp  \text{  where  }\phi_c \in \mathcal{O},\, \delta \phi_\perp \in \text{N}_{\phi_c} \mathcal{O},\,\,\|\delta \phi_\perp\|_{L^2} \lesssim \frac{\hbar}{\Delta} \\ \\
 P_{\text{micro}}[\phi] & \text{if  } \left\|\left(\frac{\delta S(\phi^A,\phi^B,E)}{\delta \phi^A}+\frac{\delta S(\phi^A,\phi^B,E)}{\delta \phi^B}\right)_{\phi^A=\phi^B=\phi}\right\|_{L^2} \gg \frac{\hbar}{\Delta}
\end{cases}\,.
\end{equation}
\end{widetext}

\noindent This formula emphasizes that there is an additional scar contribution on and near $\mathcal{O}$. Note that there are some regimes for which our scar formula does not directly apply.  The most interesting such regime is given by $\|\delta \phi_\perp\|_{L^2} \gtrsim \frac{\hbar}{\Delta}$ and $\left\|\left(\frac{\delta S(\phi^A,\phi^B,E)}{\delta \phi^A}+\frac{\delta S(\phi^A,\phi^B,E)}{\delta \phi^B}\right)_{\phi^A=\phi^B=\phi}\right\|_{L^2} \lesssim \frac{\hbar}{\Delta}$.  This regime corresponds to orbits which are nearly-periodic (i.e.~they begin and end at $\phi$ but have slightly different initial and final momenta), but are not nearby an exactly periodic orbit.  There is a way to account for such orbits, and this is discussed at the ends of Appendices~\ref{sec:bogomolny} and~\ref{app:derivation}.

While~\eqref{E:mainresult1} and~\eqref{E:piecewise1} tell us how periodic orbits in a (complex) scalar field theory augment the microcanonical ensemble via quantum scars, it is incumbent on us to find examples in which such periodic orbits arise.  In the next section we study a particularly nice set of examples, namely a wide class of complex scalar field theories which furnish periodic orbits in the form of non-topological solitons called Q-clouds~\cite{alford1988q}, which are perturbatively unstable counterparts of Q-balls~\cite{coleman1986q}.

\section{Q-balls and Q-clouds}
\label{sec:qballs}

Solitons are localized field configurations that can arise in nonlinear theories. While static scalar solitons can be stabilized by topological effects, in this work we will be concerned with non-static solitons stabilized by a $\textsf{U}(1)$ charge. Such solitons are known as Q-balls~\cite{coleman1986q} when they are classically stable, and Q-clouds~\cite{alford1988q} when they are classically unstable.  We briefly review these below.

Working in $d+1$ spacetime dimensions, we consider a complex scalar field with $\textsf{U}(1)$ symmetry and a nonlinear potential, as specified by the Lagrangian ${\cal L}=\partial_{\mu}\Phi^*\partial^{\mu}\Phi-U(|\Phi|^2)$.  In order for a soliton to not dissipate, it must be a stationary point of the energy functional at fixed $\textsf{U}(1)$ charge. Note that for a potential that has a global minimum at $U(0)=0$, corresponding to the vacuum $\Phi=0$, the perturbative spectrum consists of particles of mass $m^2=U''(0)$. To support new kinds of excitations, we consider a polynomial potential $U$ that has $U(\Phi_0)<\frac{1}{2}m^2|\Phi_0|^2$ at some $\Phi_0\neq 0$, i.e.~the potential grows less quickly than a quadratic mass term.  In particular, we leverage a potential of the form
\begin{equation}
\label{eq:sigma4potential}
    U(|\Phi|^2)=m^2|\Phi|^2-\frac{f}{2}|\Phi|^4+O(|\Phi|^6)\,,
\end{equation}
where $f>0$ and the $O(|\Phi|^6)$ terms serve to stabilize the potential, although their precise form will not matter for our analyses.  This potential furnishes spherically-symmetric, time-periodic solutions $\Phi(x,t) = e^{i \omega t}\,\sigma(x)$ for a range of $\omega$~\cite{coleman1986q, alford1988q}.  The function $\sigma$ only depends on $r = |x|$, so we will often write it as $\sigma(r)$.  The residual equations of motion for $\sigma(r)$ are
\begin{equation}
\label{E:radialEOM1}
    \sigma''=-\frac{d+1}{r}\,\sigma'-\omega^2\sigma+\sigma\,U'(\sigma^2)\,,
\end{equation}
which has a frequency-dependent damping term in the effective potential $U_{\text{eff}}=\frac{1}{2}(\omega^2\sigma^2-U)$.

\begin{figure*}[t!]
\includegraphics[width=17cm]{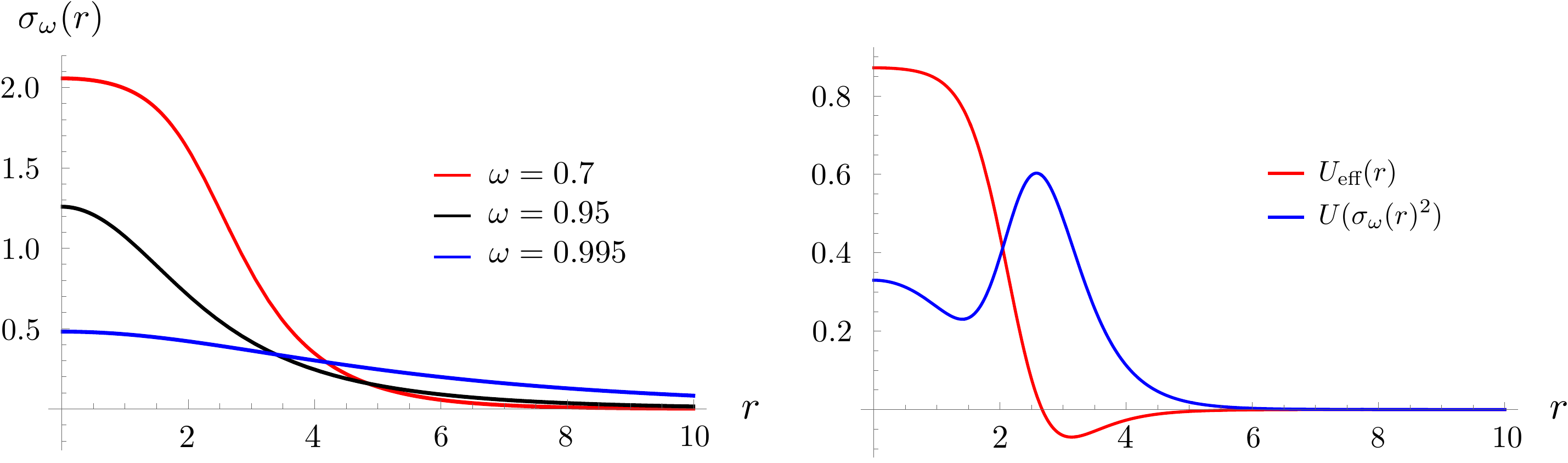}
\caption{(Left) Plot of Q-cloud profiles $\sigma_\omega(r)$ for different values of $\omega=0.7, 0.95, 0.995$. (Right) Plots of the potential $U(\sigma_\omega(r)^2)$ and the effective potential $U_{\text{eff}}(r)=\frac{1}{2}(\omega^2\sigma_\omega^2(r)-U(\sigma_\omega(r)^2))$ for $\omega=0.7$.}
\label{fig:qballs}
\end{figure*}

These periodic solutions $e^{i \omega t}\,\sigma(r)$ are called Q-balls if they are perturbatively stable~\cite{coleman1986q}, and Q-clouds if they are perturbatively unstable~\cite{coleman1986q, alford1988q}.  While both kinds of periodic orbits provide corrections to the microcanonical ensemble, we choose to focus on the unstable periodic orbits since they will give rise to quantum scars in particular.  We emphasize that it is a novel feature of quantum scars that they are supported by unstable classical solutions.  The unstable periodic orbits of interest in the scalar theory, namely the Q-clouds, occur as $\omega^2$ approaches $m^2$ from below.  Along the lines of Alford's analysis~\cite{alford1988q}, we define $\xi^2 = m^2 - \omega^2$ to keep track of the closeness between $m^2$ and $\omega^2$.  In the regime $\xi \to 0^+$, plugging the scaling ansatz $\sigma_\xi(r) = \xi w(\xi r)$ into~\eqref{E:radialEOM1} we obtain
\begin{equation}
w'' - w + f w^3 = 0
\end{equation}
which is $\xi$-independent.  This means that $\sigma_\xi(r)$ flattens out in the $\xi \to 0^+$ limit, and becomes diffuse like a `cloud', as can be seen in Figure \ref{fig:qballs}.  In this regime, the Q-cloud decays at large $r$ as $\sim \xi \, e^{- \xi r}$, and the $\textsf{U}(1)$
 charge tends to infinity as $Q \sim \xi^{-1}$.

\section{Moduli space of Q-cloud solutions in an energy window}
\label{sec:moduli}

As we explained above, our periodic orbits of interest are often arranged in moduli spaces $\mathcal{O}$ with connected components $\mathcal{O}_i$.  Here we show that there is a single connected component $\mathcal{O}_{\text{Q-cloud}}$ associated with Q-clouds in 3+1 dimensions which sit in an energy window $[E - \varepsilon/2, E + \varepsilon/2]$, and have periods less than $\sim \hbar/\varepsilon$.  Moreover, we find that $\mathcal{O}_{\text{Q-cloud}}$ is a 5-dimensional submanifold of $\mathcal{F}$: three of the dimensions correspond to spatial translations, one corresponds to time translations, and one corresponds to deforming a Q-cloud into another Q-cloud with slightly different energy.

Our basic strategy will be as follows.  Suppose $\Phi_\omega$ is a Q-cloud solution with period $T = 2\pi/\omega$ in our energy range.  We will establish that any $\tilde{\Phi}$ which is (i) a small deformation of $\Phi_\omega$, (ii) periodic in time, and (iii) has energy in the appropriate window, is itself a Q-cloud solution.  Note that Q-cloud solutions are localized in space, meaning that they must decay to zero at spatial infinity.  For (i), it is sufficient to measure `smallness' via $\left(\int_{\mathbb{R}^3} d^3 x \int_0^{\hbar/\varepsilon} \!\! dt \, |\Phi_\omega(x,t) - \tilde{\Phi}(x,t)|^2\right)^{1/2}$.

Since $\Phi_\omega(x,t) = e^{i \omega t} \sigma_\omega(x)$, where we have added a subscript to $\sigma(x)$ to make explicit its dependence on $\omega$, we can write $\tilde{\Phi}(x,t)$ as
\begin{equation*}
\tilde{\Phi}(x,t) = e^{i (\omega + \delta \omega)t}(\sigma_\omega(x) + \delta \phi_1(x)) + \sum_{n \not = 1} \! e^{i (\omega + \delta \omega)n t} \delta \phi_n(x)\,,
\end{equation*}
which has period $2\pi/(\omega + \delta \omega)$.  Note that we have not included perturbations that would cause the period to become multiplicatively larger as $T \to \xi \,T$ for $\xi \gtrsim 2$; such orbits are exponentially suppressed in the scar formula relative to solutions with period nearly equal to $T$.  For $\delta \omega$ sufficiently small, there is another Q-cloud solution with the same period, namely $\Phi_{\omega + \delta \omega} = e^{i (\omega + \delta \omega)t} \sigma_{\omega + \delta \omega}(x)$.  Writing $\sigma_{\omega + \delta \omega}(x) = \sigma_\omega(x) + \delta \sigma_\omega(x)$, let us redefine $\delta \phi_1 \to \delta \phi_1 - \delta\sigma_\omega$ in our equation for $\tilde{\Phi}(x,t)$ above to obtain
\begin{align}
\label{E:perturbedaround}
\tilde{\Phi}(x,t) &= \Phi_{\omega + \delta \omega}(x,t) + \sum_{n \in \mathbb{Z}} \! e^{i (\omega + \delta \omega)n t} \delta \phi_n(x)\,.
\end{align}
We can view this as a perturbed version of the Q-cloud solution $\Phi_{\omega + \delta \omega}$.

For $\tilde{\Phi}(x,t)$ to be a solution of the equations of motion, to first order in perturbation theory we require
\begin{align}
\label{E:perturb1}
(\omega^2n^2+\nabla^2\!-\!U'\!-\!\sigma_\omega^2 U'')\delta\phi_n-\sigma_\omega^2 U'' \, \delta\phi^*_{-n+2}&=0\\
\label{E:perturb2}
(\omega^2n^2+\nabla^2\!-\!U'\!-\!\sigma_\omega^2 U'')\delta\phi^*_{-n}\!-\sigma_\omega^2 U''\, \delta\phi_{n+2}&=0\,.
\end{align}
Evidently this system of equations couples the $\delta \phi_n$'s with one another.

Fortunately, the system of equations can be mostly decoupled. Observe that for large $|x|$, we have $\sigma_\omega \to 0$, $U' \to m^2$, and $U'' \to 0$.  Then in this regime
\begin{equation}
\label{E:laplacianconstraint1}
- \nabla^2 \delta \phi_n = m^2 \left( \frac{\omega^2}{m^2}\, n^2 - 1\right) \delta \phi_n\,,
\end{equation}
where $\omega/m < 1$. The above is an eigenvalue equation of the form $- \nabla^2 \delta \phi_n = \lambda \, \delta \phi_n$, which has oscillatory solutions for $\lambda \geq 0$, corresponding to $|n| > 1$.  But such solutions do not decay at infinity, and so we can rule them out and set $\delta \phi_{n \not = 0,\pm 1} = 0$.  Plugging $\delta \phi_{n \not = 0,\pm 1} = 0$ back into~\eqref{E:perturb1} and~\eqref{E:perturb2}, we see that $\delta \phi_0 = 0$ as well.  This leaves us with $\delta \phi_{\pm 1}$, which satisfy the system of equations
\begin{align}
(\omega^2+\nabla^2-U'-\sigma_\omega^2 U'')\delta\phi_{-1}&=0\\
\label{eq:addedmodeseqn}
(\omega^2+\nabla^2-U'-2\sigma_\omega^2 U'')(\delta\phi_1+\delta\phi^*_{1})&=0\\
\label{eq:subtractedmodeseqn}
(\omega^2+\nabla^2-U')(\delta\phi_1-\delta\phi^*_{1})&=0\,.
\end{align}
In Appendix \ref{app:moduli} we fully characterize the solutions to these equations in an appropriate regime of couplings.  Our approach is to decompose the fields into spherical harmonics which gives us infinitely many decoupled, time-independent Schr\"{o}dinger-type equations.  We show that the equations for all of the harmonics with $\ell > 1$ do not have normalizable solutions; we then solve the remaining $\ell = 0,1$ equations through a combination of analytics and numerics.  The solutions to the $\ell = 0,1$ equations reveal that $\delta \phi$ is an infinitesimal spacetime translation of the original solution $\Phi_{\omega + \delta \omega}$.

Since according to~\eqref{E:perturbedaround} our perturbations are around the Q-cloud solution $\Phi_{\omega + \delta \omega}$, there is implicitly another 1-parameter family of perturbations corresponding to changing the $\omega$ of the Q-cloud solutions.  This brings us to five parameters in total.  Within this five parameter family, no two periodic orbits overlap.  Repeating the entire analysis starting with $\Phi_{- \omega}$, we simply find the time-reversed solutions of our previous analysis.  Since the image of the solutions and their time reversals coincide, we obtain again the same moduli space $\mathcal{O}_{\text{Q-cloud}}$.  This means that the moduli space is doubly covered by orbits; this is due to the fact that our relativistic scalar field theory has $t \to - t$ symmetry.  This double covering is already accounted for in~\eqref{E:mainresult1} with a multiplicative factor of $2$.

\section{Discussion}
\label{sec:discuss}

In this work we have generalized Bogomolny's analysis~\cite{bogomolny1988smoothed} of Heller's quantum scars~\cite{heller1984bound} from the setting of ordinary quantum mechanics to quantum field theory.  We focused on the case of scalar fields, and complex scalars fields in particular, for which we could find explicit periodic solitons in the form of Q-clouds which furnish the scar formula.

We emphasize that the Q-cloud solutions, which contribute quantum scars, are classically unstable; it is unusual to have unstable solutions which contribute semiclassically to non-decay processes. In his original work on Q-clouds~\cite{alford1988q}, Alford expressed this theoretical prior, describing Q-clouds as being ``of little concern to quantum field theorists'' due to their instability.  Our work suggests that this assertion is incorrect.

Going forward, it would be desirable to find other field-theoretic examples of quantum scars, or variants such as perturbation-induced quantum scars~\cite{keski2017controllable, keski2019effects, keski2019quantum}, along the lines of our analysis.  Perhaps it is possible to find examples where the functional determinants are computable, even if only approximately.  The evaluation of these determinants will require some regularization scheme, as is usual in quantum field theory; the results will have physical import for ascertaining the enhancement or suppression of quantum scars in the field-theoretic setting.  It may be valuable to find a semiclassical explanation of the putative `scar states' which were recently discovered numerically in~\cite{delacretaz2022thermalization} in an interacting scalar field theory.  While we have focused on scalar fields for simplicity, our analysis should generalize to other field contents.  Along these lines, there a number of examples of very nearly-periodic orbits called oscillons~\cite{kudryavtsev1975solitonlike,bogolyubsky1976lifetime,gleiser1993pseudostable,copeland1995oscillons},  with various field contents.  Our scar formula also generalizes to the nearly-periodic setting, and so would suggest that oscillons may imprint themselves on bands of energy eigenstates.  In fact, the recent work~\cite{levkov2022effective} shows that many different kinds of oscillons have an effective field theory description in terms of scalar field Q-balls and Q-clouds, so perhaps the effective field theory analysis could be combined with our analysis in the present paper.

It would be particularly interesting to consider quantum scars in quantum field theories with discrete spectra (e.g.~by considering field theories on compact manifolds).  In this context, a projector onto an energy band would contain a finite number of eigenstates, and so if the projector is scarred then the constituent eigenstates would also exhibit some degree of scarring.

As mentioned in the introduction, a motivating example for our work is the ``quantum many-body scars'' of~\cite{turner2018weak,turner2018quantum,choi2019emergent,ho2019periodic, bernien2017probing}.  To understand whether or not these are quantum scars in the original sense of Heller~\cite{heller1984bound}, we generalized Heller and Bogomolny's style of analysis to quantum field theory.  This brings us closer to understanding the recent ``quantum many-body scars'' which occur in Rydberg atom arrays, described by the PXP spin chain and its variants.  If we could formulate a version of the PXP model in the continuum, say via some IR coarse-graining or collective field theory approach, then we could attempt to directly apply the techniques in the present paper.  While it may be possible to apply our methods to the spin chain model directly (i.e.~instead of working in the continuum), the classical interpretation of putative periodic orbits may be less clear.

Another application of our results would be to gravity, and in particular AdS/CFT.  One might suspect that periodic orbits (e.g.~involving particles orbiting black holes) on the gravity side of the correspondence would indicate the presence of scarred eigenstates on the CFT side.  There have been initial explorations in~\cite{Dodelson:2022eiz, maxfield2022holographic}, chiefly using CFT techniques.  It would be interesting to use our path integral methods to gain a more refined semiclassical understanding.

\vspace*{6pt}
\noindent {\bf Acknowledgments.}\quad 
We thank Soonwon Choi, Luca Delacretaz, Melissa Franklin, Eric Heller, Felipe Hern\'{a}ndez, Daniel Jafferis, Eric Heller, Misha Lukin, Joonas Keski-Rahkonen, Semon Rezchkikov, and Frank Wilczek for valuable discussions. JC is supported by a Junior Fellowship from the Harvard Society of Fellows, the Black Hole Initiative, as well as in part by the Department of Energy under grant {DE}-{SC0007870}.

\bibliographystyle{unsrtnat}
\bibliography{refs}

\pagebreak
\onecolumngrid
\appendix

\section{Review of Bogomolny's formula}
\label{sec:bogomolny}
Here we review Bogomolny's scar formula in few-body quantum mechanics~\cite{bogomolny1988smoothed}, and provide a comprehensive but self-contained derivation. We obtain the scar formula via a series of stationary phase approximations: first we use the expression for the Van Vleck propagator, or semiclassical Green's function. Then we Fourier transform from time to energy variables, performing another stationary phase approximation in the process. Finally, we perform both position and energy averaging to arrive at Bogomolny's scar formula. A detailed derivation of the Van Vleck propagator and Gutzwiller trace formula can be found in textbooks like~\citeapp{haake2010q, ChaosBook}, and we utilize aspects of these analyzes.

\subsection{Van Vleck Propagator in ordinary quantum mechanics}

The semiclassical Van Vleck propagator takes the form of a sum over all classical paths between $\textbf{q}^A$ and $\textbf{q}^B$, where $\textbf{q}^A$ is the initial endpoint of the path and $\textbf{q}^B$ is the final endpoint. For $d$ degrees of freedom,
\begin{equation}
\label{E:quantumprop1}
    \langle \textbf{q}^B | U(t) | \textbf{q}^A \rangle \approx \sum_{\text{paths }c}\frac{1}{\sqrt{(2\pi i\hbar)^d}}\left|\det\left(\frac{\partial^2 S_c(\textbf{z}^A,\textbf{z}^B,t)}{\partial \textbf{z}^A\partial \textbf{z}^B}\right)\right|_{\substack{\textbf{z}^A = \textbf{q}^A \\ \textbf{z}^B = \textbf{q}^B}}^{1/2}\exp\left[\frac{i}{\hbar}S_c(\textbf{q}^A,\textbf{q}^B,t)-i\nu_c\frac{\pi}{2}\right].
\end{equation}
Here $S_c(\textbf{q}^A,\textbf{q}^B,t)$ denotes the action with endpoints $\textbf{q}^A$, $\textbf{q}^B$ evaluated on the classical path $c$.  The reason for the $c$ subscript is that $S_c(\textbf{q}^A,\textbf{q}^B,t)$ can be multivalued on account of having multiple classical solutions which go from $\textbf{q}^A$ to $\textbf{q}^B$ in a time $t$.  As such, we can regard the subscript $c$ as indicating a choice of branch of $S_c(\textbf{q},\textbf{q}',t)$, namely the branch for which the classical path labelled by $c$ is assigned the correct action.  Also, $\nu_c$ is the Maslov index, which explicitly keeps track of the number of negative eigenvalues of $\partial^2 S_c/(\partial \textbf{q}^A\partial \textbf{q}^B)$.

Next we transform our Green's function from time to energy variables,
\begin{equation}
\label{E:quantumGreens1}
    G(\textbf{q}^A, \textbf{q}^B, E)=\frac{1}{i\hbar}\int_0^{\infty}\!\! dt\,e^{iEt/\hbar}\langle \textbf{q}^B | U(t) | \textbf{q}^A \rangle.
\end{equation}
Plugging~\eqref{E:quantumprop1} into~\eqref{E:quantumGreens1} and taking the stationary phase approximation in $t$ forces
\begin{equation}
    E+\frac{\partial S_c(\textbf{q}^A,\textbf{q}^B,t)}{\partial t}=0\,.
\end{equation}
Here $\frac{\partial S_c(\textbf{q}^A,\textbf{q}^B,t)}{\partial t} = -H(\textbf{q}^A, \textbf{q}^B, t)$ is the energy of the classical trajectory beginning at $\textbf{q}^A$ at time zero and ending at $\textbf{q}^B$ at time $t$.  We can invert this to solve for $t$; we call the solution $t_c$.  Here $t_c = t_c(\textbf{q}^A, \textbf{q}^B, E)$ is the time it takes to traverse a classical solution beginning at $\textbf{q}^A$ and ending at $\textbf{q}^B$ having energy $E$, where the solution is restricted to the $c$ branch of Hamilton's principal function.  Performing a stationary phase approximation in $t$ for the Green's function, we find
\begin{align}
    G(\textbf{q}^A, \textbf{q}^B, E)&\approx\frac{1}{i\hbar} \sum_{\text{paths }c}\frac{1}{\sqrt{(2\pi i\hbar)^{d-1}}}\left|\det\left(\frac{\partial^2 S_c(\textbf{q}^A,\textbf{q}^B,t)}{\partial t^2}\right)\right|_{t = t_c}^{-1/2}\left|\det\left(\frac{\partial^2 S_c(\textbf{z}^A,\textbf{z}^B,t_c)}{\partial \textbf{z}^A\partial \textbf{z}^B}\right)\right|_{\substack{\textbf{z}^A = \textbf{q}^A \\ \textbf{z}^B = \textbf{q}^B}}^{1/2} \nonumber \\
    & \qquad \qquad \qquad \qquad \qquad \qquad \qquad \qquad \qquad \qquad \qquad \qquad \times \exp\left[\frac{i}{\hbar}(S_c(\textbf{q}^A,\textbf{q}^B,t_c)+E t_c) -i\nu_c \frac{\pi}{2}\right]\,,
\end{align}
where now $\nu_c$ accounts for a possible additional phase.  Next, we perform the Legendre transform from $t$ to $E$ coordinates,
\begin{equation}
    S_c(\textbf{q}^A, \textbf{q}^B, E)=S_c(\textbf{q}^A,\textbf{q}^B,t_c(\textbf{q}^A, \textbf{q}^B,E))+E \,t_c(\textbf{q}^A, \textbf{q}^B, E)\,,
\end{equation}
to obtain
\begin{align}
    G(\textbf{q}^A, \textbf{q}^B, E)&\approx\frac{1}{i\hbar} \sum_{\text{paths }c}\frac{1}{\sqrt{(2\pi i\hbar)^{d-1}}}\left|\det\left(\frac{\partial^2 S_c(\textbf{q}^A,\textbf{q}^B,t)}{\partial t^2}\right)\right|_{t = t_c}^{-1/2}\left|\det\left(\frac{\partial^2 S_c(\textbf{z}^A,\textbf{z}^B,t_c)}{\partial \textbf{z}^A\partial \textbf{z}^B}\right)\right|_{\substack{\textbf{z}^A = \textbf{q}^A \\ \textbf{z}^B = \textbf{q}^B}}^{1/2}\nonumber \\
    & \qquad \qquad \qquad \qquad \qquad \qquad \qquad \qquad \qquad \qquad \qquad \qquad \qquad \times \exp\left[\frac{i}{\hbar}S_c(\textbf{q}^A,\textbf{q}^B,E) -i\nu_c \frac{\pi}{2}\right]\,.
\end{align}
Above, $S_c(\textbf{q}^A, \textbf{q}^B, E) = \int_0^{t_c} dt \,\textbf{P}_c(t) \cdot \dot{\textbf{Q}}_c(t)$ is the `abbreviated action' of a classical solution $(\textbf{Q}_c(t), \textbf{P}_c(t))$ with energy $E$, starting at $\textbf{q}^A$ and ending at $\textbf{q}^B$.  Here $t_c = t_c(\textbf{q}^A, \textbf{q}^B, E)$ is the time extent of the orbit.

For our next simplification, we rewrite the amplitude factor using the following matrix,
\begin{equation}
    \left(
    \begin{array}{cc}
    \frac{\partial^2 S_c}{\partial\textbf{z}^A\,\partial\textbf{z}^B} & \frac{\partial^2 S_c}{\partial\textbf{z}^A\,\partial E'}\\
    \frac{\partial^2 S_c}{\partial\textbf{z}^B\,\partial E'} & \frac{\partial^2 S_c}{\partial E'^2}
    \end{array}
    \right)\Bigg|_{\substack{\textbf{z}^A = \textbf{q}^A \\ \textbf{z}^B = \textbf{q}^B \\ E' = E\,\,}} =
    \left(
    \begin{array}{cc}
    -\frac{\partial\textbf{p}^A}{\partial\textbf{z}^B} & -\frac{\partial\textbf{p}^A}{\partial E'}\\
    \frac{\partial t_c}{\partial\textbf{z}^B} & \frac{\partial t_c}{\partial E'}
    \end{array}
    \right)\Bigg|_{\substack{\textbf{z}^A = \textbf{q}^A \\ \textbf{z}^B = \textbf{q}^B \\ E' = E\,\,}}\,,
\end{equation}
whose determinant corresponds to the change of variables from $(-\textbf{p}^A, t)$ to $(\textbf{q}^B, E)$. By the chain rule for determinants, the Jacobian for the change of variables can be rewritten as
\begin{align}
    -\det\left(\frac{\partial(\textbf{p}^A, t)}{\partial(\textbf{z}^B, E')}\right)\Bigg|_{\substack{\textbf{z}^A = \textbf{q}^A \\ \textbf{z}^B = \textbf{q}^B\\ E' = E\,\,}}&=-\det\left(\frac{\partial(\textbf{p}^A, t)}{\partial(\textbf{z}^B, t)}\frac{\partial(\textbf{z}^B, t)}{\partial(\textbf{z}^B, E')}\right)\Bigg|_{\substack{\textbf{z}^A = \textbf{q}^A \\ \textbf{z}^B = \textbf{q}^B\\ E' = E\,\,}}
    =\left(\det\frac{\partial\textbf{p}^A}{\partial\textbf{q}^B}\right)\Bigg|_{\substack{\textbf{z}^A = \textbf{q}^A \\ \textbf{z}^B = \textbf{q}^B}}\left(\frac{\partial^2 S_c}{\partial t^2}\right)^{-1}\Bigg|_{t=t_c}\,.
\end{align}
Next we use the fact that $H(\textbf{q}, \textbf{p})=E$ is fixed to see that
\begin{align}
\label{E:qAdot1}
   \frac{\partial}{\partial\textbf{z}^B}\,H(\textbf{z}^A, \textbf{p}^A(\textbf{z}^A, \textbf{z}^B, E))\Bigg|_{\substack{\textbf{z}^A = \textbf{q}^A \\ \textbf{z}^B = \textbf{q}^B}}&=0= \frac{\partial \textbf{p}^A}{\partial \textbf{z}^B} \cdot \frac{\partial H}{\partial \textbf{p}^A}\Bigg|_{\substack{\textbf{z}^A = \textbf{q}^A \\ \textbf{z}^B = \textbf{q}^B}}\quad \Longrightarrow \quad \frac{\partial^2 S_c}{\partial \textbf{z}^B\partial \textbf{z}^A}\Bigg|_{\substack{\textbf{z}^A = \textbf{q}^A \\ \textbf{z}^B = \textbf{q}^B}} \cdot \dot{\textbf{q}}^A=0 \\
   \frac{\partial}{\partial\textbf{z}^A}\,H(\textbf{z}^B, \textbf{p}^B(\textbf{z}^A, \textbf{z}^B, E))\Bigg|_{\substack{\textbf{z}^A = \textbf{q}^A \\ \textbf{z}^B = \textbf{q}^B}}&=0=\frac{\partial H}{\partial \textbf{p}^B} \cdot \frac{\partial \textbf{p}^B}{\partial \textbf{z}^A}\Bigg|_{\substack{\textbf{z}^A = \textbf{q}^A \\ \textbf{z}^B = \textbf{q}^B}}\quad \Longrightarrow \quad \dot{\textbf{q}}^B \cdot \frac{\partial^2 S_c}{\partial \textbf{z}^B\partial \textbf{z}^A}\Bigg|_{\substack{\textbf{z}^A = \textbf{q}^A \\ \textbf{z}^B = \textbf{q}^B}}=0\,.
\end{align}
It is convenient to decompose fluctuations around $\textbf{q}^A$ and $\textbf{q}^B$ in the following manner:  using similar notation as above, consider a classical trajectory $\textbf{Q}_c(t)$, indexed by $c$, which goes from $\textbf{q}^A$ to $\textbf{q}^B$ in a time $t$.  We can decompose a fluctuation $\delta \textbf{q}^A$ around $\textbf{q}^A$ as $\delta \textbf{q}_\parallel^A + \delta \textbf{q}_\perp^A$, where $\delta \textbf{q}_\parallel^A$ is along the classical trajectory and $\delta \textbf{q}_\perp^A$ is orthogonal to it.  We adopt a similar notation for fluctuations around $\textbf{q}^B$.
With this notation at hand, note that the $\dot{\textbf{q}}^A$ on the far-right in~\eqref{E:qAdot1} is in fact $\textbf{Q}_c'(0) =: \dot{\textbf{q}}_c^A$, which is parallel to the orbit.  As such,
\begin{equation}
    0 = \frac{\partial^2 S_c(\textbf{z}^A, \textbf{z}^B, E)}{\partial \textbf{z}^B\partial \textbf{z}^A}\Bigg|_{\substack{\textbf{z}^A = \textbf{q}^A \\ \textbf{z}^B = \textbf{q}^B}}\cdot \dot{\textbf{q}}^A_c = \frac{\partial^2 S_c(\textbf{z}^A, \textbf{z}^B, E)}{\partial \textbf{z}^B\partial \textbf{z}_\parallel^A}\Bigg|_{\substack{\textbf{z}^A = \textbf{q}^A \\ \textbf{z}^B = \textbf{q}^B}}\,\|\dot{\textbf{q}}^A_c\|_2 \quad \Longrightarrow \quad \frac{\partial^2 S_c(\textbf{z}^A, \textbf{z}^B, E)}{\partial \textbf{z}^B\partial \textbf{z}_\parallel^A}\Bigg|_{\substack{\textbf{z}^A = \textbf{q}^A \\ \textbf{z}^B = \textbf{q}^B}} = 0\,.
\end{equation}
Similarly defining $\textbf{Q}_c'(t) =: \dot{\textbf{q}}_c^B$ we have
\begin{equation}
    0 = \dot{\textbf{q}}^B_c \cdot \frac{\partial^2 S_c(\textbf{z}^A, \textbf{z}^B, E)}{\partial \textbf{z}^B\partial \textbf{z}^A}\Bigg|_{\substack{\textbf{z}^A = \textbf{q}^A \\ \textbf{z}^B = \textbf{q}^B}} = \|\dot{\textbf{q}}^B_c \|_2 \,\frac{\partial^2 S_c(\textbf{z}^A, \textbf{z}^B, E)}{\partial \textbf{z}_\parallel^B\partial \textbf{z}^A}\Bigg|_{\substack{\textbf{z}^A = \textbf{q}^A \\ \textbf{z}^B = \textbf{q}^B}} \quad \Longrightarrow \quad \frac{\partial^2 S_c(\textbf{z}^A, \textbf{z}^B, E)}{\partial \textbf{z}_\parallel^B\partial \textbf{z}^A}\Bigg|_{\substack{\textbf{z}^A = \textbf{q}^A \\ \textbf{z}^B = \textbf{q}^B}} = 0\,.
\end{equation}
These identities allow us to rewrite the determinant as
\begin{equation}
    \det\left(
    \begin{array}{ccc}
    0 & 0 & \frac{\partial^2 S_c}{\partial\textbf{z}^A_\parallel\partial E'}\\
    0 & \frac{\partial^2 S_c}{\partial\textbf{z}^A_\perp\partial \textbf{z}^B_\perp} & \frac{\partial^2 S_c}{\partial\textbf{z}^A_\perp\partial E'}\\
    \frac{\partial^2 S_c}{\partial\textbf{z}^B_\parallel\partial E'} & \frac{\partial^2 S_c}{\partial\textbf{z}^B_\perp\partial E'} & \frac{\partial ^2 S_c}{\partial E'^2}\\
    \end{array}
    \right)_{\substack{\textbf{z}^A = \textbf{q}^A \\ \textbf{z}^B = \textbf{q}^B\\ E' = E\,\,}}=
    - \left(\frac{\partial t_c}{\partial \textbf{z}_\parallel^A}\, \frac{\partial t_c}{\partial \textbf{z}_\parallel^B}\,\det\left(\frac{\partial^2 S_c}{\partial\textbf{z}^A_\perp\partial \textbf{z}^B_\perp}\right)\,\right)_{\substack{\textbf{z}^A = \textbf{q}^A \\ \textbf{z}^B = \textbf{q}^B}},
\end{equation}
where we used the fact that
\begin{equation}
    \frac{\partial^2 S_c}{\partial\textbf{z}^A_\parallel\partial E'}=\frac{\partial t_c}{\partial\textbf{z}^A_\parallel}\,, \qquad \frac{\partial^2 S_c}{\partial\textbf{z}^B_\parallel\partial E'}=\frac{\partial t_c}{\partial\textbf{z}^B_\parallel}\,.
\end{equation}
To understand these expressions, note that for any function $F = F(\textbf{z}_A, \textbf{z}_B)$ we have
\begin{equation}
\left.\frac{\partial F}{\partial \textbf{z}_\parallel^A}\right|_{\substack{\textbf{z}^A = \textbf{q}^A \\ \textbf{z}^B = \textbf{q}^B}} := \frac{\dot{\textbf{q}}_c^A}{| \dot{\textbf{q}}_c^A |_2^{1/2}} \cdot \left.\frac{\partial F}{\partial \textbf{z}^A}\right|_{\substack{\textbf{z}^A = \textbf{q}^A \\ \textbf{z}^B = \textbf{q}^B}}\,, \qquad \left.\frac{\partial F}{\partial \textbf{z}_\parallel^B}\right|_{\substack{\textbf{z}^A = \textbf{q}^A \\ \textbf{z}^B = \textbf{q}^B}} := \frac{\dot{\textbf{q}}_c^B}{| \dot{\textbf{q}}_c^B |_2^{1/2}} \cdot \left.\frac{\partial F}{\partial \textbf{z}^B}\right|_{\substack{\textbf{z}^A = \textbf{q}^A \\ \textbf{z}^B = \textbf{q}^B}}\,.
\end{equation}
Accordingly, we have
\begin{equation}
\left.\frac{\partial t_c}{\partial \textbf{z}_\parallel^A}\right|_{\substack{\textbf{z}^A = \textbf{q}^A \\ \textbf{z}^B = \textbf{q}^B}} = \frac{1}{| \dot{\textbf{q}}_c^A|_2^{1/2}}\,, \qquad \left.\frac{\partial t_c}{\partial \textbf{z}_\parallel^B}\right|_{\substack{\textbf{z}^A = \textbf{q}^A \\ \textbf{z}^B = \textbf{q}^B}} = \frac{1}{| \dot{\textbf{q}}_c^B|_2^{1/2}}\,.
\end{equation}
Thus the Green's function may be rewritten as
\begin{align}
\label{eq:energypropagator}
    G(\textbf{q}^A, \textbf{q}^B, E)\approx\frac{1}{i\hbar} \sum_{\text{paths }c}\frac{1}{\sqrt{(2\pi i\hbar)^{d-1}}}\frac{1}{|\dot{\textbf{q}}_c^A|^{1/2}|\dot{\textbf{q}}_c^B|^{1/2}}\left|\det\left(\frac{\partial^2 S_c(\textbf{z}^A,\textbf{z}^B,E)}{\partial \textbf{z}_\perp^A\partial \textbf{z}_\perp^B}\right)\right|_{\substack{\textbf{z}^A = \textbf{q}^A \\ \textbf{z}^B = \textbf{q}^B}}^{1/2}\exp\left[\frac{i}{\hbar}S_c(\textbf{q}^A,\textbf{q}^B,E)-i\nu_c \frac{\pi}{2}\right],
\end{align}
where we used $\frac{\partial S_c(\textbf{z}^A,\textbf{q}^B,t)}{\partial \textbf{z}_\perp^A}\Big|_{\substack{\textbf{z}^A = \textbf{q}^A }}=\frac{\partial S_c(\textbf{q}^A,\textbf{z}^B,E)}{\partial \textbf{z}_\perp^A}\Big|_{\textbf{z}^A = \textbf{q}^A}$ as well as $\frac{\partial S_c(\textbf{q}^A,\textbf{z}^B,t)}{\partial \textbf{z}_\perp^B}\Big|_{\substack{\textbf{z}^B = \textbf{q}^B }}=\frac{\partial S_c(\textbf{q}^A,\textbf{z}^B,E)}{\partial \textbf{z}_\perp^B}\Big|_{\textbf{z}^B = \textbf{q}^B}$.  We note that our stationary phase approximation in $t$ given by~\eqref{eq:energypropagator} is only valid for $\textbf{q}^A \not = \textbf{q}^B$; in particular, there is an extra term proportional to $\delta(\textbf{q}^A - \textbf{q}^B)$ which vanishes for $\textbf{q}^A \not = \textbf{q}^B$. This means we have to be more careful about our approximations below when we compute the trace of the Green's function wherein we set $\textbf{q}^A = \textbf{q}^B$.

\subsection{Trace of the Van Vleck Propagator}

Let us commence with taking the trace over the endpoints of the Van Vleck propagator.
The starting point is the following approximation, which is valid in the semiclassical regime~\cite{copson2004asymptotic, haake2010q}:
\begin{equation}
\label{E:asymptoticapprox1}
    \int_{t_0}^{\infty}dt\,A(t)e^{iB(t)/\hbar}\approx \frac{i A(t_0)e^{i B(t_0)/\hbar}}{B'(t_0)} + [\text{stationary phase contributions}]\,.
\end{equation}
We can apply this to the trace of the Green's function, and write the result as
\begin{equation}
\int d^d\textbf{q} \,G(\textbf{q},\textbf{q},E) \approx \int d^d \textbf{q}\,G_0(\textbf{q},\textbf{q},E) + \int d^d \textbf{q}\,G_{\text{osc}}(\textbf{q},\textbf{q},E) = G_0(E) + G_{\text{osc}}(E)\,,
\end{equation}
where $G_0(E)$ corresponds to the first term on the right-hand side of~\eqref{E:asymptoticapprox1}, and $G_{\text{osc}}(E)$ corresponds to the stationary phase terms.  For $G_0(E)$ we will take the $t_0 \to 0^+$ limit of
\begin{align}
\frac{1}{i\hbar}\int_{t_0}^{\infty} dt \int d^d\textbf{q}\,\langle \textbf{q}|e^{-i(H-E)t/\hbar}|\textbf{q} \rangle
&\approx \int d^d\textbf{q}\frac{\langle \textbf{q}|e^{-i(H-E)t_0/\hbar}|\textbf{q}\rangle}{E-H(\textbf{q},\textbf{q},t_0)}\\
&=\int d^d\textbf{q}\,d^d\textbf{q}'\,\frac{\delta(\textbf{q}-\textbf{q}')}{E-H(\textbf{q},\textbf{q}', t_0)}\langle \textbf{q}|e^{-i(H-E)t_0/\hbar}|\textbf{q}'\rangle\\
&=\int \frac{d^d\textbf{q}\,d^d\textbf{p}}{(2\pi\hbar)^d}\,d^d\textbf{q}'\,\frac{e^{-i\textbf{p}\cdot(\textbf{q}-\textbf{q}')/\hbar}}{E-H(\textbf{q},\textbf{q}', t_0)}\langle \textbf{q}|e^{-i(H-E)t_0/\hbar}|\textbf{q}'\rangle\\
&=\int \frac{d^d\textbf{q}\,d^d\textbf{p}}{(2\pi\hbar)^d}\,d^d\textbf{q}'\,\frac{e^{-i\textbf{p}\cdot(\textbf{q}-\textbf{q}')/\hbar}}{E-H(\textbf{q},\textbf{q}', t_0)} \, \mathcal{N}_{t_0} \, e^{i (S(\textbf{q},\textbf{q}',t_0) - E t_0)/\hbar}\,,
\end{align}
where $\mathcal{N}_{t_0}$ is a normalization depending on $t_0$.  The $\textbf{q}'$ integral admits a stationary phase approximation, which sets $\textbf{p} = - \partial_{\textbf{q}'} S(\textbf{q},\textbf{q}', t_0)$.  This forces the variable $\textbf{p}$ to be identified with the momentum, and so the above approximately equals
\begin{equation}
\int \frac{d^d\textbf{q}\,d^d\textbf{p}}{(2\pi\hbar)^d}\,d^d\textbf{q}'\,\frac{e^{-i\textbf{p}\cdot(\textbf{q}-\textbf{q}')/\hbar}}{E-H(\textbf{q},\textbf{p})} \, \mathcal{N}_{t_0} \, e^{i (S(\textbf{q},\textbf{q}',t_0) - E t_0)/\hbar} = \int \frac{d^d\textbf{q}\,d^d\textbf{p}}{(2\pi\hbar)^d}\,d^d\textbf{q}'\,\frac{e^{-i\textbf{p}\cdot(\textbf{q}-\textbf{q}')/\hbar}}{E-H(\textbf{q},\textbf{p})} \, \langle \textbf{q}| e^{- i (H - E) t_0/\hbar} |\textbf{q}'\rangle\,.
\end{equation}
Finally taking the limit $t_0 \to 0^+$, we have $\langle \textbf{q}| e^{- i (H - E) t_0/\hbar} |\textbf{q}'\rangle \to \delta(\textbf{q}-\textbf{q}')$ and so the integral becomes
\begin{align}
\int \frac{d^d\textbf{q}\,d^d\textbf{p}}{(2\pi\hbar)^d}\,d^d\textbf{q}'\,\frac{e^{-i\textbf{p}\cdot(\textbf{q}-\textbf{q}')/\hbar}}{E-H(\textbf{q},\textbf{p})} \, \delta(\textbf{q}-\textbf{q}') &=\int \frac{d^d\textbf{q}\,d^d\textbf{p}}{(2\pi\hbar)^d}\,\frac{1}{E-H(\textbf{q},\textbf{p})}\,.
\end{align}
In summary, we have found
\begin{equation}
G_0(E) \approx \int \frac{d^d\textbf{q}\,d^d\textbf{p}}{(2\pi\hbar)^d}\,\frac{1}{E-H(\textbf{q},\textbf{p})}\,.
\end{equation}

Next we consider the $G_{\text{osc}}(E)$ term which comes from performing a stationary phase analysis of the trace of the Green's function.  We have
\begin{equation}
    \int d^d \textbf{q} \,G_{\text{osc}}(\textbf{q},\textbf{q},E) = \frac{1}{i\hbar} \sum_{\text{orbits }c}\int d^d\textbf{q}\,\frac{1}{\sqrt{(2\pi i\hbar)^{d-1}}}\frac{1}{|\dot{\textbf{q}}|}\left|\det\left(\frac{\partial^2 S_c(\textbf{q}^A,\textbf{q}^B,E)}{\partial \textbf{q}_\perp^A\partial \textbf{q}_\perp^B}\right)\right|^{1/2}_{\textbf{q}^A=\textbf{q}^B = \textbf{q}}\exp\left[\frac{i}{\hbar}S_c(\textbf{q},\textbf{q},E)-i\nu_c \frac{\pi}{2}\right].
\end{equation}
Since the initial and final points of the trajectory are taken to be $\textbf{q}$, we have set $\dot{\textbf{q}}_A = \dot{\textbf{q}}_B = \dot{\textbf{q}}$ in the formula.
Applying the stationary phase approximation in $\textbf{q}$ yields
\begin{equation}
    0=\left(\frac{\partial S_c(\textbf{q}^A, \textbf{q}^B, E)}{\partial \textbf{q}^A}+\frac{\partial S_c(\textbf{q}^A, \textbf{q}^B, E)}{\partial \textbf{q}^B}\right)_{\textbf{q}^A=\textbf{q}^B=\textbf{q}}=-\textbf{p}^A+\textbf{p}^B.
\end{equation}
Thus the stationary points belong to periodic orbits beginning and ending at $(\textbf{q}^A,\textbf{p}^A)=(\textbf{q}^B,\textbf{p}^B)$.  Suppose that $(\textbf{q}_c, \textbf{p}_c)$ is a point along a periodic orbit.  Then letting $\textbf{q} = \textbf{q}_c + \delta \textbf{q}_\perp$ be a small fluctuation about $\textbf{q}_c$ perpendicular to the periodic orbit, we have the expansion
\begin{equation}
    S_c(\textbf{q},\textbf{q},E)\approx \left[S_c(\textbf{q}_c,\textbf{q}_c,E)+\frac{1}{2}\,\delta \textbf{q}_\perp \cdot \left(\frac{\partial^2 S_c(\textbf{q}^A,\textbf{q}^B,E)}{\partial \textbf{q}_\perp^A\partial \textbf{q}_\perp^A}+2\frac{\partial^2 S_c(\textbf{q}^A,\textbf{q}^B,E)}{\partial \textbf{q}_\perp^A\partial \textbf{q}_\perp^B}+\frac{\partial^2 S_c(\textbf{q}^A,\textbf{q}^B,E)}{\partial \textbf{q}_\perp^B \partial \textbf{q}_\perp^B}\right)_{\textbf{q}^A=\textbf{q}^B=\textbf{q}_c} \cdot \delta \textbf{q}_\perp\right]\,.
\end{equation}
We then use this to write the trace of the Green's function as
\begin{align}
    &\int d^d\textbf{q}\,G_{\text{osc}}(\textbf{q}, \textbf{q}, E)\approx\frac{1}{i\hbar} \sum_{\text{orbits }c}\frac{1}{\sqrt{(2\pi i\hbar)^{d-1}}}\int_0^{T_c} dt\int d^{d-1} \delta \textbf{q}_\perp \left|\det\left(\frac{\partial^2 S_c(\textbf{q}^A,\textbf{q}^B,E)}{\partial \textbf{q}_\perp^A\partial \textbf{q}_\perp^B}\right)\right|^{1/2}_{\textbf{q}^A = \textbf{q}^B = \textbf{Q}_c(t)}\nonumber\\
    & \times\exp\left(\frac{i}{\hbar}\!\left[S_c(\textbf{Q}_c(t),\textbf{Q}_c(t),E)\!+\!\frac{1}{2}\,\delta \textbf{q}_\perp\!\cdot\!\left(\!\frac{\partial^2 S_c(\textbf{q}^A,\textbf{q}^B,E)}{\partial \textbf{q}_\perp^A\partial \textbf{q}_\perp^A}\!+\!2\,\frac{\partial^2 S_c(\textbf{q}^A,\textbf{q}^B,E)}{\partial \textbf{q}_\perp^A\partial \textbf{q}_\perp^B}\!+\!\frac{\partial^2 S_c(\textbf{q}^A,\textbf{q}^B,E)}{\partial \textbf{q}_\perp^B\partial \textbf{q}_\perp^B}\!\right)_{\textbf{q}^A=\textbf{q}^B=\textbf{Q}_c(t)} \!\!\!\cdot \delta \textbf{q}_\perp\right]-i\nu_c \frac{\pi}{2}\right),
\end{align}
where here we are summing over periodic orbits $c$ with period $T_c$.  Integrating over the fluctuations $\delta \textbf{q}_\perp$ perpendicular to the orbit $\textbf{Q}_c(t)$ at time $t$, we are left with
\begin{align}
\label{E:resultwithmonodromy1}
\int d^d\textbf{q}\,G_{\text{osc}}(\textbf{q}, \textbf{q}, E)\approx\frac{1}{i\hbar} \sum_{\text{orbits }c}\int_0^{T_c} dt \frac{1}{\sqrt{|\det(\textsf{M}(\textbf{Q}_c(t)) - \mathds{1})|}} \, \exp\left(\frac{i}{\hbar} S_c(\textbf{Q}_c(t),\textbf{Q}_c(t),E) - i \nu_c \frac{\pi}{2}\right)
\end{align}
where $\textsf{M} = \textsf{M}(\textbf{Q}_c(t))$ is the monodromy matrix for the closed orbit $\textbf{Q}_c(t)$~\cite{haake2010q}, and $\nu_c$ has been augmented to accommodate for the number of negative eigenvalues of the determinant prefactor.  A key property of the monodromy matrix for a closed orbit is that its spectrum is independent of where it is evaluated along the closed orbit, namely $\text{spec}(\textsf{M}(\textbf{Q}_c(t))) = \text{spec}(\textsf{M}(\textbf{Q}_c(t')))$ for any $t$ and $t'$.  Moreover, the term $S_c(\textbf{Q}_c(t),\textbf{Q}_c(t),E)$ appearing in the exponential in~\eqref{E:resultwithmonodromy1} likewise satisfies $S_c(\textbf{Q}_c(t),\textbf{Q}_c(t),E) = S_c(\textbf{Q}_c(t'),\textbf{Q}_c(t'),E)$, since the action only depends on the total trajectory.  We will often write the action as $S_c(\textbf{q}_c, \textbf{q}_c, E)$ where $\textbf{Q}(0) = \textbf{q}_c$ is taken to be a `representative' point on the closed orbit $\textbf{Q}_c(t)$.

In light of these above facts about time-dependencies, we observe that the integrand of the $t$ integral in~\eqref{E:resultwithmonodromy1} actually does not depend on $T$ at all; as such, we can perform the $t$ integral to find
\begin{equation}
\int d^d\textbf{q}\,G_{\text{osc}}(\textbf{q}, \textbf{q}, E)\approx\frac{1}{i\hbar} \sum_{\text{orbits }c} \frac{T_c}{\sqrt{|\det(\textsf{M}(\textbf{Q}_c) - \mathds{1})|}} \, \exp\left(\frac{i}{\hbar} S_c(\textbf{q}_c,\textbf{q}_c,E) - i \nu_c \frac{\pi}{2}\right)\,.
\end{equation}
Finally, adding $G_{\text{osc}}$ to $G_0$, we obtain
\begin{align}
\label{E:easiertrace1}
    &\int d^d\textbf{q}\,G(\textbf{q}, \textbf{q}, E)\approx\frac{1}{(2\pi\hbar)^d}\int d^d\textbf{q}\,d^d\textbf{p}\,\frac{1}{E-H(\textbf{q},\textbf{p})}+\frac{1}{i\hbar} \sum_{\text{orbits }c} \frac{T_c}{\sqrt{|\det(\textsf{M}(\textbf{Q}_c) - \mathds{1})|}} \, \exp\left(\frac{i}{\hbar} S_c(\textbf{q}_c,\textbf{q}_c,E) - i \nu_c \frac{\pi}{2}\right)\,.
\end{align}
This is the Gutzwiller trace formula.

\subsection{Bogomolny's scar formula}
\label{subsec:Bogomolnyderivation1}

The scar formula of Bogomolny~\cite{bogomolny1988smoothed}, which involves both position and energy averaging, ultimately computes the expression
\begin{equation}
\label{eq:smear}
\langle|\psi(\textbf{q})|^2\rangle_{E,\Delta}=\frac{\sum_n\langle|\psi_n(\textbf{q})|^2\rangle_{\Delta}\delta_{\varepsilon}(E-E_n)}{\sum_n\delta_{\varepsilon}(E-E_n)}\,.
\end{equation}
Here the $\psi_n(\textbf{q})$ are energy eigenfunctions, the expression $\langle|\psi_n(\textbf{q})|^2\rangle_{\Delta}$ indicates position averaging over a window of size $\Delta$, and
\begin{equation}
\delta_\varepsilon(x) := \frac{1}{\pi} \frac{\varepsilon}{\varepsilon^2 + x^2}
\end{equation}
is a $\delta$-function which is smeared out to have width $\varepsilon$.  Thus~\eqref{eq:smear} is an average over position-smeared eigenfunctions in the energy window $[E-\varepsilon/2, E+\varepsilon/2]$.

Our goal is to write $\langle|\psi(\textbf{q})|^2\rangle_{E,\Delta}$ in terms of the Green's function $G(\textbf{q}^A, \textbf{q}^B, E)$.  To see how to do this, we first observe a few properties of the Green's function.  To simplify the analysis slightly (and in a manner that will be appropriate for the field-theoretic setting), we assume that for any triple $\textbf{q}^A, \textbf{q}^B, E$ with $\textbf{q}^A \not = \textbf{q}^B$, there is only a single orbit starting at $\textbf{q}^A$ and ending at $\textbf{q}^B$ having energy $E$.  For $\textbf{q}^A = \textbf{q}^B$ and fixed $E$, we assume that if there are any non-trivial orbits (i.e.~not a fixed point of the dynamics) beginning and ending at $\textbf{q}^A = \textbf{q}^B$ with energy $E$, then there must be exactly two such orbits which are time-reverses of one another.  In particular, it is sufficient to assume that the classical system has time-reversal symmetry.  These conditions imply that there is a single branch of $S_c(\textbf{q}^A,\textbf{q}^B,E)$ which correctly reproduces the actions of all the classical trajectories.  As such, we will drop the $c$ subscript from $S_c(\textbf{q}^A,\textbf{q}^B,E)$ henceforth.  It is easy to reintroduce the multi-valued structure if needed.

First note that the Green's function solves the equation
\begin{equation}
    (E+i\varepsilon-H(\hat{\textbf{q}}_B,\hat{\textbf{p}}_B))\,G(\textbf{q}^A, \textbf{q}^B, E)=\delta(\textbf{q}^B-\textbf{q}^A)\,,
\end{equation}
and so the solution is
\begin{equation}
    G(\textbf{q}^A, \textbf{q}^B, E+i\varepsilon)=\sum_n\frac{\psi_n(\textbf{q}^A)\psi_n^*(\textbf{q}^B)}{E+i\varepsilon-E_n}\,.
\end{equation}
It follows immediately that
\begin{equation}
\label{E:goodnum}
    -\frac{1}{\pi}\,\text{Im}\,G(\textbf{q},\textbf{q},E+i\varepsilon) = \sum_n|\psi_n(\textbf{q})|^2\delta_{\varepsilon}(E-E_n)\,,
\end{equation}
and moreover the trace gives
\begin{equation}
\label{E:gooddenom}
    -\frac{1}{\pi}\,\text{Im}\int d^d\textbf{q}\,G(\textbf{q},\textbf{q},E+i\varepsilon) = \sum_n\delta_{\varepsilon}(E-E_n)\,.
\end{equation}
Putting aside for the moment our $\Delta$-spatial smearing,~\eqref{E:goodnum} and~\eqref{E:gooddenom} reproduce the numerator and denominator of~\eqref{eq:smear}.

We essentially already computed the denominator term $\sum_n\delta_{\varepsilon}(E-E_n)$ in the subsection above, albeit for $\varepsilon \to 0$.  Working instead at small but finite $\varepsilon$, we obtain
\begin{align}
\sum_n\delta_{\varepsilon}(E-E_n) &\approx  \frac{1}{(2\pi\hbar)^d}\int d^d\textbf{q}\,d^d\textbf{p}\,\frac{1}{E + i \varepsilon-H(\textbf{q},\textbf{p})}\nonumber \\
& \qquad \qquad \qquad +\frac{1}{i\hbar} \sum_{\text{orbits }c} \frac{T_c}{\sqrt{|\det(\textsf{M}(\textbf{Q}_c) - \mathds{1})|}} \, \exp\left(\frac{i}{\hbar} S(\textbf{q}_c,\textbf{q}_c,E) - i \nu_c \frac{\pi}{2}\right) \, \exp(- \varepsilon T_c/\hbar)\,.
\end{align}
Notice the presence of the $\exp(- \varepsilon T_c/\hbar)$ which suppresses orbits with large periods; this term was not present in~\eqref{E:easiertrace1} where $\varepsilon$ was zero.

We now turn to the numerator of (\ref{eq:smear}), which has a Gaussian smearing of width $\Delta$.  Suppose that $T_{\text{max}}$ is the period of the longest orbit with energy in $[E-\varepsilon/2, E + \varepsilon/2]$.  Moreover, suppose that our classical Hamiltonian $H(\textbf{q},\textbf{p})$ describes non-relativistic particles of $m$.  Then we take
\begin{equation}
\Delta = \sqrt{\frac{\hbar T_{\text{max}}}{m}} \left(\frac{\hbar}{E\,T_{\text{max}}}\right)^{\gamma} \quad \text{for any } \quad \frac{1}{4} < \gamma < \frac{1}{2}\,,
\end{equation}
which we will justify shortly.  A key assumption of our analysis is that we are in a semiclassical regime where $E\,T_{\text{max}}/\hbar$ is large.  The main role of the $\Delta$-spatial smearing is to pick out periodic orbits (or very nearly periodic orbits) and suppress the contribution of non-periodic orbits to the scar formula.  Explicitly, we compute
\begin{equation}
\label{E:explicitly1}
    \sum_n\langle|\psi_n(\textbf{q})|^2\rangle_{\Delta}\delta_{\varepsilon}(E-E_n)=-\frac{1}{\pi}\,\text{Im}\left\{\frac{1}{(2\pi \Delta^2)^{d/2}}\int d^d \textbf{z}\,G(\textbf{z},\textbf{z},E+i\varepsilon)\,e^{-\frac{1}{2\Delta^2} (\textbf{q}-\textbf{z})^2}\right\}.
\end{equation}
Let us consider the argument of the $\text{Im}\{\,\cdot\,\}$ and analyze the integral.  We have
\begin{align}
\label{E:RHStobound0}
&\frac{1}{(2\pi \Delta^2)^{d/2}}\int d^d \textbf{z}\,G(\textbf{z},\textbf{z},E+i\varepsilon)\,e^{-\frac{1}{2\Delta^2} (\textbf{q}-\textbf{z})^2} \nonumber \\
& \qquad \approx \frac{1}{(2\pi \Delta^2)^{d/2}}\frac{1}{(2\pi \hbar)^d}\int d^d \textbf{z}\, d^d \textbf{p}\,\frac{1}{E + i \varepsilon - H(\textbf{z},\textbf{p})}\,e^{-\frac{1}{2\Delta^2} (\textbf{q}-\textbf{z})^2} \nonumber \\
& \qquad \qquad + \frac{2}{(2\pi \Delta^2)^{d/2}} \frac{1}{i\hbar} \frac{1}{\sqrt{(2\pi i\hbar)^{d-1}}}\int d^d \textbf{z}\,\frac{1}{|\dot{\textbf{z}}|}\left|\det\left(\frac{\partial^2 S(\textbf{z}^A,\textbf{z}^B,E)}{\partial \textbf{z}_\perp^A\partial \textbf{z}_\perp^B}\right)\right|_{\textbf{z}_A = \textbf{z}_B = \textbf{z}}^{1/2}e^{\frac{i}{\hbar}S(\textbf{z},\textbf{z},E)-i\nu(\textbf{z},\textbf{z},E)\frac{\pi}{2} - \frac{1}{\hbar} \varepsilon T(\textbf{z},\textbf{z},E)}\,e^{-\frac{1}{2\Delta^2} (\textbf{q}-\textbf{z})^2}\,.
\end{align}
The term in the last line has a factor of $2$ to account for the time-reverse of orbits beginning and ending at the same $\textbf{z}$, with energy $E$.  We have also dropped multiplicative $(1 + O(\varepsilon/E))$ corrections on the right-hand side.  For each integral on the right-hand side, the only part of the integration domain which substantially contributes is
\begin{equation}
\|\textbf{q} - \textbf{z}\|_2 \lesssim \Delta\,.
\end{equation}
Since $\Delta < \sqrt{\frac{\hbar T_{\text{max}}}{m}} \left(\frac{\hbar}{E\,T_{\text{max}}}\right)^{1/4}$ for $E\,T_{\text{max}}/\hbar$ large, we can make the approximation
\begin{equation}
\frac{1}{(2\pi \Delta^2)^{d/2}}\frac{1}{(2\pi \hbar)^d}\int d^d \textbf{z}\, d^d \textbf{p}\,\frac{1}{E + i \varepsilon - H(\textbf{z},\textbf{p})}\,e^{-\frac{1}{2\Delta^2} (\textbf{q}-\textbf{z})^2} \approx \frac{1}{(2\pi \hbar)^d}\int d^d \textbf{p}\,\frac{1}{E + i \varepsilon - H(\textbf{q},\textbf{p})}\,.
\end{equation}

For the second integral on the right-hand side of~\eqref{E:RHStobound0}, we can make a similar approximation:
\begin{align}
&\frac{2}{(2\pi \Delta^2)^{d/2}} \frac{1}{i\hbar} \frac{1}{\sqrt{(2\pi i\hbar)^{d-1}}}\int d^d \textbf{z}\,\frac{1}{|\dot{\textbf{z}}|}\left|\det\left(\frac{\partial^2 S(\textbf{z}^A,\textbf{z}^B,E)}{\partial \textbf{z}_\perp^A\partial \textbf{z}_\perp^B}\right)\right|_{\textbf{z}_A = \textbf{z}_B = \textbf{z}}^{1/2}e^{\frac{i}{\hbar}S(\textbf{z},\textbf{z},E)-i\nu(\textbf{z},\textbf{z},E)\frac{\pi}{2} - \frac{1}{\hbar} \varepsilon T(\textbf{z},\textbf{z},E)}\,e^{-\frac{1}{2\Delta^2} (\textbf{q}-\textbf{z})^2} \nonumber \\
&\approx  \frac{2}{i\hbar} \frac{1}{\sqrt{(2\pi i\hbar)^{d-1}}} \frac{1}{|\dot{\textbf{q}}|}\left|\det\left(\frac{\partial^2 S(\textbf{q}^A,\textbf{q}^B,E)}{\partial \textbf{q}_\perp^A\partial \textbf{q}_\perp^B}\right)\right|_{\textbf{q}^A = \textbf{q}^B = \textbf{q}}^{1/2} e^{-i\nu(\textbf{q},\textbf{q},E)\frac{\pi}{2} - \frac{1}{\hbar} \varepsilon T(\textbf{q},\textbf{q},E)} \cdot \frac{1}{(2\pi \Delta^2)^{d/2}}\int d^d \textbf{z} \, e^{\frac{i}{\hbar}S(\textbf{z},\textbf{z},E)}\,e^{-\frac{1}{2\Delta^2} (\textbf{q}-\textbf{z})^2}\,.
\end{align}
Focusing on the residual integral on the right-hand side, it is natural to expand $S(\textbf{z},\textbf{z},E)$ around $\textbf{z} = \textbf{q}$.  We have
\begin{align}
& \frac{1}{(2\pi \Delta^2)^{d/2}}\int d^d \textbf{z} \, e^{\frac{i}{\hbar}S(\textbf{z},\textbf{z},E)}\,e^{-\frac{1}{2\Delta^2} (\textbf{q}-\textbf{z})^2} \nonumber \\
\approx\,& \frac{1}{(2\pi \Delta^2)^{d/2}} \int d^d \textbf{z} \, e^{- \frac{1}{2\Delta^2}(\textbf{q}-\textbf{z})^2 + \frac{i}{\hbar} \left(S(\textbf{q},\textbf{q},E) + (\textbf{z}-\textbf{q}) \cdot \left(\frac{\partial S(\textbf{z}^A, \textbf{z}^B, E)}{\partial \textbf{z}^A} + \frac{\partial S(\textbf{z}^A, \textbf{z}^B, E)}{\partial \textbf{z}^B} \right)_{\textbf{z}^A = \textbf{z}^B = \textbf{q}} + \frac{1}{2} (\textbf{z}-\textbf{q})\cdot \textbf{A}(\textbf{q}) \cdot (\textbf{z}-\textbf{q}) + \cdots\right)}\,,
\end{align}
where
\begin{equation}
\left(\frac{\partial S(\textbf{z}^A,\textbf{z}^B,E)}{\partial \textbf{z}^A}+\frac{\partial S(\textbf{z}^A,\textbf{z}^B,E)}{\partial \textbf{z}^B}\right)_{\textbf{z}^A=\textbf{z}^B=\textbf{q}} = -\textbf{p}^A + \textbf{p}^B
\end{equation}
is the difference between the initial and final momenta of the trajectory beginning and ending at $\textbf{q}$, and we define the $d \times d$ matrix $A_{ij}(\textbf{q})$ by
\begin{equation}
\label{E:firstAdef1}
A_{ij}(\textbf{q}):= \left(\frac{\partial^2 S(\textbf{z}^A,\textbf{z}^B,E)}{\partial z^A_i \partial z^A_j}+2\,\frac{\partial^2 S(\textbf{z}^A,\textbf{z}^B,E)}{\partial z^A_i \partial z^B_j}+\frac{\partial^2 S(\textbf{z}^A,\textbf{z}^B,E)}{\partial z^B_i \partial z^B_j}\right)_{\textbf{z}^A=\textbf{z}^B=\textbf{q}}\,.
\end{equation}
Defining
\begin{equation}
\label{E:forinstance1}
\| \textbf{A}(\textbf{q}) \| := \frac{m}{\hbar T_{\text{max}}} \sup_{\|\textbf{v}\|_2 = \sqrt{\hbar T_{\text{max}}/m}} \,|\textbf{v} \cdot \textbf{A}(\textbf{q})\cdot \textbf{v}|\,,
\end{equation}
we show in Appendix~\ref{app:boundedness} that we have
\begin{equation}
\| \textbf{A}(\textbf{q})\|  \lesssim \frac{m}{T_{\text{max}}} \sqrt{\frac{E\,T_{\text{max}}}{\hbar}}\,.
\end{equation}
Note that $\|\textbf{A}\| \propto \max\{|\lambda_{\text{max}}|, |\lambda_{\text{min}}|\}$, where $\lambda_{\text{max}}$ and $\lambda_{\text{min}}$ are the largest and smallest eigenvalues of $\textbf{A}$, respectively.  We assume that the norms of the higher-order derivatives of the abbreviated action evaluated on periodic orbits are similarly bounded by $\frac{m}{T_{\text{max}}} \sqrt{\frac{E\,T_{\text{max}}}{\hbar}}$.
Since we have imposed $\Delta < \sqrt{\frac{\hbar T_{\text{max}}}{m}} \left(\frac{\hbar}{E\,T_{\text{max}}}\right)^{1/4}$, the term
$\frac{i}{2\hbar} (\textbf{q}-\textbf{z})\cdot \textbf{A}(\textbf{q})\cdot (\textbf{q}-\textbf{z})$ as well as the higher-order terms contribute negligibly to the integral in the regime that $E\,T_{\text{max}}/\hbar$ is large.  Then we are left with
\begin{equation}
\approx \frac{1}{(2\pi \Delta^2)^{d/2}} \int d^d \textbf{z} \, e^{- \frac{1}{2\Delta^2}(\textbf{q}-\textbf{z})^2 + \frac{i}{\hbar} \left(S(\textbf{q},\textbf{q},E) + (\textbf{q}-\textbf{z}) \cdot (-\textbf{p}^A + \textbf{p}^B) \right)}\,.
\end{equation}
Performing the $\textbf{z}$-integral, we find
\begin{equation}
\label{E:QMinplaceof1}
\exp\!\left(\frac{i}{\hbar} S(\textbf{q},\textbf{q},E) - \frac{1}{2} \frac{\Delta^2}{\hbar^2} \left(\textbf{p}^B - \textbf{p}^A\right)^2 \right)\,.
\end{equation}
But this is only sizable when
\begin{equation}
\label{E:onlysizableQM1}
\|\textbf{p}^B - \textbf{p}^A\|_2 \lesssim \frac{\hbar}{\Delta}\,,
\end{equation}
which is a nontrivial bound since $\Delta > \sqrt{\frac{\hbar T_{\text{max}}}{m}} \left(\frac{\hbar}{E\,T_{\text{max}}}\right)^{1/2}$.  To unpack this, let us rewrite $\Delta$ as
\begin{equation}
\Delta = \frac{\hbar}{\overline{p}} \left(\frac{E\,T_{\text{max}}}{\hbar}\right)^{\frac{1}{2}- \gamma} \quad \text{for any } \quad \frac{1}{4} < \gamma < \frac{1}{2}\,,
\end{equation}
where
\begin{equation}
\overline{p}^2 := m \, E
\end{equation}
can be regarded as squared momentum associated with $E$.  Then we can rewrite~\eqref{E:onlysizableQM1} as
\begin{equation}
\label{E:onlysizableQM2}
\|\textbf{p}^B - \textbf{p}^A\|_2 \lesssim\,\, \overline{p}\,\left(\frac{\hbar}{E\,T_{\text{max}}}\right)^{\frac{1}{2} - \gamma}\,.
\end{equation}
In this regime, we see that~\eqref{E:QMinplaceof1} can be replaced with 
\begin{equation}
\label{E:QMhere1}
\exp\!\left(\frac{i}{\hbar} S(\textbf{q},\textbf{q},E)\right)\,.
\end{equation}
Note that there is an intermediate regime where $\|\textbf{p}^B - \textbf{p}^A\|_2 = c \, \frac{\hbar}{\Delta}$, where $c \sim 1$.  In this regime $\exp\left(-\frac{1}{2} \frac{\Delta^2}{\hbar^2} \left(\textbf{p}^B - \textbf{p}^A\right)^2 \right)$ only gives rise to moderate amount of decay.  We will not consider this regime in our analysis here, i.e.~we will assume $c \ll 1$ so that we can use~\eqref{E:QMhere1} instead of~\eqref{E:QMinplaceof1}.

Since the above formula assumes that $\|\textbf{p}^B - \textbf{p}^A\|_2 \lesssim \frac{\hbar}{\Delta}$ or equivalently
\begin{equation}
\label{E:ineqcondition1}
\left\| \left(\frac{\partial S(\textbf{z}^A,\textbf{z}^B,E)}{\partial \textbf{z}^A}+\frac{\partial S(\textbf{z}^A,\textbf{z}^B,E)}{\partial \textbf{z}^B}\right)_{\textbf{z}^A=\textbf{z}^B=\textbf{q}}\right\|_2 \lesssim \frac{\hbar}{\Delta}\,,
\end{equation}
it is natural to ask: for which $\textbf{q}$ does this inequality hold?  A partial answer is that if we let $\mathcal{O}$ be the union of the images of all \textit{exactly} periodic solutions of the classical equations of motion, then the inequality in~\eqref{E:ineqcondition1} occurs for $\textbf{q}$ in the vicinity of $\mathcal{O}$.  However, there can be other points $\textbf{q}$ which achieve the desired inequality but are not near $\mathcal{O}$.  If the domain of $\textbf{q}$ was compact, then we could simply make $\hbar$ small enough to avoid the latter possibility.  (In the billiards setting of Bogomolny~\cite{bogomolny1988smoothed}, this compactness condition holds.)  If the domain of $\textbf{q}$ is not compact, it could be the case that there are infinitely many points where the left-hand side of~\eqref{E:ineqcondition1} is arbitrarily small.  This is not allowed if the left-hand side of~\eqref{E:ineqcondition1} is uniformly bounded, but in general we do not have this guarantee.  Let us avoid this possibility for now and focus only on the former setting where we look at points near $\mathcal{O}$.

To proceed, we make some reasonable assumptions about $\mathcal{O}$.  Suppose that it is a disjoint union of smooth, connected manifolds of non-zero codimension, where the minimum over the pairwise distances between the manifolds is lower bounded by a non-zero constant.  Consider one of these manifolds which comprises a single connected component, calling it $\mathcal{M}$, with dimension $k$ where $1 \leq k < d$.  Let us zoom in on some $\textbf{q}$ in $\mathcal{M}$, so that locally $\mathcal{M}$ looks like a $k$-hyperplane.

Now we pick some $\textbf{q}$ near $\mathcal{M}$ for which (i) $\textbf{q} = \textbf{q}_c + \delta \textbf{q}$ where $\textbf{q}_c \in \mathcal{M}$, and (ii) $\|\delta \textbf{q}\|_2 \lesssim \frac{\hbar}{\Delta}$.  In fact, this implies the inequality in~\eqref{E:ineqcondition1} for the following reason.  We have
\begin{equation}
\label{E:approxeq1}
\left(\frac{\partial S(\textbf{z}^A,\textbf{z}^B,E)}{\partial \textbf{z}^A}+\frac{\partial S(\textbf{z}^A,\textbf{z}^B,E)}{\partial \textbf{z}^B}\right)_{\textbf{z}^A=\textbf{z}^B=\textbf{q}} \approx \delta \textbf{q}\cdot \textbf{A}(\textbf{q}_c)
\end{equation}
because $\left(\frac{\partial S(\textbf{z}^A,\textbf{z}^B,E)}{\partial \textbf{z}^A}+\frac{\partial S(\textbf{z}^A,\textbf{z}^B,E)}{\partial \textbf{z}^B}\right)_{\textbf{z}^A=\textbf{z}^B=\textbf{q}_c} = 0$, and since the higher derivatives of $S$ at $\textbf{q}_c$ are tensorially contracted with $\delta \textbf{q}$'s which leads to suppression in the semiclassical regime.  The norm of the right-hand side is upper bounded by $\|\delta \textbf{q}\|_2 \, \|\textbf{A}(\textbf{q}_c)\| \lesssim \overline{p} \left(\frac{\hbar}{E\,T_{\text{max}}}\right)^{1/4}$ because $ \|\textbf{A}(\textbf{q}_c)\| \leq \frac{m}{T_{\text{max}}} \sqrt{\frac{E\,T_{\text{max}}}{\hbar}}$.  But then~\eqref{E:approxeq1} implies
\begin{equation}
\left\| \left(\frac{\partial S(\textbf{z}^A,\textbf{z}^B,E)}{\partial \textbf{z}^A}+\frac{\partial S(\textbf{z}^A,\textbf{z}^B,E)}{\partial \textbf{z}^B}\right)_{\textbf{z}^A=\textbf{z}^B=\textbf{q}} \right\|_2 \lesssim \frac{\hbar}{\Delta}\,,
\end{equation}
which indeed reproduces the inequality in~\eqref{E:ineqcondition1} as claimed.

In the aforementioned regime of (i) and (ii), we have
\begin{equation}
\label{E:nearby1}
\exp\!\left(\frac{i}{\hbar} S(\textbf{q},\textbf{q},E)\right) \approx \exp\!\left(\frac{i}{\hbar}S(\textbf{q}_c, \textbf{q}_c, E) + \frac{i}{2\hbar}\,\delta \textbf{q}\cdot \textbf{A}(\textbf{q}_c)\cdot\delta\textbf{q}\right)
\end{equation}
where the higher-order terms are dropped since $\| \delta \textbf{q}\|_2 \lesssim \frac{\hbar}{\Delta}$.  Let us make a few observations about the above equation.  First, we observe that since the gradient $\left(\frac{\partial S(\textbf{z}^A,\textbf{z}^B,E)}{\partial \textbf{z}^A}+\frac{\partial S(\textbf{z}^A,\textbf{z}^B,E)}{\partial \textbf{z}^B}\right)_{\textbf{z}^A=\textbf{z}^B=\textbf{q}}$ is constant (i.e.~zero) along $\mathcal{M}$, tangent vectors to $\mathcal{M}$ are in the kernel of the Hessian $\textbf{A}(\textbf{q}_c)$ for any $\textbf{q}_c \in \mathcal{M}$.  Decomposing $\mathbb{R}^d \simeq T_{\textbf{q}_c} \mathcal{M} \oplus \text{N}_{\textbf{q}_c}\mathcal{M}$ where $\text{N}_{\textbf{q}_c}\mathcal{M}$ is the normal bundle to $\mathcal{M}$ at $\textbf{q}_c$, we can can orthogonally decompose $\delta \textbf{q}$ as $\delta \textbf{q} = \delta \textbf{q}_{\parallel} + \delta \textbf{q}_\perp$ so that the right-hand side of~\eqref{E:nearby1} becomes
\begin{equation}
\label{E:nearby2}
\exp\!\left(\frac{i}{\hbar}S(\textbf{q}_c, \textbf{q}_c, E) + \frac{i}{2\hbar}\,\delta \textbf{q}_\perp\cdot \textbf{A}(\textbf{q}_c)\cdot\delta\textbf{q}_\perp\right)\,.
\end{equation}
However, note that $\textbf{q}$ only equals $\textbf{q}_c + \delta \textbf{q}_\perp$ if $\delta \textbf{q}_\parallel = 0$.  But we can find a point $\textbf{q}_c'$ on $\mathcal{M}$ which is the point on the manifold closest to $\textbf{q}$; this will mean that $\textbf{q} = \textbf{q}_c' + \delta \textbf{q}'_\perp$ since $\delta \textbf{q}_\parallel' = 0$.  We will have that the right-hand side of~\eqref{E:nearby1} equals
\begin{equation}
\label{E:nearby3}
\exp\!\left(\frac{i}{\hbar}S(\textbf{q}_c', \textbf{q}_c', E) + \frac{i}{2\hbar}\,\delta \textbf{q}_\perp'\cdot \textbf{A}(\textbf{q}_c')\cdot\delta\textbf{q}_\perp'\right)\,.
\end{equation}
This is our desired expression.

For self-consistency, we would like to check that within the regime of validity of our approximations,~\eqref{E:nearby2} is close to~\eqref{E:nearby3}.  First, we observe that since $\left(\frac{\partial S(\textbf{z}^A,\textbf{z}^B,E)}{\partial \textbf{z}^A}+\frac{\partial S(\textbf{z}^A,\textbf{z}^B,E)}{\partial \textbf{z}^B}\right)_{\textbf{z}^A=\textbf{z}^B=\textbf{q}} = 0$ for $\textbf{q}$ in $\mathcal{M}$, we must have $S(\textbf{q}_c,\textbf{q}_c,E) = S(\textbf{q}_c',\textbf{q}_c',E)$.  This can be readily seen by the fundamental theorem of calculus: if $\textbf{r}(t)$ is a differentiable path satisfying $\textbf{r}(0) = \textbf{q}_c$ and $\textbf{r}(1) = \textbf{q}_c'$, then $S(\textbf{q}_c', \textbf{q}_c', E) - S(\textbf{q}_c, \textbf{q}_c, E) = \int_0^1 dt \, \textbf{r}'(t) \cdot \nabla S(\textbf{r}(t),\textbf{r}(t),E) = 0$.  Next, we note that since $\delta \textbf{q}$ and $\delta \textbf{q}'$ only differ in the directions along the tangent space to the manifold, we have $\delta \textbf{q}_\perp = \delta \textbf{q}_\perp'$.  Finally, since the third and higher order derivatives of $S$ are assumed to be at most $\frac{m}{T_{\text{max}}} \sqrt{\frac{E\,T_{\text{max}}}{\hbar}}$ in norm, it follows that the Hessian does not significantly vary when we change $\textbf{q}_c'$ to $\textbf{q}_c$ since they are at most a distance of $\frac{\hbar}{\Delta}$ apart; in particular, the change in the Hessian is subleading relative to the terms shown in the exponential above.  Thus we can replace $\textbf{A}(\textbf{q}_c')$ with $\textbf{A}(\textbf{q}_c)$.

In summary, we have the formula
\begin{align}
& \frac{1}{(2\pi \Delta^2)^{d/2}}\int d^d \textbf{z} \, e^{\frac{i}{\hbar}S(\textbf{z},\textbf{z},E)}\,e^{-\frac{1}{2\Delta^2} (\textbf{q}-\textbf{z})^2} \nonumber \\ \nonumber \\
& \qquad  \approx \begin{cases} e^{\frac{i}{\hbar}S(\textbf{q}_c, \textbf{q}_c, E) + \frac{i}{2\hbar}\,\delta \textbf{q}_\perp \cdot \textbf{A}(\textbf{q}_c) \cdot \delta \textbf{q}_\perp} & \text{if  }\textbf{q} = \textbf{q}_c + \delta \textbf{q}_\perp\text{  where  }\textbf{q}_c \in \mathcal{O},\, \delta \textbf{q}_\perp \in \text{N}_{\textbf{q}_c} \mathcal{O},\,\,\|\delta \textbf{q}_\perp\|_2 \lesssim \frac{\hbar}{\Delta} \\ \\
0 & \text{if  } \left\|\left(\frac{\partial S(\textbf{z}^A,\textbf{z}^B,E)}{\partial \textbf{z}^A}+\frac{\partial S(\textbf{z}^A,\textbf{z}^B,E)}{\partial \textbf{z}^B}\right)_{\textbf{z}^A=\textbf{z}^B=\textbf{q}}\right\|_2 \gg \frac{\hbar}{\Delta}
\end{cases}\,.
\end{align}
Note that there are some domains of $\textbf{q}$ not covered by the above, namely $\left\|\left(\frac{\partial S(\textbf{z}^A,\textbf{z}^B,E)}{\partial \textbf{z}^A}+\frac{\partial S(\textbf{z}^A,\textbf{z}^B,E)}{\partial \textbf{z}^B}\right)_{\textbf{z}^A=\textbf{z}^B=\textbf{q}}\right\|_2 \approx \frac{\hbar}{\Delta}$, and also $\|\delta \textbf{q}_\perp \|_2 \gtrsim \frac{\hbar}{\Delta}$ but $\left\|\left(\frac{\partial S(\textbf{z}^A,\textbf{z}^B,E)}{\partial \textbf{z}^A}+\frac{\partial S(\textbf{z}^A,\textbf{z}^B,E)}{\partial \textbf{z}^B}\right)_{\textbf{z}^A=\textbf{z}^B=\textbf{q}}\right\|_2 \lesssim \frac{\hbar}{\Delta}$.  The first type of domain is not very interesting, and just corresponds to a regime in which we need to use~\eqref{E:QMinplaceof1} instead of~\eqref{E:QMhere1} since we are considering points $\textbf{q}$ which are a bit too far away from exactly periodic orbits.  The second type of domain is more interesting, and corresponds to certain kinds of nearly periodic orbits which are not nearby any exactly periodic orbits.  We will comment on this further below.

Finally, let us put everything together to get our equation for quantum scars.  Defining
\begin{equation}
\label{E:microdef1}
P_{\text{micro}}(\textbf{q}) := \frac{\int \frac{d^d \textbf{p}}{(2\pi \hbar)^d} \,\delta_\varepsilon(E - H(\textbf{q},\textbf{p}))}{\int d^d \textbf{z}\,\frac{d^d \textbf{p}}{(2\pi \hbar)^d} \,\delta_\varepsilon(E - H(\textbf{z},\textbf{p}))}
\end{equation}
and also
\begin{align}
\label{E:Pscar1}
&\delta P_{\text{scar}}(\textbf{q}_c, \delta \textbf{q}_\perp) := -\frac{2}{\pi \hbar \int d^d \textbf{z}\,\frac{d^d \textbf{p}}{(2\pi \hbar)^d} \,\delta_\varepsilon(E - H(\textbf{z},\textbf{p}))} \,\text{Im} \Bigg\{\frac{1}{i} \frac{1}{\sqrt{(2\pi i\hbar)^{d-1}}} \frac{1}{|\dot{\textbf{q}}|}\left|\det\left(\frac{\partial^2 S(\textbf{q}^A,\textbf{q}^B,E)}{\partial \textbf{q}_\perp^A\partial \textbf{q}_\perp^B}\right)\right|_{\textbf{q}^A = \textbf{q}^B = \textbf{q}}^{1/2} \nonumber \\
& \qquad \qquad \qquad \qquad \qquad \qquad \times \exp\left[- \frac{\varepsilon}{\hbar} \, T(\textbf{q}_c, \textbf{q}_c, E)- i \nu(\textbf{q}_c, \textbf{q}_c, E)\frac{\pi}{2} + \frac{i}{\hbar}\left(S(\textbf{q}_c, \textbf{q}_c, E) + \frac{1}{2}\, \delta \textbf{q}_\perp \cdot \textbf{A}(\textbf{q}_c) \cdot \delta \textbf{q}_\perp\right)\right]\Bigg\}\,,
\end{align}
we have our desired equation
\begin{equation}
\label{E:desired1}
\boxed{\langle |\psi(\textbf{q})|^2\rangle_{E,\Delta} \approx \begin{cases} P_{\text{micro}}(\textbf{q}) + \delta P_{\text{scar}}(\textbf{q}_c, \delta \textbf{q}_\perp) & \text{if  }\textbf{q} = \textbf{q}_c + \delta \textbf{q}_\perp\text{  where  }\textbf{q}_c \in \mathcal{O},\, \delta \textbf{q}_\perp \in \text{N}_{\textbf{q}_c} \mathcal{O},\,\,\|\delta \textbf{q}_\perp\|_2 \lesssim \frac{\hbar}{\Delta} \\ \\
P_{\text{micro}}(\textbf{q}) & \text{if  } \left\|\left(\frac{\partial S(\textbf{z}^A,\textbf{z}^B,E)}{\partial \textbf{z}^A}+\frac{\partial S(\textbf{z}^A,\textbf{z}^B,E)}{\partial \textbf{z}^B}\right)_{\textbf{z}^A=\textbf{z}^B=\textbf{q}}\right\|_2 \gg \frac{\hbar}{\Delta}
\end{cases}}
\end{equation}
The $\approx$ means that the formula includes multiplicative corrections $\big( 1 + O\big(\frac{\varepsilon}{E}\,,(\!\frac{\hbar}{E \,T_{\text{max}}}\!)^\gamma\big)\,\big)$, where the formula is further subject to our assumptions listed above.  We will summarize these assumptions here.  We have assumed that: (i) for each triplet $\textbf{q}^A, \textbf{q}^B, E$ with $\textbf{q}^A \not = \textbf{q}^B$, there is a single classical trajectory starting at $\textbf{q}^A$, ending at $\textbf{q}^B$, and having energy $E$; if there are any non-trivial orbits beginning at ending at $\textbf{q}^A = \textbf{q}^B$ with energy $E$, then there must be exactly two which are time-reverses of one another; (ii) $\| \textbf{A}(\textbf{q}) \| \leq \frac{m}{T_{\text{max}}} \sqrt{\frac{E\,T_{\text{max}}}{\hbar}}$ for $\textbf{q} \in \mathcal{O}$ and similarly for the third order and higher order derivatives of $S$; (iii) $\mathcal{O}$ is a disjoint union of smooth manifolds (which by definition do not have any self-intersections, otherwise they would not be smooth); and (iv) the only values of $\textbf{q}$ for which $\left\|\left(\frac{\partial S(\textbf{z}^A,\textbf{z}^B,E)}{\partial \textbf{z}^A}+\frac{\partial S(\textbf{z}^A,\textbf{z}^B,E)}{\partial \textbf{z}^B}\right)_{\textbf{z}^A=\textbf{z}^B=\textbf{q}}\right\|_2 \lesssim \frac{\hbar}{\Delta}$ are those that are close to $\mathcal{O}$. 

We end this section with a comment about certain kinds of \textit{nearly} periodic orbits.  Suppose that we found a collection of nearly periodic orbits whose image in position space forms a manifold $\mathcal{O}_{\text{nearly}}$, but such that none of the nearly period orbits sit near any exactly periodic orbits.  By `nearly periodic' we mean that the orbits begin and end at the same spatial point and satisfy $\left\| \left(\frac{\partial S(\textbf{z}^A,\textbf{z}^B,E)}{\partial \textbf{z}^A}+\frac{\partial S(\textbf{z}^A,\textbf{z}^B,E)}{\partial \textbf{z}^B}\right)_{\textbf{z}^A=\textbf{z}^B=\textbf{q}}\right\|_2 \lesssim \frac{\hbar}{\Delta}$.  Then, for $\textbf{q}$'s on or near $\mathcal{O}_{\text{nearly}}$, we could write an essentially identical expression for $\langle |\psi(\textbf{q})|^2\rangle_{E,\Delta}$ as the boxed scar formula above.

\section{Derivation of quantum scar formula for a quantum field theory}
\label{app:derivation}

In this section we generalize Bogomolny's scar formula~\cite{bogomolny1988smoothed} to the quantum field theory setting.  Our analysis here closely mirrors that of Appendix~\ref{sec:bogomolny} above, but is reformulated to work in field theory.  Along the way we will require field-theoretic versions of the Van Vleck propagator and Gutzwiller trace formula; we derive these results by adapting the analyses in~\citeapp{haake2010q, ChaosBook}.

\subsection{Notation in field theory}

Here we will consider complex scalar fields in $d+1$ spacetime dimensions with spatial profiles $\phi(x) = \frac{1}{\sqrt{2}}(\phi_1(x) + i \phi_2(x))$.  It will often be convenient to treat this as a two-component object $\boldsymbol{\phi} := (\phi_1(x), \phi_2(x))$.  We will use a similar bolded notation for $\boldsymbol{\pi} := (\pi_1(x), \pi_2(x))$. 

It is useful to introduce two different kinds of norms.  First we have a more ordinary $2$-norm on $\mathbb{R}^2$, given by
\begin{equation}
\| (a,b) \|_2 = \sqrt{a^2 + b^2}\,.
\end{equation}
For instance, we will sometime write expressions like
\begin{equation}
\| \boldsymbol{\phi}^A(x) - \boldsymbol{\phi}^B(x) \|_2^2 = (\phi_1^A(x) - \phi_1^B(x))^2 + (\phi_2^A(x) - \phi_2^B(x))^2\,.
\end{equation}

Another very useful norm for our analysis is a type of $L^2$ norm for two-component functions $\boldsymbol{f}(x) = (f_1(x), f_2(x))$, namely
\begin{equation}
    \|\boldsymbol{f}\|_{L^2} := \left(\int d^dx\,\left(f_1(x)^2+f_2(x)^2\right)\right)^{1/2}.
\end{equation}
This norm has an associated inner product, given by
\begin{equation}
\langle \boldsymbol{f}, \boldsymbol{g} \rangle_{L^2} := \int d^d x \, (f_1(x) g_1(x) + f_2(x) g_2(x))\,.
\end{equation}
This inner product will be so useful that in the field theory analysis below, we will simply define the dot product between two vector-valued functions as
\begin{equation}
\boldsymbol{f} \cdot \boldsymbol{g} := \langle \boldsymbol{f}, \boldsymbol{g}\rangle_{L_2}\,.
\end{equation}
Similarly, if we have some kernel $A_{ij}(x,y)$ for $i,j = 1,2$, then we use the notation
\begin{equation}
\boldsymbol{f} \cdot \boldsymbol{A} \cdot \boldsymbol{g} := \int d^d x\,d^d y \sum_{i,j=1}^2 f_i(x) A_{ij}(x,y) g_j(y)
\end{equation}
and relatedly
\begin{equation}
\boldsymbol{A} \cdot \boldsymbol{g} := \int d^d y \sum_{j=1}^2 A_{ij}(x,y) \, g_j(y)\,.
\end{equation}
We have an analogous expression for $\boldsymbol{f} \cdot \boldsymbol{A}$.

Next we require some functional derivative notation, which will interface well with our $L^2$ inner product notation.  Given a functional $\mathcal{F}[\boldsymbol{f}(x)]$, we use the functional derivative notation $\frac{\delta \mathcal{F}}{\delta \boldsymbol{f}}$ to denote the infinite-dimensional vector with components $\frac{\delta \mathcal{F}}{\delta f_i(x)}$ for $i=1,2$ and $x \in \mathbb{R}^d$.  This is analogous to vector derivative notation in finite dimensions.  If $\mathcal{G}[\boldsymbol{g}(x)]$ is another functional, then using our dot product notation we have
\begin{equation}
\frac{\delta \mathcal{F}}{\delta \boldsymbol{f}} \cdot \frac{\delta \mathcal{G}}{\delta \boldsymbol{g}} = \int d^d x \, \sum_{i=1}^2 \frac{\delta \mathcal{F}}{\delta f_i(x)} \frac{\delta \mathcal{G}}{\delta g_i(x)}\,.
\end{equation}
These notations will make our quantum field-theoretic derivations look more like the ordinary quantum mechanical setting.

\subsection{Van Vleck propagator in quantum field theory}
The derivation of the scar formula in quantum field theory proceeds analogously to the derivation in the ordinary quantum mechanical setting in Appendix~\ref{sec:bogomolny}. We start by deriving the Van Vleck propagator, which we subsequently transform from time to energy variables. The scar formula is then obtained by a functional smearing over field configurations in conjunction with an energy average over a window $[E - \varepsilon/2, E + \varepsilon/2]$.  As explained above, restrict our attention to complex scalar field theories in $d+1$ spacetime dimensions; an analysis for real-valued scalar field theories is essentially identical.  We further assume our complex scalar field theory has time-reversal symmetry.  

We begin by approximating the field-theoretic propagator
\begin{equation}
\label{E:fieldtheoreticprop1}
    \langle \boldsymbol{\phi}^B | U(t) | \boldsymbol{\phi}^A \rangle = \int_{\boldsymbol{\Phi}(x,0)= \boldsymbol{\phi}^A(x)}^{\boldsymbol{\Phi}(x, t)=\boldsymbol{\phi}^B(x)}[d\boldsymbol{\Phi}]\,\exp\!\left(\frac{i}{\hbar}S[\boldsymbol{\Phi}]\right)\,.
\end{equation}
Some explanation of our terminology is required.  In field theory, the propagator usually denotes the 2-point function of fields (usually in the ground state), e.g.~$\langle \Omega| \hat{\boldsymbol{\Phi}}(x,t) \hat{\boldsymbol{\Phi}}(y,0) |\Omega\rangle$ where we have written the operators in the Heisenberg picture.  This 2-point function depends on the initial and final times, as well as $x$ and $y$.  By contrast, we will be interested in $K[t, \phib^B; \, 0,\phib^A] := \langle \boldsymbol{\phi}^B | U(t) | \boldsymbol{\phi}^A \rangle$ which not only depends on the initial and final times, but is also a \textit{functional} of the field configurations $\phib^A(x)$ and $\phib^B(x)$.  Our goal is to develop the Van Vleck formula for $K[t, \phib^B; \, 0,\phib^A]$, and thereafter a Gutzwiller trace formula based upon it.  A useful reference for the Van Vleck propagator in the ordinary quantum mechanical setting is~\cite{littlejohn1992van}, which emphasizes a symplectic geometry point of view.  Below we will use field-theoretic generalizations of parts of~\cite{littlejohn1992van}.

It is perhaps most intuitive to motivate the Van Vleck propagator in field theory in an anachronistic way, starting with the path integral.  (Indeed, the Van Vleck propagator in ordinary quantum mechanics predates Feynman's path integral for ordinary quantum mechanics by two decades.)  We observe that~\eqref{E:fieldtheoreticprop1} can be evaluated in the path integral framework using a stationary phase approximation as
\begin{equation}
\label{E:pathintegralapproxsum1}
\langle \phib^B | U(t) |\phib^A \rangle \approx \sum_{\text{paths }c} \frac{1}{\sqrt{\mathcal{D}_c}} \, e^{\frac{i}{\hbar} \, S_c(\phib^A, \phib^B, t)}\,,
\end{equation}
where we are summing over classical paths $c$ which begin at $\phib^A$ at time zero and end at $\phib^B$ at time $t$.  Above, $S_c(\phib^A, \phib^B, t)$ denotes the classical action of the path $c$, i.e.~Hamilton's principal function (here it is a functional), and $1/\sqrt{\mathcal{D}_c}$ denotes a corresponding functional determinant factor.  Let us regard the $c$ in $S_c(\phib^A, \phib^B, t)$ as designating a choice of `branch' of Hamilton's principal function; the function is multivalued on account of having multiple classical paths which begin at $\phib^A$ and end at $\phib^B$ in the same amount of time.  Our desired generalization of the Van Vleck formula comes about by establishing that
\begin{equation}
\label{E:Dkerneltoestablish1}
\frac{1}{\sqrt{\mathcal{D}_c[\phib^B, t\,; \phib^A, 0]}} = \frac{1}{(2\pi i \hbar)^{\mathcal{V}}} \left(\det \frac{\delta^2 S_c(\chib^A, \chib^B, t)}{\delta \chib^A \delta \chib^B}\right)_{\substack{\chib^A = \phib^A \\ \chib^B = \phib^B}}^{1/2} \, e^{- i \nu_c \frac{\pi}{2}}
\end{equation}
where $\mathcal{V}$ is the number of points in space and $\nu_c$ is the Maslov index.  Let us analyze a single term in the sum in~\eqref{E:pathintegralapproxsum1}, e.g.~for a fixed branch $c$ of Hamilton's principal function which contains a single classical path going from $\phib^A$ to $\phib^B$ in a time $t$.  If we consider the classical path up to an infinitesimal amount of time $\delta t$, then the endpoint will be some $\widetilde{\phib}^B$.  Then we have
\begin{equation}
\label{E:deltatapprox1}
    \frac{1}{\sqrt{\mathcal{D}_{c,\delta t}}} \, e^{\frac{i}{\hbar} \, S_c(\phib^A, \widetilde{\phib}^B, \delta t)} \approx \left(\frac{1}{2\pi i\hbar \,\delta t}\right)^{{\cal V}}\exp\left[\frac{i}{\hbar}\!\left(\frac{1}{2\,\delta t}\,\| \widetilde{\boldsymbol{\phi}}^B - \boldsymbol{\phi}^A\|_2^2 -\left(\frac{1}{2}\,\nabla \widetilde{\phib}^B \cdot \nabla \widetilde{\phib}^B+ V(\widetilde{\phib}^B)\right)\delta t\right)\right]\,.
\end{equation}
Letting $\pib(x)$ be the momentum conjugate to $\phib(x)$, for infinitesimal times we have
\begin{align}
  \pi_i^A(x)&=- \frac{\delta S_c(\phib^A, \widetilde{\phib}^B, \delta t)}{\delta\phi^A_i(x)}=\frac{1}{\delta t}(\widetilde{\phi}^B_i(x)-\phi^A_i(x))\,,\\
    \frac{\delta\pi_i^A(x)}{\delta\widetilde{\phi}^B_j(y)}&=\frac{1}{\delta t}\delta_{ij}\delta(x-y)\,,\\
    \label{E:accountfor2v1}
    \det\frac{\delta\pi_i^A(x)}{\delta\widetilde{\phi}^B_j(y)}&=\left(\frac{1}{\delta t}\right)^{2{\cal V}}\,,
\end{align}
where the factor of two in the exponent of~\eqref{E:accountfor2v1} accounts for the fact that the complex scalar has two components. The infinitesimal-time propagator in~\eqref{E:deltatapprox1} can then be written as
\begin{equation}
    \frac{1}{\sqrt{\mathcal{D}_{c,\delta t}}} \, e^{\frac{i}{\hbar} S_c(\phib^A, \widetilde{\phib}^B, \delta t)} = \frac{1}{(2\pi i\hbar)^{{\cal V}}}\left|\det\frac{\delta \pib^A}{\delta\widetilde{\phib}^B}\right|^{1/2}e^{\frac{i}{\hbar} S_c(\phib^A, \widetilde{\phib}^B, \delta t)}
\end{equation}
where equality here is meant in the sense of the infinitesimal $\delta t$ limit.  We have opted to put absolute values around the determinant term since it is already positive.  We would like to generalize this formula to non-infinitesimal times.  Let us suggestively write
\begin{align}
 \frac{1}{\sqrt{\mathcal{D}_{c,t}}} \, e^{\frac{i}{\hbar} S_c(\phib^A, \phib^B, t)} =  F_c(\widetilde{\phib}^B,  \delta t\,;\,\phib^B, t)\,\frac{1}{\sqrt{\mathcal{D}_{c,\delta t}}} \, e^{\frac{i}{\hbar} \left(S_c(\widetilde{\phib}^B, \phib^B, t - \delta t) + S_c(\phib^A, \widetilde{\phib}^B, \delta t)\right)}
\end{align}
or equivalently
\begin{align}
\label{E:topluginshortly1}
 \frac{1}{\sqrt{\mathcal{D}_{c,t}}} \, e^{\frac{i}{\hbar} S_c(\phib^A, \phib^B, t)} =  F_c(\widetilde{\phib}^B,  \delta t\,;\,\phib^B, t)\,\frac{1}{(2\pi i\hbar)^{{\cal V}}}\left|\det\frac{\delta\pib^A}{\delta\widetilde{\phib}^B}\right|^{1/2}e^{\frac{i}{\hbar} S_c(\phib^A, \phib^B, t)}\,.
\end{align}
In particular, we know how the phase transforms as we increase $\delta t$ to $t$: we simply accumulate phase by integrating the action of the classical path from $\phib^A$ to $\phib^B$ in a time $t$ which is indexed by the branch $c$ of Hamilton's principal function. 
The remaining mystery is the $F_c(\widetilde{\phib}^B,  \delta t\,;\,\phib^B, t)$ function.  To see what it is, we note that the amplitude prefactor $\frac{1}{\sqrt{\mathcal{D}_{c,t}}}$ transforms like the square root of a density.  In fact, defining
\begin{equation}
\rho_c[\phib, t] := \frac{1}{|\mathcal{D}_c[\phib^A, 0\,;\,\phib, t]|}
\end{equation}
for $\phib^A$ fixed and $\phib$ on the $c$ branch of Hamilton's principal function, and evolving $\frac{1}{\sqrt{D_{c,t}}} \, e^{\frac{i}{\hbar}S_c[\phib^A, \phib, t]}$ according to the functional Schr\"{o}dinger evolution, we find in the semiclassical limit that
\begin{align}
H\!\left[\phib, \frac{\delta S_c[\phib^A, \phib, t]}{\delta \phib}\right] + \frac{\partial S_c[\phib^A, \phib, t]}{\partial t} &= 0\,,\\
\frac{\partial \rho_c[\phib, t]}{\partial t} + \rho_c[\phib, t]\,\frac{\delta}{\delta \phib}\cdot v_c[\phib, t] + v_c[\phib, t] \cdot \frac{\delta}{\delta \phib} \,\rho_c[\phib,t] &= 0\,, \qquad v_c[\phib, t] = \frac{\delta H[\phib, \pib]}{\delta \pib}\Bigg|_{\pi = \frac{\delta S_c[\phib^A, \phib, t]}{\delta \phib}}\,.
\end{align}
The first equation is just the Hamilton-Jacobi equation, which is solved by Hamilton's principal function (i.e.~the integrated action along a classical path beginning at $\phib^A$ at time zero and ending at $\phib$ at time $t$).  The second equation is a transport equation for $\rho_c[\phib, t]$, where $v_c[\phib, t]$ acts as a velocity field in infinite dimensions.  Such a transport equation is solved by
\begin{equation}
\rho_c[\phib', t'] = \left|\det \frac{\delta \phib[\phib',t,t'] }{\delta \phib'}\right|\rho_c[\phib[\phib', t,t'], t]\,,
\end{equation}
where $\phib[\phib', t, t']$ means the following:  we consider the point in phase space $\left(\phib', \frac{\delta S_c[ \phib^A, \chib, t]}{\delta \chib}\Big|_{\chib = \phib'}\right)$ where $\frac{\delta S_c[\phib^A, \chib, t]}{\delta \chib}\Big|_{\chib = \phib'}$ is the momentum, and then evolve the point in phase space backwards from $t'$ to $t$; the resulting point in field space has its field configuration (i.e.~we ignore the momentum) as $\phib[\phib', t, t']$.  But this means that our mysterious $F_c(\widetilde{\phib}^B,  \delta t\,;\,\phib^B, t)$ is just
\begin{align}
F_c(\widetilde{\phib}^B,  \delta t\,;\,\phib^B, t) = \left(\det \frac{\delta \widetilde{\phib}^B[\phib^B, \delta t,  t]}{\delta \phib^B}\right)^{1/2}
\end{align}
since $\left|F_c(\widetilde{\phib}^B,  \delta t\,;\,\phib^B, t)\right|^2 = \left|\det \frac{\delta \widetilde{\phib}^B}{\delta \phib^B}\right|$.  One subtlety with the above is that we need to decide which square root of $\det \frac{\delta \widetilde{\phib}^B }{\delta \phib^B}$ we are to take; depending on the number of negative eigenvalues of $\frac{\delta \widetilde{\phib}^B }{\delta \phib^B}$, the choice of square root can lead to overall factors of $\pm 1$ or $\pm i$.  Such a factor is packaged into the Maslov index $\nu_c$ as
\begin{align}
F_c(\widetilde{\phib}^B,  \delta t\,;\,\phib^B, t) = \left|\det \frac{\delta \widetilde{\phib}^B[\phib^B, \delta t,  t]}{\delta \phib^B}\right|^{1/2} e^{- i \nu_c \frac{\pi}{2}}\,.
\end{align}
The value of $\nu_c$ is prescribed by a more detailed analysis, but we will not delve into this further here.  In the ordinary quantum mechanics setting, a very useful reference is~\cite{littlejohn1992van}.  

In any case,~\eqref{E:topluginshortly1} becomes
\begin{align}
 \frac{1}{\sqrt{\mathcal{D}_{c,t}}} \, e^{\frac{i}{\hbar} S_c(\phib^A, \phib^B, t)} &=  \frac{1}{(2\pi i\hbar)^{{\cal V}}}\left|\det \frac{\delta \widetilde{\phib}^B }{\delta \phib^B}\right|^{1/2}\left|\det\frac{\delta\pib^A}{\delta\widetilde{\phib}^B}\right|^{1/2}e^{\frac{i}{\hbar} S_c(\phib^A, \phib^B, t) - i \nu_c \frac{\pi}{2}} \nonumber \\
 &= \frac{1}{(2\pi i\hbar)^{{\cal V}}} \left|\det\frac{\delta\pib^A}{\delta \phib^B}\right|^{1/2}e^{\frac{i}{\hbar} S_c(\phib^A, \phib^B, t)- i \nu_c \frac{\pi}{2}} \nonumber \\
 &= \frac{1}{(2\pi i\hbar)^{{\cal V}}} \left|\det \frac{\delta^2 S_c(\chib^A, \chib^B, t)}{\delta \chib^A \delta \chib^B}\right|_{\substack{\chib^A = \phib^A \\ \chib^B = \phib^B}}^{1/2} e^{\frac{i}{\hbar} S_c(\phib^A, \phib^B, t)- i \nu_c \frac{\pi}{2}}\,.
\end{align}
The above indeed establishes~\eqref{E:Dkerneltoestablish1}, and so summing over $c$'s the full Van Vleck propagator in quantum field theory is just
\begin{align}
\label{E:desiredQFTVV1}
\langle \phib^B| U(t) |\phib^A\rangle \approx \sum_{\text{paths }c}  \frac{1}{(2\pi i\hbar)^{{\cal V}}} \left|\det \frac{\delta^2 S_c(\chib^A, \chib^B, t)}{\delta \chib^A \delta \chib^B}\right|_{\substack{\chib^A = \phib^A \\ \chib^B = \phib^B}}^{1/2} e^{\frac{i}{\hbar} S_c(\phib^B, \phib^A, t) - i \nu_c \frac{\pi}{2}}
\end{align}
where the $\approx$ is in the sense of a semiclassical limit.

Next let us transform our equation for the quantum field-theoretic Van Vleck propagator in~\eqref{E:desiredQFTVV1} from time variables to energy variables:
\begin{equation}
\label{E:timetoenergyQFT1}
    G(\phib^A, \phib^B, E)=\frac{1}{i\hbar}\int_0^{\infty}dt\,e^{iEt/\hbar}\langle\phib^B|U(t)|\phib^A\rangle\,.
\end{equation}
We consider the stationary phase approximation in $t$, which picks out fixed energy paths:
\begin{equation}
\label{E:statphaseQFT1}
    E+\frac{\partial S_c(\phib^A, \phib^B, t)}{\partial t}=0\quad \Longrightarrow \quad  E=H(\phib^A, \phib^B, t_c)\,.
\end{equation}
Here $t_c$ is the time extent of the path along the $c$ branch of Hamilton's principal function which starts at $\phib^A$, ends at $\phib^B$, and has energy $E$.  Then it is useful to define Legendre transform of Hamilton's principal function as
\begin{equation}
S_c(\phib^A, \phib^B, E):= S_c(\phib^A, \phib^B, t_c(\phib^A, \phib^B, E)) + E \, t_c(\phib^A, \phib^B, E)
\end{equation}
where as before $S_c(\textbf{q}^A, \textbf{q}^B, E) = \int_0^{t_c} dt \int d^d x \sum_{i=1}^2 \Pi_{i,c}(x,t) \Phi_{i,c}(x,t)$ is the `abbreviated action' of a classical solution $(\boldsymbol{\Phi}_c(x,t), \boldsymbol{\Pi}_c(x,t))$ with energy $E$, beginning at $\phib^A$ and terminating at $\phib^B$.

Performing the time integral in~\eqref{E:timetoenergyQFT1} and using the stationary phase approximation in~\eqref{E:statphaseQFT1}, the Green's function in energy variables becomes
\begin{align}
\label{E:GreenQFTfirst1}
    G(\phib^A,\phib^B,E) &\approx \frac{1}{i\hbar}\sum_{\text{paths }c}\frac{1}{(2\pi i\hbar)^{{\cal V}-(1/2)}}\left|\det\left( \frac{\delta^2S_c(\chib^A, \chib^B, E)}{\delta \chib^A\delta\chib^B}\right)\right|_{\substack{\chib^A = \phib^A \\ \chib^B = \phib^B}}^{1/2}\left|\det\left(\frac{\partial^2S_c(\phib^A, \phib^B, t)}{\partial t^2}\right)\right|_{t = t_c}^{-1/2} \nonumber \\
    & \qquad \qquad \qquad \qquad \qquad \qquad \qquad \qquad \qquad \qquad \qquad \qquad \qquad \times \exp\left[\frac{i}{\hbar}S_c(\phib^A, \phib^B, E)-i\nu_c \frac{\pi}{2}\right],
\end{align}
where $\nu_c$ accounts for an additional possible phase. As in the ordinary quantum mechanics case, we next want to repackage the amplitude factor. We start by writing the matrix
\begin{equation}
\label{E:writingmatrix1}
\left(
\begin{array}{cc}
\frac{\delta^2S_c}{\delta\chib^A\delta\chib^B} & \frac{\delta \partial S_c}{\delta\chib^A\partial E'}\\
\frac{\delta \partial S_c}{\delta\chib^B\partial E'} & \frac{\partial^2S_c}{\partial E'^2}
\end{array}
\right)\Bigg|_{\substack{\chib^A = \phib^A \\ \chib^B = \phib^B\\ E' = E\,\,}}=
\left(
\begin{array}{cc}
-\frac{\delta\pib^A}{\delta\chib^B} & -\frac{\partial\pib^A}{\partial E'}\\
\,\,\frac{\delta t}{\delta\chib^B} & \,\,\frac{\partial t}{\partial E'}
\end{array}
\right)\Bigg|_{\substack{\chib^A = \phib^A \\ \chib^B = \phib^B\\ E' = E\,\,}}\,,
\end{equation}
which has a determinant corresponding to the change of variables from $(-\pib^A, t)$ to $(\phib^B, E)$. Using the chain rule for determinants, this can be rewritten as
\begin{align}
-\det\frac{\partial(\pib^A, t)}{\partial(\chib^B, E')}\Bigg|_{\substack{\chib^A = \phib^A \\ \chib^B = \phib^B\\ E' = E\,\,}} &=-\det\left(\frac{\partial(\pib^A, t)}{\partial(\chib^B, t)}\frac{\partial(\chib^B, t)}{\partial(\chib^B, E')}\right)\Bigg|_{\substack{\chib^A = \phib^A \\ \chib^B = \phib^B\\ E' = E\,\,}} 
=\left(\det\frac{\delta\pib^A}{\delta\phib^B}\right)\left(\frac{\partial^2 S_c}{\partial t^2}\right)^{-1}\Bigg|_{t = t_c}\,.
\end{align}
To further simply this determinant, we use the fact that the energy $H(\phib,\pib)=E$ is fixed, namely
\begin{align}
\label{E:qAdot2}
\frac{\delta}{\delta \chib^B}\,H(\chib^A, \pib^A[\chib^A, \chib^B, E])\Bigg|_{\substack{\chib^A = \phib^A \\ \chib^B = \phib^B}}&=0=  \frac{\delta\pib^A}{\delta\chib^B} \cdot \frac{\delta H}{\delta \pib^A}\Bigg|_{\substack{\chib^A = \phib^A \\ \chib^B = \phib^B}}\quad \Longrightarrow \quad \frac{\delta^2 S_c}{\delta\chib^B \delta\chib^A}\Bigg|_{\substack{\chib^A = \phib^A \\ \chib^B = \phib^B}} \cdot \dot{\phib}^A =0\\
\frac{\delta}{\delta \chib^A}\,H(\chib^B, \pib^B[\chib^A, \chib^B, E])\Bigg|_{\substack{\chib^A = \phib^A \\ \chib^B = \phib^B}}&=0=  \frac{\delta H}{\delta \pib^B} \cdot \frac{\delta\pib^B}{\delta\chib^A}\Bigg|_{\substack{\chib^A = \phib^A \\ \chib^B = \phib^B}}\quad \Longrightarrow \quad \dot{\phib}^B \cdot \frac{\delta^2 S_c}{\delta\chib^B \delta\chib^A}\Bigg|_{\substack{\chib^A = \phib^A \\ \chib^B = \phib^B}}=0\,.
\end{align}
As in the ordinary quantum mechanical setting, we now decompose fluctuations around $\phib^A$ and $\phib^B$. Let us focus on a classical trajectory $\boldsymbol{\Phi}_c(x,t)$ which goes from $\phib^A$ to $\phib^B$ in an amount of time $t$.  Fluctuations $\delta \phib^A$ around $\phib^A$ can be decomposed as $\delta \phib_{\parallel}^A + \delta \phib_\perp^A$.  Here $\delta \phib_\parallel^A$ is parallel to the classical trajectory and $\delta \phib_\perp^A$ is orthogonal to the classical trajectory. As such, $\delta \phib_\parallel^A$ corresponds to a single coordinate direction in field space, and $\delta \phib_\perp^A$ corresponds to all of the residual orthogonal coordinate directions in field space.  (Later on when we define the moduli space $\mathcal{O}$, we will redefine $\delta \phib_{\parallel}^A$ as being in the coordinate directions in field space tangent to $\mathcal{O}$ and $\delta \phib_\perp^A$ as being in the coordinate directions in field space normal to $\mathcal{O}$.)  Our notation for fluctuations around $\phib^B$ follows analogously.  Using this notation, the $\dot{\phib}^A$ on the far-right of~\eqref{E:qAdot2} is equivalent to $\partial_s\boldsymbol{\Phi}_c(x,s)|_{s = 0} =: \dot{\phib}_{c}^A(x)$.  Clearly $\dot{\phib}_{c}^A$ is parallel to the orbit since it is a tangent vector.  Then we have
\begin{equation}
    0 = \frac{\delta^2 S_c}{\delta\chib^B \delta\chib^A}\Bigg|_{\substack{\chib^A = \phib^A \\ \chib^B = \phib^B}} \cdot \dot{\phib}^A_{c} = \frac{\delta^2 S_c}{\delta\chib^B \delta\chib^A_{\parallel}}\Bigg|_{\substack{\chib^A = \phib^A \\ \chib^B = \phib^B}} \cdot \dot{\phib}^A_{c} \quad \Longrightarrow \quad \frac{\delta^2 S_c(\chib^A, \chib^B, E)}{\delta\chib^B \delta\chib^A_{\parallel}}\Bigg|_{\substack{\chib^A = \phib^A \\ \chib^B = \phib^B}} = 0\,.
\end{equation}
In the same vein, letting $\partial_s\boldsymbol{\Phi}_{c}(x,s)|_{s = t} =: \dot{\phib}_{c}^B(x)$ gives us
\begin{equation}
    0 = \dot{\phib}^B_{c} \cdot \frac{\delta^2 S_c}{\delta\chib^B \delta\chib^A}\Bigg|_{\substack{\chib^A = \phib^A \\ \chib^B = \phib^B}} =  \dot{\phib}^B_{c} \cdot \frac{\delta^2 S_c}{\delta\chib^B_{\parallel} \delta\chib^A}\Bigg|_{\substack{\chib^A = \phib^A \\ \chib^B = \phib^B}} \quad \Longrightarrow \quad \frac{\delta^2 S_c(\chib^A, \chib^B, E)}{\delta\chib^B_{\parallel} \delta\chib^A}\Bigg|_{\substack{\chib^A = \phib^A \\ \chib^B = \phib^B}} = 0\,.
\end{equation}
Then the left-hand side of~\eqref{E:writingmatrix1} can be further decomposed as
\begin{equation}
\label{eq:detmatrix}
\left.\left(
\begin{array}{ccc}
\frac{\delta^2 S_c}{\delta\chib_{\parallel}^A\delta\chib_{\parallel}^B} & \frac{\delta^2 S_c}{\delta\chib_{\parallel}^A\delta\chib_{\perp}^B} & \frac{\delta \partial S_c}{\delta\chib_{\parallel}^A\partial E'}\\
\frac{\delta^2 S_c}{\delta\chib_{\perp}^A\delta\chib_{\parallel}^B} & \frac{\delta^2 S_c}{\delta\chib_{\perp}^A\delta\chib_{\perp}^B} & \frac{\delta \partial S_c}{\delta\chib_{\parallel}^A\partial E'}\\
\frac{\partial \delta S_c}{\partial E'\delta\chib_{\parallel}^B} & \frac{\partial \delta  S_c}{\partial E'\delta\chib_{\perp}^B} & \frac{\partial^2 S_c}{\partial E'^2}
\end{array}
\right)\right|_{\substack{\chib^A = \phib^A \\ \chib^B = \phib^B\\ E' = E\,\,}}=
\left.\left(
\begin{array}{ccc}
0 & 0 & \frac{\delta \partial S_c}{\delta\chib_{\parallel}^A\partial E'}\\
0 & \frac{\delta^2 S_c}{\delta\chib_{\perp}^A\delta\chib_{\perp}^B} & \frac{\delta \partial S_c}{\delta\chib_{\perp}^A\partial E'}\\
\frac{\partial \delta S_c}{\partial E'\delta\chib_{\parallel}^B} & \frac{\partial \delta S_c}{\partial E'\delta\chib_{\perp}^B} & \frac{\partial^2 S_c}{\partial E'^2}
\end{array}
\right)\right|_{\substack{\chib^A = \phib^A \\ \chib^B = \phib^B\\ E' = E\,\,}}\,.
\end{equation}
We can write the determinant of the above in a compact manner, using several identities.  First note that
\begin{equation}
    \frac{\delta \partial S_c}{\delta \chib^A_\parallel\partial E'}=\frac{\delta t_c}{\delta\chib^A_\parallel}\,, \quad \frac{\partial \delta S_c}{\partial E' \delta \chib^B_\parallel}=\frac{\delta t_c}{\delta\chib^B_\parallel}\,.
\end{equation}
To understand these expressions, note that for any functional $F = F[\chib_A, \chib_B]$ we have
\begin{equation}
\left.\frac{\delta F}{\delta \chib_\parallel^A}\right|_{\substack{\chib^A = \phib^A \\ \chib^B = \phib^B}} := \frac{\dot{\phib}_c^A}{\| \dot{\phib}_c^A \|_2^{1/2}} \cdot \left.\frac{\delta F}{\delta \chib^A}\right|_{\substack{\chib^A = \phib^A \\ \chib^B = \phib^B}}\,, \qquad \left.\frac{\delta F}{\delta \chib_\parallel^B}\right|_{\substack{\chib^A = \phib^A \\ \chib^B = \phib^B}} := \frac{\dot{\phib}_c^B}{\| \dot{\phib}_c^B \|_2^{1/2}} \cdot \left.\frac{\delta F}{\delta \chib^B}\right|_{\substack{\chib^A = \phib^A \\ \chib^B = \phib^B}}\,.
\end{equation}
Accordingly, we have
\begin{equation}
\left.\frac{\delta t_c}{\delta \chib_\parallel^A}\right|_{\substack{\chib^A = \phib^A \\ \chib^B = \phib^B}} = \frac{1}{\| \dot{\phib}_c^A\|_2^{1/2}}\,, \qquad \left.\frac{\delta t_c}{\delta \chib_\parallel^B}\right|_{\substack{\chib^A = \phib^A \\ \chib^B = \phib^B}} = \frac{1}{\| \dot{\phib}_c^B \|_2^{1/2}}\,.
\end{equation}
Then the determinant terms in~\eqref{E:GreenQFTfirst1} become
\begin{equation}
- \frac{1}{\big\|\dot{\phib}_c^A\big\|_{L^2}} \frac{1}{\big\|\dot{\phib}_c^B\big\|_{L^2}}\,\det \left(\frac{\delta^2 S_c}{\delta\chib_{\perp}^A\delta\chib_{\perp}^B}\right)_{\substack{\chib^A=\phib^A\\\chib^B=\phib^B}}.
\end{equation}
Thus we have rewritten the Green's function in energy variables as
\begin{equation}
G(\phib^A, \phib^B, E) \approx \frac{1}{i\hbar}\sum_{\text{orbits }c}\frac{1}{(2\pi i\hbar)^{{\cal V}-(1/2)}}\frac{1}{\big\|\dot{\phib}_c^A\big\|_{L^2} \big\|\dot{\phib}_c^B\big\|_{L^2}}\left|\det\left(\frac{\delta^2 S_c(\chib^A, \chib^B, E)}{\delta\chib_{\perp}^A\delta\chib_{\perp}^B}\right)\right|_{\substack{\chib^A = \phib^A \\ \chib^B = \phib^B}}^{1/2}\exp\left[\frac{i}{\hbar}S_c(\phib^A, \phib^B, E)-i\nu_c \frac{\pi}{2}\right]\,,
\end{equation}
where the $\nu_c$ terms accounts for the phase of the determinant.  Absorbing the Gaussian normalization factor into the determinant, we can rewrite the above equation as
\begin{equation}
\label{eq:qftenergyprop}
G(\phib^A, \phib^B, E) \approx \frac{1}{i\hbar}\sum_{\text{orbits }c}\frac{1}{\big\|\dot{\phib}_c^A\big\|_{L^2} \big\|\dot{\phib}_c^B\big\|_{L^2}}\left|\det\left(\frac{1}{2\pi i \hbar}\frac{\delta^2 S_c(\chib^A, \chib^B, E)}{\delta\chib_{\perp}^A\delta\chib_{\perp}^B}\right)\right|_{\substack{\chib^A = \phib^A \\ \chib^B = \phib^B}}^{1/2}\exp\left[\frac{i}{\hbar}S_c(\phib^A, \phib^B, E)-i\nu_c\frac{\pi}{2}\right]\,.
\end{equation}

\subsection{Trace of the Van Vleck propagator in quantum field theory}
Consider the equation~\eqref{E:timetoenergyQFT1} for the Van Vleck propagator in energy variables, which we worked to write in a very explicit form in~\eqref{eq:qftenergyprop} by using a stationary phase approximation.  Note that this analysis neglects $\delta[\phib^A - \phib^B]$-type contributions at $\phib^A = \phib^B$ which will be important when we imminently take the trace of the Van Vleck propagator.  In particular, we must be careful to consider both the contributions to~\eqref{E:timetoenergyQFT1} from the stationary paths (i.e.~\eqref{eq:qftenergyprop}), as well as the contribution from the $t \to 0^+$ endpoint of the time integral. We again use the following approximation, which is valid in the semiclassical regime:
\begin{equation}
    \int_{t_0}^{\infty}dt\,A(t)e^{iB(t)/\hbar}\approx \frac{i A(t_0)\,e^{i B(t_0)/\hbar}}{B'(t_0)}+[\text{stationary phase contributions}].
\end{equation}
Thus the trace of the propagator will contain two terms,
\begin{equation}
    \int[d\phib]\,G(\phib, \phib, E)\approx \int[d\phib]\,G_0(\phib, \phib, E)+\int[d\phib]\,G_{\text{osc}}(\phib, \phib, E)=G_0(E)+G_{\text{osc}}(E).
\end{equation}
Here the first term $G_0(E)$ corresponds to the contribution from the $t_0 \to 0^+$ endpoint, whereas the second term $G_{\text{osc}}(E)$ contains the stationary phase contributions.  As with the ordinary quantum mechanical setting, for $G_0(E)$ we consider the $t_0 \to 0^+$ limit of the following:
\begin{align}
\frac{1}{i\hbar}\int_{t_0}^{\infty}dt\int [d\phib]\langle \phib|e^{-i(H-E)t/\hbar}|\phib\rangle &\approx\int [d\phib]\frac{\langle \phib|e^{-i(H-E)t_0/\hbar}|\phib\rangle}{E-H(\phib,\phib,t_0)}\\
&=\int [d\phib]\,[d\phib']\,\frac{\delta(\phib-\phib')}{E-H(\phib,\phib', t_0)}\langle \phib|e^{-i(H-E)t_0/\hbar}|\phib'\rangle\\
&=\int \frac{[d\phib]\,[d\pib]}{(2\pi\hbar)^{\cal V}}\,[d\phib']\,\frac{e^{-i\pib(\phib-\phib')/\hbar}}{E-H(\phib,\phib', t_0)}\langle \phib|e^{-i(H-E)t_0}|\phib'\rangle\\
&=\int \frac{[d\phib]\,[d\pib]}{(2\pi\hbar)^{\cal V}}\,[d\phib']\,\frac{e^{-i\pib(\phib-\phib')/\hbar}}{E-H(\phib,\phib', t_0)}{\cal N}_{t_0}e^{i(S(\phib, \phib', E)-Et_0)/\hbar}\,.
\end{align}
Here ${\cal N}_{t_0}$ is a normalization factor. Performing a stationary phase approximation in the $\phib'$ integral which sets $\pib=-\frac{\delta}{\delta \phib'} S(\phib, \phib', E)$, the $\pib$ variable is identified with the momentum.  Then we obtain
\begin{equation}
    \int \frac{[d\phib]\,[d\pib]}{(2\pi\hbar)^{\cal V}}\,[d\phib']\,\frac{e^{-i\pib(\phib-\phib')/\hbar}}{E-H(\phib,\phib', t_0)}{\cal N}_{t_0}e^{i(S(\phib, \phib', E)-Et_0)/\hbar}=\int \frac{[d\phib]\,[d\pib]}{(2\pi\hbar)^{\cal V}}\,[d\phib']\,\frac{e^{-i\pib(\phib-\phib')/\hbar}}{E-H(\phib,\pib, t_0)}\langle \phib|e^{-i(H-E)t_0}|\phib'\rangle\,.
\end{equation}
Taking $t_0\rightarrow 0^+$ limit so that $\langle \phib|e^{-i(H-E)t_0}|\phib'\rangle\rightarrow\delta(\phib-\phib')$, we have
\begin{align}
\int \frac{[d\phib]\,[d\pib]}{(2\pi\hbar)^{\cal V}}\,[d\phib']\,\frac{e^{-i\pib(\phib-\phib')/\hbar}}{E-H(\phib,\phib', t_0)}\,\delta(\phib-\phib')&=\int [d\phib]\,\left[\frac{d\pib}{2\pi\hbar}\right]\,\frac{1}{E-H(\phib,\pib)}\,.
\end{align}
Altogether, we find
\begin{equation}
G_0(E) \approx \int [d\phib]\,\left[\frac{d\pib}{2\pi\hbar}\right]\,\frac{1}{E-H(\phib,\pib)}\,.
\end{equation}
Now we compute $G_{\text{osc}}$, which involves taking the trace of the Green's function over $\phib$ as: 
\begin{align}
&\int[d\phib]\,G_{\text{osc}}(\phib, \phib, E) \nonumber \\
&\qquad =\frac{1}{i\hbar}\sum_{\text{orbits }c}\int [d\phib]\left(\frac{1}{\big\|\dot{\phib}_c^A\big\|_{L^2} \big\|\dot{\phib}_c^B\big\|_{L^2}}\left|\det\left(\frac{1}{2\pi i \hbar}\frac{\delta^2 S_c(\phib^A, \phib^B, E)}{\delta\phib_{\perp}^A\delta\phib_{\perp}^B}\right)\right|^{1/2}_{\phib^A = \phib^B = \phib}\exp\left[\frac{i}{\hbar}S_c(\phib^A, \phib^B, E)-i\nu_c \frac{\pi}{2}\right]\right).
\end{align}
Performing the stationary phase approximation in $\phi$ that is induced by the trace yields the condition
\begin{equation}
    0=\left[\frac{\delta S_c(\phib^A, \phib^B, E)}{\delta\phib^A}+\frac{\delta S_c(\phib^A, \phib^B, E)}{\delta\phib^B}\right]_{\phib^A=\phib^B=\phib}=-\pib^A+\pib^B\,.
\end{equation}
This means that the stationary $\phib$'s belong to periodic orbits with endpoints $(\phib^A, \pib^A) = (\phib^B , \pib^B )$ in phase space.  Letting $(\phib_c, \pib_c)$ be a point along a periodic orbit, and letting $\phib = \phib_c + \delta\phib_{\perp}$ where $\delta \phib_\perp$ is a small fluctuation about $\phib_c$ perpendicular to the periodic orbit, we have the expansion
\begin{align}
&S_c(\phib, \phib, E)\approx\Bigg[S_c(\phib_c, \phib_c, E)+\frac{1}{2}\, \delta\phib_{\perp} \cdot \left(\frac{\delta^2 S_c(\phib^A, \phib^B, E)}{\delta \phib^A \delta\phib^A}+2\,\frac{\delta^2 S_c(\phib^A, \phib^B, E)}{\delta \phib^A \delta\phib^B}+\frac{\delta^2 S_c(\phib^A, \phib^B, E)}{\delta \phib^B \delta\phib^B}\right)_{\phib^A = \phib^B =\phib_{c}} \!\!\!\!\cdot \delta\phib_{\perp}\Bigg]\,.
\end{align}
Thus the $G_{\text{osc}}(E)$ term evaluates to
\begin{align}
&\int [d\phib]\,G_{\text{osc}}(\phib, \phib, E)\approx\frac{1}{i\hbar}\sum_{\text{orbits }c} \int_0^{T_c} dt \int [d\delta \phib_\perp]\,\left|\det\left(\frac{1}{2\pi i \hbar}\frac{\delta^2 S_c(\phib^A, \phib^B, E)}{\delta\phib_{\perp}^A\delta\phib_{\perp}^B}\right)\right|^{1/2}_{\phib^A = \phib^B = \phib_c} \\
& \times \exp\Bigg(\frac{i}{\hbar}\Bigg[S_c(\phib_c, \phib_c, E) \nonumber \\
& \qquad \qquad \qquad + \frac{1}{2}\, \delta\phib_{\perp} \cdot \left(\frac{\delta^2 S_c(\phib^A, \phib^B, E)}{\delta \phib^A \delta\phib^A}+2\,\frac{\delta^2 S_c(\phib^A, \phib^B, E)}{\delta \phib^A \delta\phib^B}+\frac{\delta^2 S_c(\phib^A, \phib^B, E)}{\delta \phib^B \delta\phib^B}\right)_{\phib^A = \phib^B =\boldsymbol{\Phi}_{c}(x,t)} \cdot \delta\phib_{\perp}\Bigg] - i \nu_c \frac{\pi}{2}\Bigg)\,.
\end{align}
Here we have summed over periodic orbits $\boldsymbol{\Phi}_c(x,t)$ with period $T_c$, i.e.~$\boldsymbol{\Phi}_c(x,0) = \boldsymbol{\Phi}_c(x,T_c) = \phib_c$, each labelled by $c$.  Performing the integral over $\delta \phib_\perp$, we obtain
\begin{align}
\int [d\phib]\,G_{\text{osc}}(\phib, \phib, E) \approx \frac{1}{i \hbar} \sum_{\text{orbits }c} \int_0^{T_c} dt \, \frac{1}{\sqrt{|\det(\textsf{M}(\boldsymbol{\Phi}_c(x,t)) - \mathds{1})|}} \, \exp\left(\frac{i}{\hbar}\, S_c(\phib_c, \phib_c, E ) - i \nu_c \frac{\pi}{2}\right)
\end{align}
where $\textsf{M} = \textsf{M}(\boldsymbol{\Phi}_c(x,t))$ is the monodromy matrix for $\boldsymbol{\Phi}_c(x,t)$, and we have absorbed the contribution of minus signs from negative eigenvalues in the determinant into the exponential phase term $\nu_c$.  As in the ordinary quantum mechanical setting, the spectrum of $\textsf{M}(\boldsymbol{\Phi}_c(x,t))$ does not depend on $t$ and so we can perform the $t$ integral above to obtain
\begin{align}
\int [d\phib]\,G_{\text{osc}}(\phib, \phib, E) &\approx \frac{1}{i \hbar} \sum_{\text{orbits }c}  \frac{T_c}{\sqrt{|\det(\textsf{M}(\boldsymbol{\Phi}_c(x,t)) - \mathds{1})|}} \, \exp\left(\frac{i}{\hbar}\, S_c(\phib_c, \phib_c, E ) - i \nu_c \frac{\pi}{2}\right) \nonumber \\
&= \frac{1}{i \hbar} \sum_{\text{orbits }c}  \frac{T_c}{\sqrt{|\det(\textsf{M}(\boldsymbol{\Phi}_c(x,t)) - \mathds{1})|}} \, \exp\left(\frac{i}{\hbar}\, S_c(\boldsymbol{\Phi}_c(x,t), \boldsymbol{\Phi}_c(x,t), E ) - i \nu_c \frac{\pi}{2}\right)\,.
\end{align}
where in the second line we have used $S_c(\phib_c, \phib_c, E ) = S_c(\boldsymbol{\Phi}_c(x,t), \boldsymbol{\Phi}_c(x,t), E )$ since the Legendre transform of Hamilton's principal function is independent of the $t$ of $\boldsymbol{\Phi}_c(x,t)$.

Adding the $G_0$ and $G_{\text{osc}}$ terms together, we arrive at the following expression for the trace of the Green's function:
\begin{align}
\int [d\phib] \,G(\phib,\phib,E) &\approx \int [d\phib] \left[\frac{d\phib}{2\pi \hbar}\right] \,\frac{1}{E - H(\phib, \pib)} \nonumber \\
& \qquad \qquad + \frac{1}{i \hbar} \sum_{\text{orbits }c} \frac{T_c}{\sqrt{|\det(\textsf{M}(\boldsymbol{\Phi}_c(x,t)) - \mathds{1})|}} \, \exp\left(\frac{i}{\hbar}\, S_c(\boldsymbol{\Phi}_c(x,t), \boldsymbol{\Phi}_c(x,t), E ) - i \nu_c \frac{\pi}{2}\right)\,.
\end{align}

\subsection{Scar formula in quantum field theory}

Finally we consider the quantum field theory scar formula, noting that by analogy to the quantum mechanics case, we average over an energy window of size $\varepsilon$ centered at $E$, and a window $\Delta$ (defined in the $L^2$ norm) centered on $\phib$. Again we work with energy eigenstates $\Psi_n[\phib]$ with energy $E_n$ which solve the Schr\"{o}dinger equation (here $n$ is a possibly continuous index), where now $\Psi[\phib]$ is a wavefunctional. The objective is to compute the expression
\begin{equation}
\label{E:desiredQFT1}
\langle|\Psi[\phib]|^2\rangle_{E,\Delta}=\frac{\sum_n\langle|\Psi_n[\phib]|^2\rangle_\Delta \delta_{\varepsilon}(E-E_n)}{\sum_n\delta_{\varepsilon}(E-E_n)}\,.
\end{equation}
In this equation $\langle|\Psi_n(\phib)|^2\rangle_{\Delta}$ indicates position averaging with a window of size $\Delta$, and $\delta_\varepsilon(x) := \frac{1}{\pi} \frac{\varepsilon}{\varepsilon^2 + x^2}$ is a smeared $\delta$-function.

Analogous to the ordinary quantum mechanical setting, we desire to write $\langle|\Psi_n(\phib)|^2\rangle_{E,\Delta}$ using $G(\phib^A, \phib^B, E)$.  As before, we make a useful assumption: for any triple $\phib^A$, $\phib^B$, $E$ with $\phib^A \not = \phib^B$, there is only a single orbit starting at $\phib^A$ and ending at $\phib^B$ with energy $E$.  For $\phib^A = \phib^B$ and $E$ fixed, if there are any non-trivial orbits (i.e.~non-fixed points) beginning and ending at $\phib^A = \phib^B$ then there must be exactly two such orbits which are time reverses of one another.  This holds because we have imposed time-reversal symmetry.  Indeed, these assumptions mean that $S_c$ has a single branch, and so we can drop the $c$ subscript.

We proceed by recalling that the Green's function solves the equation
\begin{equation}
    (E + i\varepsilon - H(\hat{\phib}^B, \hat{\pib}^B))\,G(\phib^A, \phib^B , E) = \delta(\phib^B - \phib^A)\,,
\end{equation}
and so we can write the Green's function as
\begin{equation}
    G(\phib^A, \phib^B, E+i\varepsilon)=\sum_n\frac{\Psi_n[\phib^A]\Psi_n^*[\phib^B]}{E+i\varepsilon-E_n}\,.
\end{equation}
It follows that
\begin{equation}
\label{E:numQFT1}
    -\frac{1}{\pi}\,\text{Im}\,G(\phib,\phib,E+i\varepsilon) = \sum_n|\Psi_n[\phib]|^2\delta_{\varepsilon}(E-E_n)\,,
\end{equation}
and 
\begin{equation}
\label{E:denomQFT1}
    -\frac{1}{\pi}\,\text{Im}\int [d\phib]\,G(\phib,\phib,E+i\varepsilon) = \sum_n\delta_{\varepsilon}(E-E_n)\,.
\end{equation}
If we take the ratio of~\eqref{E:numQFT1} and~\eqref{E:denomQFT1} and perform a $\Delta$-smearing of $\phib$, then we recover the right-hand side of~\eqref{E:desiredQFT1}.  We already know how to compute~\eqref{E:denomQFT1} from the previous subsection; using our formula for $\sum_n \delta(E - E_n)$ from there and putting in $i \varepsilon$'s, we have
\begin{align}
\sum_n \delta_\varepsilon(E - E_n) &\approx \int [d\phib] \left[\frac{d\pib}{2\pi \hbar}\right] \,\frac{1}{E + i \varepsilon - H(\phib, \pib)} \nonumber \\
& \qquad \qquad + \frac{1}{i \hbar} \sum_{\text{orbits }c} \frac{T_c}{\sqrt{|\det(\textsf{M}(\boldsymbol{\Phi}_c(x,t)) - \mathds{1})|}} \, \exp\left(\frac{i}{\hbar}\, S_c(\boldsymbol{\Phi}_c(x,t), \boldsymbol{\Phi}_c(x,t), E ) - i \nu_c \frac{\pi}{2}\right) \, \exp\left(- \varepsilon T_c/\hbar\right)
\end{align}
where the $\exp(-\varepsilon T_c/\hbar)$ suppresses longer orbits.

Next let us consider the Gaussian smearing of $\phib$ with width $\Delta$ in the sense of $L^2$.  As before, we suppose that $T_{\text{max}}$ is the period of the longest orbit with energy in $[E - \varepsilon/2, E + \varepsilon/2]$.  Taking\footnote{Restoring the speed of light, $\Delta = \sqrt{\hbar \, T_{\text{max}}\,c^2} \left(\frac{\hbar}{E\,T_{\text{max}}}\right)^{\gamma}$.}
\begin{equation}
\label{E:DeltaboundQFT1}
\Delta = \sqrt{\hbar \, T_{\text{max}}} \left(\frac{\hbar}{E\,T_{\text{max}}}\right)^{\gamma} \quad \text{for any } \quad \frac{1}{4} < \gamma < \frac{1}{2}\,,
\end{equation}
we write
\begin{align}
\sum_n \langle |\Psi_n[\phib]|^2\rangle_\Delta \, \delta_\varepsilon(E-E_n) = - \frac{1}{\pi}\,\text{Im}\left\{\frac{1}{(2\pi \Delta^2)^{\mathcal{V}}} \int [d\chib]\, G(\chib, \chib, E + i \varepsilon) \, e^{- \frac{1}{2\Delta^2} \| \phib - \chib \|_{L^2}^2}\right\}
\end{align}
where the argument of $\text{Im}\{\,\cdot\,\}$ can be written approximately in the semiclassical regime $\frac{E\,T_{\text{max}}}{\hbar} \gg 1$ as
\begin{align}
\label{E:toreferbackto1}
&\frac{1}{(2\pi \Delta^2)^{\mathcal{V}}} \int [d\chib]\,\left[\frac{d\pib}{2\pi \hbar}\right]\, \frac{1}{E + i \varepsilon - H(\chib, \pib)}\, e^{- \frac{1}{2\Delta^2}\|\phib - \chib\|_{L^2}^2}  \nonumber \\
+\,\,&\frac{2}{(2\pi \Delta^2)^{\mathcal{V}}} \, \frac{1}{i\hbar}\, \int [d\chib] \, \frac{1}{\|\dot{\chib}\|_{L^2}} \left|\det\left(\frac{\delta^2 S(\chib^A, \chib^B, E)}{\delta \chib_\perp^A \delta \chib_\perp^B}\right)\right|_{\chib_A = \chib_B = \chib}^{1/2} e^{\frac{i}{\hbar}S(\chib, \chib, E) - i \nu(\chib, \chib, E) \frac{\pi}{2} - \frac{1}{\hbar} \varepsilon T(\chib, \chib, E)} e^{- \frac{1}{2\Delta^2} \|\phib - \chib\|_{L^2}^2}\,.
\end{align}
There is a factor of $2$ in the second line which accounts for the time-reversal symmetry of our system.  In particular, if there is a non-trivial orbit beginning and ending at $\chib$ with energy $E$, then there are two such orbits.

In~\eqref{E:toreferbackto1} we only receive non-negligible contributions to the integrals when
\begin{equation}
\| \phib - \chib \|_{L^2} \lesssim \Delta 
\end{equation}
on account of the Gaussian decay induced by the $\Delta$-smearing.  Since $\Delta < \sqrt{\hbar T_{\text{max}}} \, \left(\frac{\hbar}{E\,T_{\text{max}}}\right)^{1/4}$, in the semiclassical regime $\frac{E\,T_{\text{max}}}{\hbar} \gg 1$ we can approximate
\begin{equation}
\frac{1}{(2\pi \Delta^2)^{\mathcal{V}}} \int [d\chib]\,\left[\frac{d\pib}{2\pi \hbar}\right]\, \frac{1}{E + i \varepsilon - H(\chib, \pib)}\, e^{- \frac{1}{2\Delta^2}\|\phib - \chib\|_{L^2}^2}  \approx  \int \left[\frac{d\pib}{2\pi \hbar}\right]\, \frac{1}{E + i \varepsilon - H(\phib, \pib)}\,.
\end{equation}
By contrast, the second line of~\eqref{E:toreferbackto1} contains the term $e^{\frac{i}{\hbar}S(\chib, \chib, E)}$ which can have non-trivial stationary phase contributions in the semiclassical regime.  We can approximate all of the other terms in that integral by setting $\chib = \phib$ as
\begin{align}
&\frac{2}{(2\pi \Delta^2)^{\mathcal{V}}} \, \frac{1}{i\hbar}\, \int [d\chib] \, \frac{1}{\|\dot{\chib}\|_{L^2}} \left|\det\left(\frac{\delta^2 S(\chib^A, \chib^B, E)}{\delta \chib_\perp^A \delta \chib_\perp^B}\right)\right|_{\chib_A = \chib_B = \chib}^{1/2} e^{\frac{i}{\hbar}S(\chib, \chib, E) - i \nu(\chib, \chib, E) \frac{\pi}{2} - \frac{1}{\hbar} \varepsilon T(\chib, \chib, E)} e^{- \frac{1}{2\Delta^2} \|\phib - \chib\|_{L^2}^2} \nonumber \\
&\approx \frac{2}{i\hbar} \frac{1}{\|\dot{\phib}\|_{L^2}} \left|\det\left(\frac{\delta^2 S(\phib^A, \phib^B, E)}{\delta \phib_\perp^A \delta \phib_\perp^B}\right)\right|_{\phib_A = \phib_B = \phib}^{1/2} e^{-i \nu(\phib, \phib, E)\frac{\pi}{2} - \frac{1}{\hbar} \varepsilon T(\phib, \phib, E)} \cdot \frac{1}{(2\pi \Delta^2)^{\mathcal{V}}} \int [d\chib]\, e^{\frac{i}{\hbar} S(\chib, \chib, E)} e^{- \frac{1}{2 \Delta^2} \|\phib - \chib\|_{L^2}^2}\,.
\end{align}
We can analyze the $\frac{1}{(2\pi \Delta^2)^{\mathcal{V}}} \int [d\chib]\, e^{\frac{i}{\hbar} S(\chib, \chib, E)} e^{- \frac{1}{2 \Delta^2} \|\phib - \chib\|_{L^2}^2}$ integral by expanding the Legendre transform of Hamilton's principal function around $\chib = \phib$:
\begin{align}
\label{E:chibphibapprox1}
&\frac{1}{(2\pi \Delta^2)^{\mathcal{V}}} \int [d\chib]\, e^{\frac{i}{\hbar} S(\chib, \chib, E)} e^{- \frac{1}{2 \Delta^2} \|\phib - \chib\|_{L^2}^2} \nonumber \\
\approx\,\,&\frac{1}{(2\pi \Delta^2)^{\mathcal{V}}} \int [d\chib]\, e^{- \frac{1}{2 \Delta^2} \|\phib - \chib\|_{L^2}^2 + \frac{i}{\hbar}\left(S(\phib, \phib, E) + (\chib - \phib)\cdot \left(\frac{\delta S(\chib^A, \chib^B, E)}{\delta \chib^A} + \frac{\delta S(\chib^A, \chib^B, E)}{\delta \chib^B}\right)_{\chib^A = \chib^B = \phib} + \frac{1}{2}(\chib - \phib) \cdot \textbf{A}[\phib]\cdot(\chib - \phib) + \cdots\right)}\,.
\end{align}
In the second line we have
\begin{equation}
 \left(\frac{\delta S(\chib^A, \chib^B, E)}{\delta \chib^A} + \frac{\delta S(\chib^A, \chib^B, E)}{\delta \chib^B}\right)_{\chib^A = \chib^B = \phib} = - \pib^A + \pib^B,
\end{equation}
and we define the kernel
\begin{equation}
\label{E:MQFTdef1}
    A_{ij}[\phib](x,y):=\left(\frac{\delta^2 S(\chib^A, \chib^B, E)}{\delta\chi_i^A(x)\delta\chi_j^A(y)}+2\frac{\delta^2 S(\chib^A, \chib^B, E)}{\delta\chi_i^A(x)\delta\chi_j^B(y)}+\frac{\delta^2 S(\chib^A, \chib^B, E)}{\delta\chi_i^B(x)\delta\chi_j^B(y)}\right)_{\chib^A=\chib^B=\phib}\,.
\end{equation}
The eigenvalues of $\textbf{A}$ are bounded in magnitude when $E$ is finite; a proof of this fact appears in Appendix~\ref{app:boundedness}. More specifically, defining the operator norm
\begin{equation}
\label{E:QFTMopnorm1}
    \| \textbf{A}[\phib] \|:=\frac{1}{\hbar \, T_{\text{max}}}\sup_{\|\chib\|_{L^2} = \sqrt{\hbar \, T_{\text{max}}}}\,\left| \chib \cdot \textbf{A} \cdot \chib\right|\,,
\end{equation}
from Appendix~\ref{app:boundedness} we have
\begin{equation}
\| \textbf{A} \| \lesssim \frac{1}{T_{\text{max}}}\, \sqrt{\frac{E\,T_{\text{max}}}{\hbar}}\,.
\end{equation}
We will also assume that the higher derivatives of $S$ have a corresponding tensor norm upper bounded by the same quantity, although we will not prove this.  Notice that because $\Delta < \sqrt{\hbar\, T_{\text{max}}}\,\left(\frac{\hbar}{E\,T_{\text{max}}}\right)^{1/4}$, the quadratic term $\frac{i}{2\hbar} \frac{1}{2}(\chib - \phib) \cdot \textbf{A}[\phib]\cdot(\chib - \phib)$ and the higher-order terms `$\cdots$' in the exponential in the second line of~\eqref{E:chibphibapprox1} contribute negligibly to the integral in the semiclassical regime.  As such, a good approximation to~\eqref{E:chibphibapprox1} is
\begin{equation}
\label{E:QFTinplaceof1}
\exp\left(\frac{i}{\hbar} \,S(\phib, \phib, E) - \frac{1}{2}\,\frac{\Delta^2}{\hbar^2} \, \|\pib^B - \pib^A\|_{L^2}^2\right)\,.
\end{equation}
But the right-hand side is exponentially suppressed unless
\begin{equation}
\label{E:QFTtorewritewithPi1}
\| \pib^A - \pib^B \|_{L^2} \lesssim \frac{\hbar}{\Delta}\,.
\end{equation}
To better unpack this bound, we can rewrite~\eqref{E:DeltaboundQFT1} as
\begin{equation}
\Delta = \frac{\hbar}{\overline{\pi}} \left(\frac{E\,T_{\text{max}}}{\hbar}\right)^{\frac{1}{2} - \gamma} \quad \text{for any } \quad \frac{1}{4} < \gamma < \frac{1}{2}\,,
\end{equation}
where
\begin{equation}
\overline{\pi}^2 := E\,.
\end{equation}
With this notation, we can express~\eqref{E:QFTtorewritewithPi1} as
\begin{equation}
\| \pib^A - \pib^B \|_{L^2} \lesssim \overline{\pi}\,\left(\frac{\hbar}{E\,T_{\text{max}}}\right)^{\frac{1}{2} - \gamma}\,.
\end{equation}
Then in the semiclassical regime we can approximate~\eqref{E:chibphibapprox1} by
\begin{equation}
\label{E:simpleQFTexponential1}
\exp\left(\frac{i}{\hbar}\,S(\phib, \phib, E)\right)\,.
\end{equation}
As in the ordinary quantum mechanical setting, there is an intermediate regime where $\|\pib^B - \pib^A\|_2 = c \, \frac{\hbar}{\Delta}$ for $c \sim 1$.  Within this regime $\exp\left(-\frac{1}{2} \frac{\Delta^2}{\hbar^2} \|\pib^B - \pib^A\|_{L^2}^2 \right)$ provides a moderate amount of decay.  We will not consider this regime further, and restrict ourselves to $c \ll 1$ so that we can use~\eqref{E:simpleQFTexponential1} in place of~\eqref{E:QFTinplaceof1}.

As we have seen, a crucial condition in the simplification above is $\| \pib^A - \pib^B \|_{L^2} \lesssim \frac{\hbar}{\Delta}$, or equivalently
\begin{equation}
\label{E:QFTineqtoreferto1}
\left\| \left(\frac{\delta S(\chib^A, \chib^B, E)}{\delta \chib^A} + \frac{\delta S(\chib^A, \chib^B, E)}{\delta \chib^B}\right)_{\chib^A = \chib^B = \phib} \right\|_{L^2} \lesssim \frac{\hbar}{\Delta}.
\end{equation}
As explained in the ordinary quantum mechanical setting in Appendix~\ref{subsec:Bogomolnyderivation1}, these inequalities are satisfied for exactly periodic solutions to the classical equations of motion, as well as nearby nearly-periodic orbits (i.e.~beginning and ending at the same $\phib$ but having different initial and final momenta $\pib^A$ and $\pib^B$).  There may be other nearly-periodic orbits which are not in the vicinity of any exactly periodic orbit, and these can satisfy the inequalities as well.  However, these latter kind of orbits are difficult to find analytically, and so we confine our attention to exactly periodic orbits and nearby nearly-periodic orbits.

As before, let $\mathcal{O}$ be the union of the images of all \textit{exactly} periodic solutions to the equations of motion.  If $\mathcal{F}$ is the space of complex fields $\phi : \mathbb{R}^d \to \mathbb{C}$, then $\mathcal{O}$ can be regarded as a subset of $\mathcal{F}$.  We equip $\mathcal{F}$ with the $L^2$ inner product we have defined previously.  Let us assume that $\mathcal{O}$ is a disjoint union of smooth, connected manifolds of finite dimension (and hence infinite codimension since they sit in an infinite-dimensional space), and that the minimum pairwise distance between these manifolds (using the $L^2$ norm as a distance) is upper bounded by some positive constant.  We denote each component of $\mathcal{O}$ by $\mathcal{O}_i$.  Let us pick a particular $\mathcal{O}_i$ to study and call it $\mathcal{M}$, and suppose that it has dimension $k$.

We further pick some $\phib$ nearby the submanifold $\mathcal{M}$ such that (i) $\phib = \phib_c + \delta \phib$ for $\phib_c \in \mathcal{M}$, and (ii) $\| \delta \phib \|_{L^2} \lesssim \frac{\hbar}{\Delta}$.  Such a $\phib$ will satisfy~\eqref{E:QFTineqtoreferto1} since $\left(\frac{\delta S(\chib^A, \chib^B, E)}{\delta \chib^A} + \frac{\delta S(\chib^A, \chib^B, E)}{\delta \chib^B}\right)_{\chib^A = \chib^B = \phib} \approx \delta \phib \cdot \textbf{A}[\phib_c]$ and $\|\delta \phib \cdot \textbf{A}[\phib_c]\|_{L^2} \leq \|\delta \phib\|_{L^2} \|\textbf{A}[\phib_c]\| \lesssim \frac{\hbar}{\Delta}$.

For $\phib$ in the regime of (i) and (ii) above, we can perform an expansion in the exponent of~\eqref{E:simpleQFTexponential1} as
\begin{equation}
\label{E:simpleQFTexponential2}
\exp\left(\frac{i}{\hbar}\,S(\phib, \phib, E)\right) \approx \exp\left(\frac{i}{\hbar}\,S(\phib_c, \phib_c, E) + \frac{i}{2\hbar}\,\delta \phib \cdot \textbf{A}[\phib_c] \cdot \delta \phib \right)\,.
\end{equation}
We have dropped the terms at higher than quadratic order since we have assumed that they are upper bounded by $\frac{1}{T_{\text{max}}} \sqrt{\frac{E\,T_{\text{max}}}{\hbar}}$ in the tensor norm and also $\|\delta \phib\|_{L^2} \lesssim \frac{\hbar}{\Delta}$.  Let us decompose $\mathcal{F} := T_{\phib_c} \mathcal{M} \oplus \text{N}_{\phib_c}\mathcal{M}$, which enables us to orthogonally decompose $\delta \phib = \delta \phib_{\parallel} + \delta \phib_{\perp}$ where $\delta \phib_{\parallel} \in T_{\phi_c} \mathcal{M}$ and $\delta \phib_{\perp} \in \text{N}_{\phi_c} \mathcal{M}$.  Note that since $\dim \mathcal{M} = k$, we have $\dim T_{\phib_c} \mathcal{M} = k$ and $\dim \text{N}_{\phib_c} \mathcal{M} = \infty$.  Since $\left\| \left(\frac{\delta S(\chib^A, \chib^B, E)}{\delta \chib^A} + \frac{\delta S(\chib^A, \chib^B, E)}{\delta \chib^B}\right)_{\chib^A = \chib^B = \phib_c} \right\|_{L^2} \lesssim \frac{\hbar}{\Delta}$ is constant (namely zero) along $\mathcal{M}$, for any $\phib_c \in \mathcal{M}$ we have that $T_{\phib_c}\mathcal{M}$ is in the nullspace of $\textbf{A}[\phib_c]$.  Accordingly, the right-hand side of~\eqref{E:simpleQFTexponential2} equals
\begin{equation}
\label{E:simpleQFTexponential3}
 \exp\left(\frac{i}{\hbar}\,S(\phib_c, \phib_c, E) + \frac{i}{2\hbar}\,\delta \phib_\perp \cdot \textbf{A}[\phib_c] \cdot \delta \phib_\perp \right)\,.   
\end{equation}
Given a point $\phib = \phib_c + \delta \phib$ satisfying the assumptions of (i) and (ii) above, we can opt to find a $\phib_c'$ such that $\phib = \phib_c' + \delta \phib'$ where $\delta \phib'$ lies only in the normal bundle $\text{N}_{\phib'} \mathcal{M}$, i.e.~$\delta \phib' = \delta \phib_\perp'$.  In particular, we can choose $\phib_c'$ to be the point on $\mathcal{M}$ closest to $\phib$.  Identical arguments as those in Appendix~\ref{subsec:Bogomolnyderivation1} tell us that
\begin{equation}
\label{E:simpleQFTexponential4}
 \exp\left(\frac{i}{\hbar}\,S(\phib_c', \phib_c', E) + \frac{i}{2\hbar}\,\delta \phib_\perp' \cdot \textbf{A}[\phib_c'] \cdot \delta \phib_\perp' \right)
\end{equation}
and~\eqref{E:simpleQFTexponential3} agree in semiclassical limit.

We end this subsection by collecting all of our results and writing down the scar formula for quantum field theory.  We have found that
\begin{align}
& \frac{1}{(2\pi \Delta^2)^{\mathcal{V}}}\int [d\chib] \, e^{\frac{i}{\hbar}S(\chib, \chib,E)}\,e^{-\frac{1}{2\Delta^2} \|\phib-\chib\|_{L^2}^2} \nonumber \\ \nonumber \\
& \qquad  \approx \begin{cases} e^{\frac{i}{\hbar}S(\phib_c, \phib_c, E) + \frac{i}{2\hbar}\,\delta \phib_\perp \cdot \textbf{A}[\phib_c] \cdot \delta \phib_\perp} & \text{if  }\phib = \phib_c + \delta \phib_\perp\text{  where  }\phib_c \in \mathcal{O},\, \delta \phib_\perp \in \text{N}_{\phib_c} \mathcal{O},\,\,\|\delta \phib_\perp\|_{L^2} \lesssim \frac{\hbar}{\Delta} \\ \\
0 & \text{if  } \left\|\left(\frac{\delta S(\chib^A,\chib^B,E)}{\delta \chib^A}+\frac{\delta S(\chib^A,\chib^B,E)}{\delta \chib^B}\right)_{\chib^A=\chib^B=\textbf{q}}\right\|_{L^2} \gg \frac{\hbar}{\Delta}
\end{cases}\,.
\end{align}
The two cases above do not cover two different types of domains of $\phib$\,: one domain such that $\left\|\left(\frac{\delta S(\chib^A,\chib^B,E)}{\delta \chib^A}+\frac{\delta S(\chib^A,\chib^B,E)}{\delta \chib^B}\right)_{\chib^A=\chib^B=\textbf{q}}\right\|_{L^2} \approx \frac{\hbar}{\Delta}$, and another domain such that $\|\delta \phib_\perp \|_{L^2} \gtrsim \frac{\hbar}{\Delta}$ but $\left\|\left(\frac{\delta S(\chib^A,\chib^B,E)}{\delta \chib^A}+\frac{\delta S(\chib^A,\chib^B,E)}{\delta \chib^B}\right)_{\chib^A=\chib^B=\phib}\right\|_2 \lesssim \frac{\hbar}{\Delta}$.  Here the first type of domain is not physically interesting, since it corresponds to $\phib$ which are just far away enough from exactly periodic orbits such that we need to use~\eqref{E:QFTinplaceof1} in place of of~\eqref{E:simpleQFTexponential1}.  However, the second type of domain is quite interesting, and corresponds to nearly-periodic orbits (i.e.~periodic in $\phib$ but not in $\pib$) which are not nearby exactly periodic orbits. 

To write down the actual scar formula, we require the definitions
\begin{equation}
\label{E:microdefQFT1}
P_{\text{micro}}[\phib] := \frac{\int [d\pib] \,\delta_\varepsilon(E - H(\phib, \pib))}{\int [d\phib]\,[d\pib]\,\delta_\varepsilon(E - H(\phib, \pib))}
\end{equation}
and
\begin{align}
\label{E:PscarQFT1}
&\delta P_{\text{scar}}[\phib_c, \delta \phib_\perp] := -\frac{2}{\pi \hbar \int [d\chib]\left[\frac{d \textbf{p}}{2\pi \hbar}\right] \delta_\varepsilon(E - H(\chib,\pib))} \,\text{Im} \Bigg\{\frac{1}{i} \frac{1}{\|\dot{\phib}\|_{L^2}}\left|\det\left(\frac{1}{2\pi i \hbar}\frac{\delta^2 S(\phib^A,\phib^B,E)}{\delta \phib_\perp^A\delta \phib_\perp^B}\right)\right|_{\phib^A = \phib^B = \phib}^{1/2} \nonumber \\
& \qquad \qquad \qquad \qquad \qquad \qquad \times \exp\left[- \frac{\varepsilon}{\hbar} \, T(\phib_c, \phib_c, E) - i \nu(\phib_c, \phib_c, E)\frac{\pi}{2} + \frac{i}{\hbar}\left(S(\phib_c, \phib_c, E) + \frac{1}{2}\, \delta \phib_\perp \cdot \textbf{A}[\phib_c] \cdot \delta \phib_\perp\right)\right]\Bigg\}\,.
\end{align}
The functional determinant term must be regularized, as is standard in quantum field theory.  The scar formula can be written as
\begin{equation}
\label{E:desiredQFTscar1}
\boxed{\langle |\Psi[\phib]|^2\rangle_{E,\Delta} \approx \begin{cases} P_{\text{micro}}[\phib] + \delta P_{\text{scar}}[\phib_c, \delta \phib_\perp] & \text{if  }\phib = \phib_c + \delta \phib_\perp\text{  where  }\phib_c \in \mathcal{O},\, \delta \phib_\perp \in \text{N}_{\phib_c} \mathcal{O},\,\,\|\delta \phib_\perp\|_{L^2} \lesssim \frac{\hbar}{\Delta} \\ \\
P_{\text{micro}}[\phib] & \text{if  } \left\|\left(\frac{\delta S(\chib^A,\chib^B,E)}{\delta \chib^A}+\frac{\delta S(\chib^A,\chib^B,E)}{\delta \chib^B}\right)_{\chib^A=\chib^B=\phib}\right\|_{L^2} \gg \frac{\hbar}{\Delta}
\end{cases}}
\end{equation}
The approximate nature of the scar formula above means that it includes multiplicative corrections $\big( 1 + O\big(\frac{\varepsilon}{E}\,,(\!\frac{\hbar}{E \,T_{\text{max}}}\!)^\gamma\big)\,\big)$.  Moreover, we recall the assumptions that went into the derivation of the quantum field theory scar formula, which are directly analogous to those we used in the ordinary quantum mechanical setting: (i) for each triplet $\phib^A, \phib^B, E$ with $\phib^A \not = \phib^B$, there is a single classical trajectory starting at $\phib^A$, ending at $\phib^B$, and having energy $E$; if there are any non-trivial orbits beginning and ending at $\phib^A = \phib^B$ with energy $E$, then there must be exactly two which are time-reverses of one another; (ii) $\| \textbf{A}(\phib) \| \leq \frac{1}{T_{\text{max}}} \sqrt{\frac{E\,T_{\text{max}}}{\hbar}}$ for $\phib \in \mathcal{O}$ and likewise for the third order and higher order derivatives of $S$; (iii) $\mathcal{O}$ is a disjoint union of smooth manifolds of finite dimension (for instance this means the manifolds do not have self-intersections); and (iv) the only values of $\phib$ for which $\left\|\left(\frac{\delta S(\chib^A,\chib^B,E)}{\delta \chib^A}+\frac{\delta S(\chib^A,\chib^B,E)}{\delta \chib^B}\right)_{\chib^A=\chib^B=\phib}\right\|_{L^2} \lesssim \frac{\hbar}{\Delta}$ are those which are close to $\mathcal{O}$. 

The comments at the end of Appendix A about nearly periodic orbits in the ordinary quantum mechanical setting also generalize to the quantum field theory setting.  In particular, if $\mathcal{O}_{\text{nearly}}$ is a manifold in field space $\mathcal{F}$ which consists of nearly periodic orbits (in the sense of $\left\|\left(\frac{\delta S(\chib^A,\chib^B,E)}{\delta \chib^A}+\frac{\delta S(\chib^A,\chib^B,E)}{\delta \chib^B}\right)_{\chib^A=\chib^B=\phib}\right\|_{L^2} \lesssim \frac{\hbar}{\Delta}$\,) which are not nearby any exactly periodic orbits, then for $\phib$'s on or near $\mathcal{O}_{\text{nearly}}$ we could write an expression for $\langle |\Psi(\phib)|^2\rangle_{E,\Delta}$ which is nearly identical to the boxed scar formula.  This may be useful for studying so-called oscillons~\cite{kudryavtsev1975solitonlike,bogolyubsky1976lifetime,gleiser1993pseudostable,copeland1995oscillons}, which are examples of nearly-periodic orbits in quantum field theories.  However, in many cases, it is hard to ascertain whether or not an oscillon is nearby (or perhaps equal to) an exactly periodic solution due to numerical precision.  Moreover, it seems difficult to classify moduli spaces of nearly periodic oscillons that are not nearby exactly periodic orbits.  These difficulties aside, the broader point is that oscillons may indeed contribute to quantum scarring in bands of eigenfunctions in quantum field theories.

\section{Dimensional Analysis}
\label{app:dimensions}
Here we discuss dimensional analysis for both the ordinary quantum mechanical scar formula as well as its quantum field theory counterpart.

The ordinary quantum mechanical scar formula in~\eqref{E:desired1} has two cases.  In the second case, we only need to consider $P_{\text{micro}}(\textbf{q})$, defined in~\eqref{E:microdef1}.  We see from the definition that $P_{\text{micro}}(\textbf{q})$ has dimensions of $1/[L]^d$, as expected.  In the first case of the scar formula~\eqref{E:desired1}, there is also a contribution from $\delta P_{\text{scar}}(\textbf{q}, \delta \textbf{q}_\perp)$.  This should also have dimensions of $1/[L]^d$, but let us check this.  We recall the formula~\eqref{E:Pscar1}, namely
\begin{align*}
&\delta P_{\text{scar}}(\textbf{q}_c, \delta \textbf{q}_\perp) := -\frac{2}{\pi \hbar \int d^d \textbf{z}\,\frac{d^d \textbf{p}}{(2\pi \hbar)^d} \,\delta_\varepsilon(E - H(\textbf{z},\textbf{p}))} \,\text{Im} \Bigg\{\frac{1}{i} \frac{1}{\sqrt{(2\pi i\hbar)^{d-1}}} \frac{1}{|\dot{\textbf{q}}|}\left|\det\left(\frac{\partial^2 S(\textbf{q}^A,\textbf{q}^B,E)}{\partial \textbf{q}_\perp^A\partial \textbf{q}_\perp^B}\right)\right|_{\textbf{q}^A = \textbf{q}^B = \textbf{q}}^{1/2} \nonumber \\
& \qquad \qquad \qquad \qquad \qquad \qquad \times \exp\left[- \frac{\varepsilon}{\hbar} \, T(\textbf{q}_c, \textbf{q}_c, E) - i \nu(\textbf{q}_c, \textbf{q}_c, E)\frac{\pi}{2} + \frac{i}{\hbar}\left(S(\textbf{q}_c, \textbf{q}_c, E) + \frac{1}{2}\, \delta \textbf{q}_\perp \cdot \textbf{A}(\textbf{q}_c) \cdot \delta \textbf{q}_\perp\right)\right]\Bigg\}\,.
\end{align*}
A short inspection shows that this term has dimension
\begin{equation}
    \frac{[E]}{\hbar^{(d+1)/2}}\frac{[T]}{[L]}\left(\frac{\hbar}{[L]^2}\right)^{(d-1)/2}=\frac{1}{[L]^d}\frac{[E][T]}{\hbar}=\frac{1}{[L]^d}\,,
\end{equation}
which gives the expected answer.

Now let us turn to the quantum field theory setting.  The corresponding scar formula is given by~\eqref{E:desiredQFTscar1}, which again has two cases.  In the second case, we only need to examine $P_{\text{micro}}[\phib]$, which is given above in~\eqref{E:microdefQFT1}.  This has dimensions of $1/[\phi]^{2\mathcal{V}}$, where $\mathcal{V}$ can be regarded as the `number of points in space'.  The factor of $2$ in the exponent accounts for $\phi$ being a complex scalar field (i.e.~it has two components).  Indeed, $1/[\phi]^{2\mathcal{V}}$ is the expected answer since $P_{\text{micro}}[\phib]$ is a probability functional in $\phib = (\phi_1, \phi_2)$.

In the first case of~\eqref{E:desiredQFTscar1}, we also need to contend with $\delta P_{\text{scar}}[\phib_c, \delta\phib_\perp]$, which is defined in~\eqref{E:microdefQFT1} above.  Since this equation is on the previous page, we do not rewrite it here.  Examining the form of the equation reveals that it has dimension
\begin{equation}
    \frac{[E]}{\hbar}\frac{[T]}{[\phi]}\left(\frac{1}{\hbar}\frac{\hbar}{[\phi]^2}\right)^{(2{\cal V}-1)/2}=\frac{1}{[\phi]^{2{\cal V}}}\,,
\end{equation}
which is the expected answer.

\section{Boundedness of Hessian for fixed energy trajectories}
\label{app:boundedness}

In this Appendix we bound the norm of the $\textbf{A}$ operator in both the ordinary quantum mechanical setting and the quantum field theory setting.  The bounds and their derivations are very similar to one another.

\subsection{Boundedness in the ordinary quantum mechanical setting}

In~\eqref{E:firstAdef1} in Appendix \ref{sec:bogomolny}, we defined the operator
\begin{equation}
\label{eq:hessianQM}
A_{ij}(\textbf{q}) := \left(\frac{\partial^2 S(\textbf{z}^A,\textbf{z}^B,E)}{\partial z^A_i \partial z^A_j}+2\,\frac{\partial^2 S(\textbf{z}^A,\textbf{z}^B,E)}{\partial z^A_i \partial z^B_j}+\frac{\partial^2 S(\textbf{z}^A,\textbf{z}^B,E)}{\partial z^B_i \partial z^B_j}\right)_{\textbf{z}^A=\textbf{z}^B=\textbf{q}}\,.
\end{equation}
In~\eqref{E:forinstance1} we further defined the operator norm $\| A_{ij}(\textbf{q}) \|$, which we reprise here in slightly more explicit notation:
\begin{equation}
    \| A_{ij}(\textbf{q}) \|:=\sup_{v_i} \left|\frac{\sum_{i,j} v_i \, A_{ij}(\textbf{q})\, v_j}{\sum_i v_i^2}\right|\,.
\end{equation}
Our goal here is to show that $\| A_{ij}(\textbf{q}) \|$ is bounded when $\textbf{q}$ is on a periodic orbit with energy $E$.  As in Appendix \ref{sec:bogomolny}, we assume that our quantum mechanical theory has canonical momentum terms with mass $m$, and that the theory is invariant under time-reversal.  Our approach to bounding the norm is to show that the three terms in~\eqref{eq:hessianQM} are individually bounded. In particular, we use the triangle inequality
\begin{equation}
\label{E:Mnormtriangle1QM}
   \left\|A_{ij}(\textbf{q})\right\| \leq \left\|\left(\frac{\partial^2 S(\textbf{z}^A, \textbf{z}^B, E)}{\partial z_i^A\partial z_j^A} + \frac{\partial^2 S(\textbf{z}^A, \textbf{z}^B, E)}{\partial z_i^B \partial z_j^B}\right)_{\textbf{z}^A=\textbf{z}^B=\textbf{q}} \right\|+2\left\|\frac{\partial^2 S(\textbf{z}^A, \textbf{z}^B, E)}{\partial z_i^A\partial z_j^B}\Bigg|_{\textbf{z}^A=\textbf{z}^B=\textbf{q}} \right\|\,.
\end{equation}
We start with the definition of the Legendre transform of Hamilton's principal function to fixed energy:
\begin{align}
\label{E:SEdef1QM}
    S(\textbf{z}^A, \textbf{z}^B, E) := S(\textbf{z}^A, \textbf{z}^B, t(\textbf{z}^A, \textbf{z}^B, E))+E\,t(\textbf{z}^A, \textbf{z}^B, E)\,.
\end{align}
Let us begin with a bound on the second term on the right-hand side of~\eqref{E:Mnormtriangle1QM}, which contains the mixed $A$ and $B$ derivatives.

\subsubsection{Mixed derivative term}

Letting $\frac{d}{dz}$ denote a total derivative (as opposed to a partial derivative), we can leverage~\eqref{E:SEdef1QM} to write
\begin{align}
\label{E:firstmixed1QM}
    \left.\frac{\partial}{\partial z^A_i} \frac{\partial }{\partial  z^B_j}\,S(\textbf{z}^A, \textbf{z}^B, E)\right|_{\textbf{z}^A=\textbf{z}^B = \textbf{q}}&= \left.\frac{d}{d z^A_i} \frac{d}{d z^B_j}\left(S(\textbf{z}^A, \textbf{z}^B, t(\textbf{z}^A, \textbf{z}^B, E))+E\,t(\textbf{z}^A, \textbf{z}^B, E)\right)\right|_{\textbf{z}^A = \textbf{z}^B} \nonumber \\
    &=\left.\frac{\partial p^B_j}{\partial z^A_i}+\frac{\partial p^B_j}{\partial t}\frac{\partial  t}{\partial z^A_i}\right|_{\textbf{z}^A=\textbf{z}^B = \textbf{q}} \\ \nonumber \\
    \label{E:secondmixed1QM}
    \left.\frac{\partial }{\partial  z^B_j}\frac{\partial }{\partial  z^A_i}\,S(\textbf{z}^A, \textbf{z}^B, E)\right|_{\textbf{z}^A=\textbf{z}^B = \textbf{q}}&= \left.\frac{d}{d z^B_j}\frac{d}{d z^A_i}\left(S(\textbf{z}^A, \textbf{z}^B, t(\textbf{z}^A, \textbf{z}^B, E))+E\,t(\textbf{z}^A, \textbf{z}^B, E)\right)\right|_{\textbf{z}^A=\textbf{z}^B = \textbf{q}} \nonumber \\
    &= \left.-\frac{\partial p^A_i}{\partial z^B_j}-\frac{\partial p^A_i}{\partial t}\frac{\partial  t}{\partial z^B_j}\right|_{\textbf{z}^A=\textbf{z}^B = \textbf{q}}\,.
\end{align}
But since $\frac{\partial}{\partial z^B_j}\frac{\partial}{\partial z^A_i}\,S(\textbf{z}^A, \textbf{z}^B, E) = \frac{\partial}{\partial z^A_i}\frac{\partial}{\partial z^B_j}\,S(\textbf{z}^A, \textbf{z}^B, E)$, we can average~\eqref{E:firstmixed1QM} and~\eqref{E:secondmixed1QM} to obtain
\begin{align}
 \left.\frac{\partial}{\partial z^A_i} \frac{\partial}{\partial z^B_j}\,S(\textbf{z}^A, \textbf{z}^B, E)\right|_{\textbf{z}^A=\textbf{z}^B = \textbf{q}}&=\frac{1}{2}\left(-\frac{\partial p_i^A}{\partial z_j^B}+\frac{\partial p_j^B}{\delta z^A_i}\right)_{\textbf{z}^A=\textbf{z}^B = \textbf{q}}+\frac{1}{2}\left(-\frac{\partial p_i^A}{\partial t}\frac{\partial t}{\partial z^B_j}+\frac{\partial p_j^B}{\partial t}\frac{\partial t}{\partial z^A_i}\right)_{\textbf{z}^A=\textbf{z}^B = \textbf{q}}\,.
\end{align}
Thus we have
\begin{align}
\label{E:mixednorm1QM}
\left\|\frac{\partial^2 S(\textbf{z}^A, \textbf{z}^B, E)}{\partial z_i^A \partial z_j^B}\Bigg|_{\textbf{z}^A= \textbf{z}^B=\textbf{q}} \right\| &= \frac{m}{\hbar \, T_{\text{max}}} \sup_{\|\delta \textbf{z}\|_{L^2}=\sqrt{\hbar \, T_{\text{max}}/m}}\frac{1}{2}\sum_{i,j}\delta z_i \left(-\frac{\partial p_i^A}{\partial z_j^B}+\frac{\partial p_j^B}{\partial z^A_i}-\frac{\partial p_i^A}{\partial t}\frac{\partial t}{\partial z^B_j}+\frac{\partial p_j^B}{\partial t}\frac{\partial t}{\partial z^A_i}\right)_{\textbf{z}^A=\textbf{z}^B= \textbf{q}}\delta z_j\,.
\end{align}
The right-hand side can be simplified.  Since $\textbf{q}$ is on a periodic orbit with energy $E$, we have $\textbf{z}^A = \textbf{z}^B$, and hence $\frac{\partial \textbf{p}_A}{\partial t} = \frac{\partial \textbf{p}_B}{\partial t}$.  Moreover, since our theory is time-reversal symmetric, $t(\textbf{z}^A, \textbf{z}^B, E) = t(\textbf{z}^B, \textbf{z}^A, E)$, and so $\left.\frac{\partial t}{\partial z^B_i}\right|_{\textbf{z}^A = \textbf{z}^B = \textbf{q}} = \left.\frac{\partial t}{\partial z^A_i(y)}\right|_{\textbf{z}^A = \textbf{z}^B = \textbf{q}}$.  Accordingly, the last two terms in the parentheses in~\eqref{E:mixednorm1QM} cancel, and so we are left with
\begin{align}
\label{E:mixednorm2QM}
\left\|\frac{\partial^2 S(\textbf{z}^A, \textbf{z}^B, E)}{\partial z_i^A \partial z_j^B}\Bigg|_{\textbf{z}^A=\textbf{z}^B=\textbf{q}} \right\| &= \frac{m}{\hbar \, T_{\text{max}}} \sup_{\|\delta \textbf{z}\|_{L^2}=\sqrt{\hbar \, T_{\text{max}}/m}}\sum_{i,j} \delta z_i\left(-\frac{\partial p_i^A}{\partial z_j^B}+\frac{\partial p_j^B}{\partial z^A_i}\right)_{\textbf{z}^A=\textbf{z}^B=\textbf{q}}\delta z_j\nonumber \\
&\leq \frac{1}{2}\left\|\left.\frac{\partial p_i^A}{\partial z_j^B}\right|_{\textbf{z}^A = \textbf{z}^B = \textbf{q}} \right\| + \frac{1}{2}\left\|\left.\frac{\partial p_j^B}{\partial z_i^A}\right|_{\textbf{z}^A = \textbf{z}^B = \textbf{q}} \right\|\,.
\end{align}
Let us bound the first of the two terms on the right-hand side of the inequality; a bound on the second term will follow by a nearly identical argument.

We have
\begin{align}
\label{E:pitobound1QM}
\left\|\left.\frac{\partial p_i^A}{\partial z_j^B}\right|_{\textbf{z}^A = \textbf{z}^B = \textbf{q}} \right\| &= \frac{m}{\hbar \, T_{\text{max}}} \sup_{\|\delta \textbf{z}\|_{L^2}=\sqrt{\hbar \, T_{\text{max}}/m}}\sum_{i,j}\delta z_i \left(\frac{\partial p_j^B}{\partial z^A_i}\right)_{\textbf{z}^A=\textbf{z}^B = \textbf{q}}\delta z_j\,.
\end{align}
Defining
\begin{equation}
\label{E:deltapijBdef1QM}
\delta p_j^B := \sum_{i} \delta z_i\left(\frac{\partial p_j^B}{\partial z^A_i}\right)_{\textbf{z}^A=\textbf{z}^B = \textbf{q}},
\end{equation}
which depends on $\delta z_i$, we can rewrite~\eqref{E:pitobound1QM} as
\begin{align}
\frac{m}{\hbar \, T_{\text{max}}} \sup_{\|\delta \textbf{z}\|_{L^2}=\sqrt{\hbar \, T_{\text{max}}/m}}\sum_{j} \delta p_j^B \,\delta z_j &\leq \frac{m}{\hbar \, T_{\text{max}}} \sup_{\|\delta \textbf{z}\|_{L^2}=\sqrt{\hbar \, T_{\text{max}}/m}}  \|\delta \textbf{p}^B\|_{L_2} \| \delta \textbf{z} \|_{L^2} \nonumber \\
&\leq \sqrt{\frac{m}{\hbar \, T_{\text{max}}}}\, \sup_{\|\delta \textbf{z}\|_{2}=\sqrt{\hbar \, T_{\text{max}}/m}}  \|\delta \textbf{p}^B\|_{2}\,,
\end{align}
where we have used the Cauchy-Schwarz inequality.  Finally, we bound $\|\delta \textbf{p}^B\|_{2}$.  To do so, we note the interpretation of the definition of $\delta \textbf{p}^B$ in~\eqref{E:deltapijBdef1QM}: it is the change in momentum at the final time (i.e.~at $B$) if we perturb our periodic orbit by $q_j \to q_j + \delta z_j$ at the initial time (i.e.~at $A$), but still insist that it ends at the same $q_j$ and has the same energy $E$.  Letting $\textbf{q}^A = \textbf{q}^B = \textbf{q}$ be the unperturbed initial position as usual, and further letting $\textbf{p}^A = \textbf{p}^B = \textbf{p}$ be the unperturbed initial and final momentum, we have
\begin{align}
    E&= \frac{1}{2m}(\textbf{p}+\delta\textbf{p}^B)^2 +U(\textbf{q})\,.
\end{align}
Letting $U_{\text{min}} := \min_{\textbf{q}} U(\textbf{q})$, we assume this value to be finite.  Without loss of generality, we also assume $U_{\text{min}} \geq 0$.  Then we obtain
\begin{equation}
\| \textbf{p} + \delta \textbf{p}^B \|_{2}^2 \leq 2 m (E - U_{\text{min}})\,.
\end{equation}
By the reverse triangle inequality, we have
\begin{equation}
    \|\delta \textbf{p}^B\|_{2}\leq\sqrt{2m(E - U_{\text{min}})}+\|\textbf{p}\|_{2} \leq 2 \sqrt{2m(E - U_{\text{min}})}\,,
\end{equation}
which is finite.  In total we have the bound
\begin{equation}
\left\|\frac{\partial^2 S(\textbf{z}^A, \textbf{z}^B, E)}{\partial z_i^A\partial z_j^B}\Bigg|_{\textbf{z}^A=\textbf{z}^B=\textbf{q}} \right\| \leq 2\sqrt{\frac{m}{\hbar \, T_{\text{max}}}}\, \sqrt{2m(E - U_{\text{min}})}\,,
\end{equation}
as we desired.

\subsubsection{Non-mixed derivative terms}

Now we turn to bounding the first term on the right-hand side of~\eqref{E:Mnormtriangle1QM}.  By similar arguments as in the subsection above, we have
\begin{align}
\left(\frac{\partial^2 S(\textbf{z}^A, \textbf{z}^B, E)}{\delta z_i^A\delta z_j^A} + \frac{\delta^2 S(\textbf{z}^A, \textbf{z}^B, E)}{\partial z_i^B\delta z_j^B}\right)_{\textbf{z}^A=\textbf{z}^B=\textbf{q}} = \left(-\frac{\partial p_i^A}{\partial z^A_j}+\frac{\partial p_i^B}{\partial z^B_j}-\frac{\partial p_i^A}{\partial t}\frac{\partial t}{\partial z^A_j}+\frac{\partial p_i^B}{\partial t}\frac{\partial t}{\partial z^B_j}\right)_{\textbf{z}^A=\textbf{z}^B = \textbf{q}}\,.
\end{align}
Noting as before that for our periodic orbit $\textbf{p}^A = \textbf{p}^B$ and by time-reversal symmetry $t(\textbf{z}_A, \textbf{z}_B, E) = t(\textbf{z}_B, \textbf{z}_A, E)$, the above simplifies to
\begin{equation}
\left(\frac{\partial ^2 S(\textbf{z}^A, \textbf{z}^B, E)}{\partial z_i^A\partial z_j^A} + \frac{\partial ^2 S(\textbf{z}^A, \textbf{z}^B, E)}{\partial z_i^B\partial z_j^B}\right)_{\textbf{z}^A=\textbf{z}^B=\textbf{q}} = \left(-\frac{\partial p_i^A}{\partial z^A_j}+\frac{\partial p_i^B}{\partial z^B_j}\right)_{\textbf{z}^A=\textbf{z}^B = \textbf{q}}\,.
\end{equation}
Taking the norm of both sides, we have
\begin{align}
\left\| \left(\frac{\partial^2 S(\textbf{z}^A, \textbf{z}^B, E)}{\partial z_i^A\partial z_j^A} + \frac{\partial ^2 S(\textbf{z}^A, \textbf{z}^B, E)}{\partial z_i^B \partial z_j^B}\right)_{\textbf{z}^A=\textbf{z}^B=\textbf{q}}\right\| &= \left\| \left(-\frac{\partial p_i^A}{\partial z^A_j}+\frac{\partial p_i^B}{\partial z^B_j}\right)_{\textbf{z}^A=\textbf{z}^B = \textbf{q}} \right\| \nonumber \\
&\leq \left\|\frac{\partial p_i^A}{\partial z^A_j}\Bigg|_{\textbf{z}^A=\textbf{z}^B = \textbf{q}}\right\| + \left\| \frac{\partial p_i^B}{\partial z^B_j} \Bigg|_{\textbf{z}^A=\textbf{z}^B = \textbf{q}}\right\|\,.
\end{align}
Let us bound the first term on the right-hand side of the inequality, since a bound on the second term will follow by the same argument.  We have
\begin{align}
\label{E:tosimplifyshortly1QM}
\left\|\frac{\partial p_i^A}{\partial z^A_j}\Bigg|_{\textbf{z}^A=\textbf{z}^B = \textbf{q}}\right\| = 
    \frac{m}{\hbar\,T_{\text{max}}} \sup_{\|\delta \textbf{z}\|_{2}=\sqrt{\hbar \, T_{\text{max}}/m}}\sum_{i,j}\delta z_i\left(\frac{\partial \pi_j^A}{\partial z^A_i}\right)_{\textbf{z}^A=\textbf{z}^B = \textbf{q}}\delta z_j\,.
\end{align}
To simplify this we define
\begin{equation}
\label{E:deltapijAdef1QM}
\delta p_j^A :=  \sum_{i} \delta z_i \left(\frac{\partial p_j^A}{\partial z^A_i}\right)_{\textbf{z}^A=\textbf{z}^B = \textbf{q}},
\end{equation}
which depends on $\delta z_i$. We find that~\eqref{E:tosimplifyshortly1QM} simplifies to
\begin{align}
\label{E:tosimplifyshortly2QM}
    \frac{m}{\hbar\,T_{\text{max}}}\sup_{\|\delta \textbf{z}\|_{2}=\sqrt{\hbar \, T_{\text{max}}/m}} \sum_{j} \delta p_j^A\,\delta z_j &\leq \frac{m}{\hbar\,T_{\text{max}}} \sup_{\|\delta \textbf{z}\|_{2}=\sqrt{\hbar \, T_{\text{max}}/m}} \| \delta \textbf{p}^A \|_{2} \|\delta \textbf{z} \|_{2} \nonumber \\
    &\leq \sqrt{\frac{m}{\hbar \, T_{\text{max}}}} \sup_{\|\delta z\|_{2}=\sqrt{\hbar \, T_{\text{max}}/m}} \| \delta \textbf{p}^A \|_{2}\,.
\end{align}
It remains to bound $\| \delta \textbf{p}^A \|_{2}$.  From the definition in~\eqref{E:deltapijAdef1QM}, we can interpret $\delta p_j^A$ as follows: it is the change in momentum at the initial time (i.e.~at $A$) if we perturb our periodic orbit by $q_j \to q_j + \delta z_j$ at the initial time (i.e.~at $A$), but still insist that we have the same energy $E$.  Accordingly,
\begin{align}
E&= \frac{1}{2m}(\textbf{p} + \delta \textbf{p}^A)^2+U(\textbf{q} + \delta \textbf{z})
\end{align}
implies
\begin{equation}
\| \textbf{p} + \delta \textbf{p}^A \|_{2}^2 \leq 2 m (E - U_{\text{min}})\,,
\end{equation}
and by an identical reverse triangle inequality argument as in the previous subsection we have
\begin{equation}
    \|\delta \textbf{p}^A\|_{2}\leq 2 \sqrt{2m(E - U_{\text{min}})}\,.
\end{equation}
Altogether we have
\begin{equation}
\left\| \left(\frac{\partial^2 S(\textbf{z}^A, \textbf{z}^B, E)}{\partial z_i^A\partial z_j^A} + \frac{\partial^2 S(\textbf{z}^A, \textbf{z}^B, E)}{\partial z_i^B \partial z_j^B}\right)_{\textbf{z}^A=\textbf{z}^B=\textbf{q}}\right\| \leq 4 \sqrt{\frac{m}{\hbar \, T_{\text{max}}}}\,\sqrt{2m(E - U_{\text{min}})}
\end{equation}
which is indeed finite.

\subsubsection{Putting the bounds together}

In total, for $\phi$ a periodic orbit with energy $E$ we have the bound
\begin{equation}
\| A_{ij}(\textbf{q}) \| \leq \frac{6 \sqrt{2 } \,m}{\sqrt{\hbar \, T_{\text{max}}}} \sqrt{E - U_{\text{min}}} \leq  \frac{6 \sqrt{2}\,m}{T_{\text{max}}} \sqrt{\frac{(E - U_{\text{min}}) T_{\text{max}}}{\hbar}}\,.
\end{equation}
Since we have assumed $U_{\text{min}} \geq 0$, we have thus shown that
\begin{equation}
\| A_{ij}(\textbf{q}) \|  \lesssim  \frac{m}{T_{\text{max}}} \sqrt{\frac{E \, T_{\text{max}}}{\hbar}}
\end{equation}
as we wanted.

\subsection{Boundedness in the quantum field theory setting}

Recall that in~\eqref{E:MQFTdef1} in Appendix \ref{app:derivation} we had the definition
\begin{equation}
\label{eq:hessian}
    A_{ij}[\phi](x,y):=\left(\frac{\delta^2 S(\chib^A, \chib^B, E)}{\delta\chi_i^A(x)\delta\chi_j^A(y)}+2\frac{\delta^2 S(\chib^A, \chib^B, E)}{\delta\chi_i^A(x)\delta\chi_j^B(y)}+\frac{\delta^2 S(\chib^A, \chib^B, E)}{\delta\chi_i^B(x)\delta\chi_j^B(y)}\right)_{\chib^A=\chib^B=\phi}\,.
\end{equation}
We equipped this operator with an operator norm $\| A_{ij}[\phi](x,y) \|$ in~\eqref{E:QFTMopnorm1}, which can be equivalently written as
\begin{equation}
    \| A_{ij}[\phi](x,y) \|:=\sup_{\chi_i(x)} \left|\frac{\int d^{d}x\,\sum_{i,j=1}^2\chi_i(x)A_{ij}[\phi](x,y)\chi_j(y)}{\int d^{d}x\,\sum_{i=1}^2 \chi_i(x)^2}\right|\,.
\end{equation}
In the present subsection, we will show that if $\phi$ is a periodic orbit with energy $E$, then the norm $\| A_{ij}[\phi](x,y) \|$ is bounded.  We take an identical approach as in the ordinary quantum mechanical setting explained above, namely by using the triangle inequality as
\begin{equation}
\label{E:Mnormtriangle1}
   \left\|A_{ij}[\phi](x,y)\right\| \leq \left\|\left(\frac{\delta^2 S(\chib^A, \chib^B, E)}{\delta\chi_i^A(x)\delta\chi_j^A(y)} + \frac{\delta^2 S(\chib^A, \chib^B, E)}{\delta\chi_i^B(x)\delta\chi_j^B(y)}\right)_{\chib^A=\chib^B=\phi} \right\|+2\left\|\frac{\delta^2 S(\chib^A, \chib^B, E)}{\delta\chi_i^A(x)\delta\chi_j^B(y)}\Bigg|_{\chib^A=\chib^B=\phi} \right\|
\end{equation}
and bounding the individual terms on the right-hand side.
As before we begin by recalling the definition of the Legendre transform of Hamilton's principal function to fixed energy, namely:
\begin{align}
\label{E:SEdef1}
    S(\chib^A, \chib^B, E) := S(\chib^A, \chib^B, t(\chib^A, \chib^B, E))+E\,t(\chib^A, \chib^B, E)\,.
\end{align}
In what follows, we will bound the second term on the right-hand side of~\eqref{E:Mnormtriangle1}.

\subsubsection{Mixed derivative term}

Similar to before, we let $\frac{D}{D\chi}$ denote a total functional derivative (as opposed to a partial functional derivative).  Using~\eqref{E:SEdef1} we have
\begin{align}
\label{E:firstmixed1}
    \left.\frac{\delta}{\delta\chi^A_i} \frac{\delta}{\delta\chi^B_j}\,S(\chib^A, \chib^B, E)\right|_{\chib^A=\chib^B = \phib}&= \left.\frac{D}{D\chi^A_i} \frac{D}{D\chi^B_j}\left(S(\chib^A, \chib^B, t(\chib^A, \chib^B, E))+E\,t(\chib^A, \chib^B, E)\right)\right|_{\chib^A = \chib^B} \nonumber \\
    &=\left.\frac{\delta\pi^B_j}{\delta\chi^A_i}+\frac{\partial\pi^B_j}{\partial t}\frac{\delta t}{\delta\chi^A_i}\right|_{\chib^A=\chib^B = \phib} \\ \nonumber \\
    \label{E:secondmixed1}
    \left.\frac{\delta}{\delta \chi^B_j}\frac{\delta}{\delta \chi^A_i}\,S(\chib^A, \chib^B, E)\right|_{\chib^A=\chib^B = \phib}&= \left.\frac{D}{D \chi^B_j}\frac{D}{D \chi^A_i}\left(S(\chib^A, \chib^B, t(\chib^A, \chib^B, E))+E\,t(\chib^A, \chib^B, E)\right)\right|_{\chib^A=\chib^B = \phib} \nonumber \\
    &= \left.-\frac{\delta\pi^A_i}{\delta\chi^B_j}-\frac{\partial\pi^A_i}{\partial t}\frac{\delta t}{\delta\chi^B_j}\right|_{\chib^A=\chib^B = \phib}\,.
\end{align}
Noting that $\frac{\delta}{\delta \chi^B_j}\frac{\delta}{\delta \chi^A_i}\,S(\chib^A, \chib^B, E) = \frac{\delta}{\delta \chi^A_i}\frac{\delta}{\delta \chi^B_j}\,S(\chib^A, \chib^B, E)$, by averaging~\eqref{E:firstmixed1} and~\eqref{E:secondmixed1} we arrive at
\begin{align}
 \left.\frac{\delta}{\delta\chi^A_i} \frac{\delta}{\delta\chi^B_j}\,S(\chib^A, \chib^B, E)\right|_{\chib^A=\chib^B = \phib}&=\frac{1}{2}\left(-\frac{\delta\pi_i^A}{\delta\chi_j^B}+\frac{\delta\pi_j^B}{\delta\chi^A_i}\right)_{\chib^A=\chib^B = \phib}+\frac{1}{2}\left(-\frac{\partial\pi_i^A}{\partial t}\frac{\delta t}{\delta\chi^B_j}+\frac{\partial\pi_j^B}{\partial t}\frac{\delta t}{\delta\chi^A_i}\right)_{\chib^A=\chib^B = \phib}
\end{align}
which gives
\begin{align}
\label{E:mixednorm1}
&\left\|\frac{\delta^2 S(\chib^A, \chib^B, E)}{\delta\chi_i^A(x)\delta\chi_j^B(y)}\Bigg|_{\chib^A=\chib^B=\phi} \right\| = \\
& \frac{1}{\hbar \, T_{\text{max}}} \sup_{\|\delta\chib\|_{L^2}=\sqrt{\hbar \, T_{\text{max}}}}\frac{1}{2}\int d^{d}x\,d^{d}y\,\sum_{i,j=1}^2\delta\chi_i(y)\left(-\frac{\delta\pi_i^A(x)}{\delta\chi_j^B(y)}+\frac{\delta\pi_j^B(x)}{\delta\chi^A_i(y)}-\frac{\partial\pi_i^A(x)}{\partial t}\frac{\delta t}{\delta\chi^B_j(y)}+\frac{\partial\pi_j^B(x)}{\partial t}\frac{\delta t}{\delta\chi^A_i(y)}\right)_{\chib^A=\chib^B=\phib}\delta\chi_j(x)\,.\nonumber
\end{align}
We will simplify the right-hand side of the above equation.  By assumption, $\phib$ belongs to a periodic orbit that has energy $E$, and accordingly $\pib^A = \pib^B$.  This implies $\frac{\partial \pib_A}{\partial t} = \frac{\partial \pib_B}{\partial t}$.  Since we have assumed our theory is time reversal symmetric, we have $t(\chib^A, \chib^B, E) = t(\chib^B, \chib^A, E)$ which gives $\left.\frac{\delta t}{\delta\chi^B_i(y)}\right|_{\chib^A = \chib^B = \phib} = \left.\frac{\delta t}{\delta\chi^A_i(y)}\right|_{\chib^A = \chib^B = \phib}$.  This means that the last two terms in the parentheses in~\eqref{E:mixednorm1} cancel one another.  The simplified result is then
\begin{align}
\label{E:mixednorm2}
\left\|\frac{\delta^2 S(\chib^A, \chib^B, E)}{\delta\chi_i^A(x)\delta\chi_j^B(y)}\Bigg|_{\chib^A=\chib^B=\phib} \right\| &= \frac{1}{\hbar \, T_{\text{max}}} \sup_{\|\delta\chib\|_{L^2}=\sqrt{\hbar \, T_{\text{max}}}}\frac{1}{2}\int d^{d}x\,d^{d}y\,\sum_{i,j=1}^2\delta\chi_i(y)\left(-\frac{\delta\pi_i^A(x)}{\delta\chi_j^B(y)}+\frac{\delta\pi_j^B(x)}{\delta\chi^A_i(y)}\right)_{\chib^A=\chib^B=\phib}\delta\chi_j(x)\nonumber \\
&\leq \frac{1}{2}\left\|\left.\frac{\delta\pi_i^A(x)}{\delta\chi_j^B(y)}\right|_{\chib^A = \chib^B = \phib} \right\| + \frac{1}{2}\left\|\left.\frac{\delta\pi_j^B(x)}{\delta\chi_i^A(y)}\right|_{\chib^A = \chib^B = \phib} \right\|\,.
\end{align}
We proceed by bounding the first term on the right-hand side of the inequality, since the second term follows by the same steps.

Explicitly we have
\begin{align}
\label{E:pitobound1}
\left\|\left.\frac{\delta\pi_i^A(x)}{\delta\chi_j^B(y)}\right|_{\chib^A = \chib^B = \phib} \right\| &= \frac{1}{\hbar \, T_{\text{max}}} \sup_{\|\delta\chib\|_{L^2}=\sqrt{\hbar \, T_{\text{max}}}}\int d^{d}x\,d^{d}y\,\sum_{i,j=1}^2\delta\chi_i(y)\left(\frac{\delta\pi_j^B(x)}{\delta\chi^A_i(y)}\right)_{\chib^A=\chib^B = \phib}\delta\chi_j(x)\,,
\end{align}
and so defining
\begin{equation}
\label{E:deltapijBdef1}
\delta \pi_j^B(x) := \int d^d y \sum_{i=1}^2 \delta\chi_i(y)\left(\frac{\delta\pi_j^B(x)}{\delta\chi^A_i(y)}\right)_{\chib^A=\chib^B = \phib},
\end{equation}
which has dependence on $\delta\chi_i(x)$,~\eqref{E:pitobound1} can be rewritten as
\begin{align}
\frac{1}{\hbar \, T_{\text{max}}} \sup_{\|\delta\chib\|_{L^2}=\sqrt{\hbar \, T_{\text{max}}}}\int d^{d}x\,\sum_{j=1}^2\delta\pi_j^B(x) \,\delta\chi_j(x) &\leq \frac{1}{\hbar \, T_{\text{max}}} \sup_{\|\delta\chib\|_{L^2}=\sqrt{\hbar \, T_{\text{max}}}}  \|\delta \pib^B\|_{L_2} \| \delta \chib \|_{L^2} \nonumber \\
&\leq \frac{1}{\sqrt{\hbar \, T_{\text{max}}}}\, \sup_{\|\delta\chib\|_{L^2}=\sqrt{\hbar \, T_{\text{max}}}}  \|\delta \pib^B\|_{L_2}\,.
\end{align}
Here we used the Cauchy-Schwarz inequality in the first line.  Now we bound $\|\delta \pib^B\|_{L_2}$, and as in the ordinary quantum mechanical setting it is convenient to take stock of the interpretation of the definition of $\delta \pib^B$ in~\eqref{E:deltapijBdef1}.  Notice that $\delta \pib^B$ is the change in momentum at the final time (i.e.~at $B$) if we perturb our periodic orbit by $\phi_j \to \phi_j + \delta \chi_j$ at the initial time (i.e.~at $A$), but require that it ends in the same field configuration $\phi_j$ and has the same energy $E$.  Using the our usual notation $\phib^A = \phib^B = \phib$ for the unperturbed initial position,  we additionally notate $\pib^A = \pib^B = \pib$ for the unperturbed initial and final momentum.  With these notations we write
\begin{align}
    E&= \int d^dx\left[\frac{1}{2}\sum_{i=1}^2(\pib+\delta\pib^B)^2+\frac{1}{2}\sum_{i=1}^2(\nabla\phi_i)^2+U\left(\sum_{i=1}^2\phi_i^2\right)\right]\,.
\end{align}
As in the quantum mechanical case, we define $U_{\text{min}} := \min_{x \in \mathbb{R}_{\geq 0}} U(x)$ which for us is always finite.  Let us further suppose without loss of generality that $U_{\text{min}} \geq 0$.  Then we have
\begin{equation}
\| \pib + \delta \pib^B \|_{L^2}^2 \leq 2 (E - U_{\text{min}})\,.
\end{equation}
Leveraging the reverse triangle inequality gives us
\begin{equation}
    \|\delta\pib^B\|_{L^2}\leq\sqrt{2(E - U_{\text{min}})}+\|\pib\|_{L^2} \leq 2 \sqrt{2(E - U_{\text{min}})}\,.
\end{equation}
We have thus obtained the total bound
\begin{equation}
\left\|\frac{\delta^2 S(\chib^A, \chib^B, E)}{\delta\chi_i^A(x)\delta\chi_j^B(y)}\Bigg|_{\chib^A=\chib^B=\phib} \right\| \leq 2\, \frac{1}{\sqrt{\hbar\,T_{\text{max}}}}\,\sqrt{2(E - U_{\text{min}})}\,.
\end{equation}

\subsubsection{Non-mixed derivative terms}

Going back to~\eqref{E:Mnormtriangle1}, we now bound the first term on its right-hand side.  Using nearly identical arguments as above, we have
\begin{align}
\left(\frac{\delta^2 S(\chib^A, \chib^B, E)}{\delta\chi_i^A(x)\delta\chi_j^A(y)} + \frac{\delta^2 S(\chib^A, \chib^B, E)}{\delta\chi_i^B(x)\delta\chi_j^B(y)}\right)_{\chib^A=\chib^B=\phib} = \left(-\frac{\delta\pi_i^A}{\delta\chi^A_j}+\frac{\delta\pi_i^B}{\delta\chi^B_j}-\frac{\partial\pi_i^A}{\partial t}\frac{\delta t}{\delta\chi^A_j}+\frac{\partial\pi_i^B}{\partial t}\frac{\delta t}{\delta\chi^B_j}\right)_{\chib^A=\chib^B = \phib}\,.
\end{align}
Since for our periodic orbit we have $\pib^A = \pib^B$ and by virtue of time-reversal symmetry we have $t(\chib_A, \chib_B, E) = t(\chib_B, \chib_A, E)$, we can simplify the form of the above equation to obtain
\begin{equation}
\left(\frac{\delta^2 S(\chib^A, \chib^B, E)}{\delta\chi_i^A(x)\delta\chi_j^A(y)} + \frac{\delta^2 S(\chib^A, \chib^B, E)}{\delta\chi_i^B(x)\delta\chi_j^B(y)}\right)_{\chib^A=\chib^B=\phib} = \left(-\frac{\delta\pi_i^A}{\delta\chi^A_j}+\frac{\delta\pi_i^B}{\delta\chi^B_j}\right)_{\chib^A=\chib^B = \phib}\,.
\end{equation}
By taking the norm of both sides of the above equation, we have
\begin{align}
\left\| \left(\frac{\delta^2 S(\chib^A, \chib^B, E)}{\delta\chi_i^A(x)\delta\chi_j^A(y)} + \frac{\delta^2 S(\chib^A, \chib^B, E)}{\delta\chi_i^B(x)\delta\chi_j^B(y)}\right)_{\chib^A=\chib^B=\phib}\right\| &= \left\| \left(-\frac{\delta\pi_i^A}{\delta\chi^A_j}+\frac{\delta\pi_i^B}{\delta\chi^B_j}\right)_{\chib^A=\chib^B = \phib} \right\| \nonumber \\
&\leq \left\|\frac{\delta\pi_i^A}{\delta\chi^A_j}\Bigg|_{\chib^A=\chib^B = \phib}\right\| + \left\| \frac{\delta\pi_i^B}{\delta\chi^B_j} \Bigg|_{\chib^A=\chib^B = \phib}\right\|\,.
\end{align}
We now bound the first term on the right-hand side of the inequality.  An identical bound the second term follows by the same steps.  We explicitly write out
\begin{align}
\label{E:tosimplifyshortly1}
\left\|\frac{\delta\pi_i^A}{\delta\chi^A_j}\Bigg|_{\chib^A=\chib^B = \phib}\right\| = 
    \frac{1}{\hbar\,T_{\text{max}}} \sup_{\|\delta\chib\|_{L^2}=\sqrt{\hbar \, T_{\text{max}}}}\int d^{d}x\,d^{d}y\,\sum_{i,j=1}^2\delta\chi_i(y)\left(\frac{\delta\pi_j^A(x)}{\delta\chi^A_i(y)}\right)_{\chib^A=\chib^B = \phib}\delta\chi_j(x)
\end{align}
and define
\begin{equation}
\label{E:deltapijAdef1}
\delta \pi_j^A(x) := \int d^d y \sum_{i=1}^2 \delta\chi_i(y)\left(\frac{\delta\pi_j^A(x)}{\delta\chi^A_i(y)}\right)_{\chib^A=\chib^B = \phib},
\end{equation}
which has dependence on $\delta\chi_i(x)$. With this notation,~\eqref{E:tosimplifyshortly1} becomes
\begin{align}
\label{E:tosimplifyshortly2}
    \frac{1}{\hbar\,T_{\text{max}}}\sup_{\|\delta\chib\|_{L^2}=\sqrt{\hbar \, T_{\text{max}}}}\int d^{d}x\sum_{j=1}^2 \delta \pi_j^A(x)\,\delta\chi_j(x) &\leq \frac{1}{\hbar\,T_{\text{max}}} \sup_{\|\delta\chi(x)\|_{L^2}=\sqrt{\hbar \, T_{\text{max}}}} \| \delta \pib^A \|_{L_2} \|\delta \chi_j \|_{L_2} \nonumber \\
    &\leq \frac{1}{\sqrt{\hbar \, T_{\text{max}}}} \sup_{\|\delta\chi(x)\|_{L^2}=\sqrt{\hbar \, T_{\text{max}}}} \| \delta \pib^A \|_{L_2}\,.
\end{align}
To finish the bound we need to treat $\| \delta \pib^A \|_{L_2}$.  Using the definition in~\eqref{E:deltapijAdef1}, let us interpret $\delta \pi_j^A$: it represents the change in momentum at the initial time (i.e.~at $A$) if we perturb our periodic orbit by $\phi_j \to \phi_j + \delta \chi_j$ at the initial time (i.e.~at $A$), but maintain the same energy $E$.  Then
\begin{align}
E&= \int d^dx\left[\frac{1}{2}\sum_{i=1}^2(\pi_i+\delta\pi_i^A)^2+\frac{1}{2}\sum_{i=1}^2(\nabla(\phi_i + \delta \chi_i))^2+U\left(\sum_{i=1}^2(\phi_i + \delta \chi_i)^2\right)\right]
\end{align}
gives us
\begin{equation}
\| \pib + \delta \pib^A \|_{L^2}^2 \leq 2 (E - U_{\text{min}})\,.
\end{equation}
Using the reverse triangle inequality argument from above we find
\begin{equation}
    \|\delta\pib^A\|_{L^2}\leq 2 \sqrt{2(E - U_{\text{min}})}\,.
\end{equation}
In total we have
\begin{equation}
\left\| \left(\frac{\delta^2 S(\chib^A, \chib^B, E)}{\delta\chi_i^A(x)\delta\chi_j^A(y)} + \frac{\delta^2 S(\chib^A, \chib^B, E)}{\delta\chi_i^B(x)\delta\chi_j^B(y)}\right)_{\chib^A=\chib^B=\phib}\right\| \leq 4\,\frac{1}{\sqrt{\hbar\,T_{\text{max}}}}\,\sqrt{2(E - U_{\text{min}})}\,.
\end{equation}

\subsubsection{Putting the bounds together}

Putting together the previous results, assuming that $\phi$ is a periodic orbit with energy $E$ we have
\begin{equation}
\| A_{ij}[\phi](x,y) \| \leq \frac{6 \sqrt{2}}{\sqrt{\hbar \, T_{\text{max}}}} \sqrt{E - U_{\text{min}}} \leq  \frac{6 \sqrt{2}}{T_{\text{max}}} \sqrt{\frac{E \, T_{\text{max}}}{\hbar}}\,.
\end{equation}
Since by assumption $U_{\text{min}} \geq 0$ we have
\begin{equation}
\| A_{ij}[\phi](x,y) \|  \lesssim  \frac{1}{T_{\text{max}}} \sqrt{\frac{E \, T_{\text{max}}}{\hbar}}
\end{equation}
as claimed in the main text.

\section{Characterization of moduli space of Q-cloud solutions in an energy window}
\label{app:moduli}
In this Appendix we elaborate on the arguments made in Section \ref{sec:moduli} where we characterized the moduli space of Q-cloud solutions.  There we showed that it is 5-dimensional, wherein any two points in the moduli space are related by some combination of spacetime translations and energy deformation.
We work with the usual Q-cloud Lagrangian in $3+1$ dimensions with a sextic potential given by 
\begin{equation}
    U(\sigma^2) := m^2\sigma^2-\frac{1}{2}f\sigma^4+g\sigma^6\,.
\end{equation}
For concreteness, we can take $m = 1$, $f = 1$, $g = 1/20$, although the parameters are flexible.

We begin by recalling some of the setup from Section~\ref{sec:moduli} in the main body of the paper. Let  $\Phi_\omega(x,t) = e^{i \omega t} \sigma_\omega(x)$ be a Q-cloud solution with period $T = 2\pi/\omega$ in our desired energy range.  Then we wanted to show that if $\tilde{\Phi}$ is (i) a small deformation of $\Phi_\omega$, (ii) time-periodic, and (iii) in our desired energy window, then $\tilde{\Phi}$ is itself a Q-cloud solution.  When we say that $\tilde{\Phi}$ is a `small deformation' of $\Phi_\omega$, we mean that
\begin{equation}
\left(\int_{\mathbb{R}^3} d^3 x \int_0^{\hbar/\varepsilon} \!\! dt \, |\Phi_\omega(x,t) - \tilde{\Phi}(x,t)|^2\right)^{1/2}
\end{equation}
is small.  This is sufficient because we are only considering periodic orbits of length at most $\sim \hbar/\varepsilon$ on account of the exponential decay factor $\exp(- \varepsilon T/\hbar)$ in the scar formula.  Recalling the discussion around~\eqref{E:perturbedaround}, we can write $\tilde{\Phi}$ as
\begin{align}
\label{E:perturbedaround2}
\tilde{\Phi}(x,t) &= \Phi_{\omega + \delta \omega}(x,t) + \sum_{n \in \mathbb{Z}} \! e^{i (\omega + \delta \omega)n t} \delta \phi_n(x)
\end{align}
or more compactly as
\begin{equation}
\tilde{\Phi}(x,t) = \Phi_{\omega + \delta \omega}(x,t) + \delta \Phi(x,t)\,,  
\end{equation}
which can be viewed as a perturbed version of the Q-cloud $\Phi_{\omega + \delta \omega}$ with period $2\pi/(\omega + \delta \omega)$.

For $\Phi_{\omega + \delta \omega}+\delta\Phi$ to be a solution to the equations of motion, $\delta\Phi$ must be a zero mode. This is equivalent to the condition
\begin{equation}
    \left(
\begin{array}{cc}
-(\Phi_{\omega + \delta \omega}^*)^2U'' & -\partial_{\mu}\partial^{\mu}-U'-|\Phi_{\omega + \delta \omega}|^2U''\\
-\partial_{\mu}\partial^{\mu}-U'-|\Phi_{\omega + \delta \omega}|^2U'' & -\Phi_{\omega + \delta \omega}^2U''
\end{array}
\right)
\!\!\left(
\begin{array}{c}
\delta\Phi\\
\delta\Phi^*
\end{array}
\right)=\left(\begin{array}{c}
0 \\ 0
\end{array}\right)\,.
\end{equation}
Next we plug in the mode expansion of $\delta \Phi$ and expand the result to first order in the $\delta \phi_n$'s and $\delta \omega$.  The $\delta \omega$ terms only contribute at second order in fluctuations, and so at first order in fluctuations we have
\begin{align}
(\omega^2n^2+\nabla^2-U'-\sigma_{\omega}^2U'')\delta\phi_n-\sigma_{\omega}^2U'' \, \delta\phi^*_{-n+2}&=0\\
(\omega^2n^2+\nabla^2-U'-\sigma_{\omega}^2U'')\delta\phi^*_{-n}-\sigma_{\omega}^2U''\, \delta\phi_{n+2}&=0\,.
\end{align}
These equations simplify significantly upon making the following observation.  For large $|x|$, we have $\sigma_\omega \to  0$, $U' \to  m^2$, and $U'' \to 0$, and so in this regime we find
\begin{align}
\label{E:largex1}
-\nabla^2\delta\phi_n&=m^2\left(\frac{\omega^2}{m^2}n^2 - 1\right)\delta\phi_n\,.
\end{align}
Here $\omega/m < 1$, and so as explained in Section~\ref{sec:moduli} if we want solutions which decay at infinity (i.e.~are non-oscilliatory), we require $\delta \phi_{n \not = 0, \pm 1} = 0$.  If we plug $\delta \phi_{n \not = 0, \pm 1} = 0$ into the zero mode equations, we find that $\delta \phi_0 = 0$.  The remaining equations are given by
\begin{align}
\label{E:moderemaining1}
(\omega^2+\nabla^2-U'-\sigma_\omega^2U'')\delta\phi_{-1}&=0\\
\label{eq:addedmodeseqn2}
(\omega^2+\nabla^2-U'-2\sigma_\omega^2U'')(\delta\phi_1+\delta\phi^*_{1})&=0\\
\label{eq:subtractedmodeseqn2}
(\omega^2+\nabla^2-U')(\delta\phi_1-\delta\phi^*_{1})&=0\,.
\end{align}

We would like to show that the \textit{only} solutions to the above three equations are given by perturbative spacetime translation zero modes. This can be done in two steps: first we show that the perturbative spacetime translation zero modes solve the above equations.  Second, we prove that these are the only solutions.  The first step is easy.  The perturbative time translation zero mode (which for Q-clouds coincides with a perturbative global $\textsf{U}(1)$ transformation) is given by
\begin{equation}
\delta \Phi(x,t) = \epsilon \cdot i \omega \,e^{i \omega t} \sigma_\omega(x)
\end{equation}
for $\epsilon$ a small parameter.  Here we have $\delta\phi_1= \epsilon \cdot i \omega \sigma_\omega$, $\delta\phi_1^*= - \epsilon \cdot i \omega \sigma_\omega$, $\delta\phi_{-1}=0$, and $\delta \phi_{-1}^* = 0$.  It is readily checked that these modes solve~\eqref{E:moderemaining1}-\eqref{eq:subtractedmodeseqn2}.  Similarly, for spatial translations we have
\begin{equation}
\delta \Phi(x,t) = \epsilon \, e^{i \omega t} \, \partial_j \sigma_\omega(x)
\end{equation}
for $j = 1,2,3$,
giving $\delta \phi_1 = \epsilon \, \partial_j \sigma_\omega$, $\delta \phi_{1}^* = \epsilon \, \partial_j \sigma_\omega$, $\delta \phi_{-1} = 0$, $\delta \phi_{-1}^* = 0$ which likewise solve~\eqref{E:moderemaining1}-\eqref{eq:subtractedmodeseqn2}.

For the second step, we need to show that there are no other solutions to~\eqref{E:moderemaining1}-\eqref{eq:subtractedmodeseqn2}.  We expand the $\delta \phi_{\pm 1}$ modes in spherical harmonics as
\begin{equation}
\delta\phi_{\pm 1}(r,\theta,\phi) =\sum_{\ell = 0}^\infty \sum_{k = -\ell}^\ell \frac{1}{r}\,\delta\varphi_{\pm 1}^{\ell k}(r)\,Y_{\ell m}(\theta, \phi) 
\end{equation}
where the $1/r$ factor is for convenience.  In these variables, we can decouple~\eqref{E:moderemaining1}-\eqref{eq:subtractedmodeseqn2} into an infinite tower of 1D Schr\"{o}dinger-type equations, namely
\begin{align}
\label{E:schro1}
\left(-\frac{d^2}{dr^2} + V_{-1}^{\ell}(r)\right)\delta \varphi_{-1}^{\ell k}(r)&=0\,, \qquad V_{-1}^{\ell}(r) =  -\omega^2+m^2-2f\sigma_\omega^2(r)+3g\sigma_\omega^4(r)+\frac{\ell(\ell+1)}{r^2} \\
\label{E:schro2}
\left(-\frac{d^2}{dr^2} + V_{+}^{\ell}(r)\right)\left(\delta \varphi_{1}^{\ell k}(r) + \delta \varphi_{1}^{*\,\ell k}(r)\right)&=0\,, \qquad \,\,V_{+}^{\ell}(r) = -\omega^2+m^2-3f\sigma_\omega^2(r)+5g\sigma_\omega^4(r)+\frac{\ell(\ell+1)}{r^2}
\\
\label{E:schro3}
\left(-\frac{d^2}{dr^2} + V_{-}^{\ell}(r)\right)\left(\delta \varphi_{1}^{\ell k}(r) - \delta \varphi_{1}^{*\,\ell k}(r)\right)&=0\,, \qquad \,\, V_{-}^{\ell}(r) =  -\omega^2+m^2-f\sigma_\omega^2(r)+g\sigma_\omega^4(r)+\frac{\ell(\ell+1)}{r^2}\,.
\end{align}
Our goal is to show that (i)~\eqref{E:schro1} has no nontrivial, normalizable solutions for all $\ell$; (ii)~\eqref{E:schro2} has only the single normalizable solution $\sigma_\omega'(r)$ at $\ell =1$, with no other nontrivial normalizable solutions for $\ell \not = 1$; and (iii)~\eqref{E:schro3} has only the single normalizable solution $r \sigma_\omega(r)$ at $\ell = 0$, with no other nontrivial normalizable solutions for $\ell \geq 1$.

We achieve these goals by numerically solving the eigenvalue problems
\begin{align}
\label{E:schro11}
\left(-\frac{d^2}{dr^2} + V_{-1}^{\ell}(r)\right)\delta \varphi_{-1}^{\ell k}(r)&= E_{-1}^\ell \delta \varphi_{-1}^{\ell k}(r)  \\
\label{E:schro21}
\left(-\frac{d^2}{dr^2} + V_{+}^{\ell}(r)\right)\left(\delta \varphi_{1}^{\ell k}(r) + \delta \varphi_{1}^{*\,\ell k}(r)\right)&= E_{+}^\ell \left(\delta \varphi_{1}^{\ell k}(r) + \delta \varphi_{1}^{*\,\ell k}(r)\right)
\\
\label{E:schro31}
\left(-\frac{d^2}{dr^2} + V_{-}^{\ell}(r)\right)\left(\delta \varphi_{1}^{\ell k}(r) - \delta \varphi_{1}^{*\,\ell k}(r)\right)&= E_{-}^\ell \left(\delta \varphi_{1}^{\ell k}(r) - \delta \varphi_{1}^{*\,\ell k}(r)\right)\,,
\end{align}
and looking for zero energy solutions corresponding to normalizable eigenstates.
We have elected to choose $m = 1$, $f = 1$, $g = 1/20$, and scan across $\omega$'s from $\omega = 0.85$ to $\omega = 0.95$.  Our results can be found in Tables \ref{tab:omega0-85}, \ref{tab:omega0-9}, and \ref{tab:omega0-95}, respectively.  These results can be understood as follows. 
\begin{itemize}
    \item For~\eqref{E:schro11} pertaining to the $V_{-1}^\ell(r)$ potential, we find two bound states with negative energy at $\ell = 0$, and a continuum of unbound states with strictly positive energy.  Thus there is no state of zero energy at $\ell = 0$.  For $\ell \geq 1$, the lowest energy state has positive energy (which increases as $\ell$ increases), and so there are no states of zero energy for $\ell \geq 1$.
    \item For~\eqref{E:schro21} corresponding to the $V_{+}^\ell(r)$ potential, at $\ell = 0$ we find three bound states with negative energy, and a continuum over unbound states at higher energies.  Accordingly there is no normalizable state of zero energy at $\ell = 0$.  At $\ell = 1$, there is a single bound state $r \sigma_\omega'(r)$ with zero energy corresponding to the spatial translation zero modes.  All of the other states in the spectrum are unbound states, and so there are no other zero energy states at $\ell = 1$.  For $\ell \geq 2$ there are no bound states at all, and hence no normalizable solutions at zero energy.
    \item For~\eqref{E:schro31} which has the $V_{-}^\ell(r)$ potential, at $\ell = 0$ there is one negative energy bound state and one zero energy bound state $r \sigma_\omega(r)$ corresponding to the time translation zero mode.  For $\ell \geq 1$, all of the energies are greater than zero.
\end{itemize}
Our numerical results are reliable until the level spacing becomes too small to distinguish between the bound and unbound states, which occurs for $\omega$ too close to $m$. For the parameters $m=1$, $f=1$, $g=1/20$, our numerics can resolve the difference between bound and unbound states up to around $\omega = 0.999$.
$$$$
\begin{table*}[h!]
\begin{tabular}{c||cccc}
\hline
\hline
 $V_{-1}^\ell(r)$ & $E_{-1}^0$ & $E_{-1}^1$ & $E_{-1}^2$ & $E_{-1}^3$  \\
\hline
\hline
$\ell=0$ & $-4.09609$ & $-1.16907$ & $\substack{\text{first unbound}\\ (0.224325)}$ & \text{unbound}  \\
$\ell=1$ & $\substack{\text{first unbound}\\ (0.295918)}$ & \text{unbound} & $\text{unbound}$ & \text{unbound}  \\
\hline
\hline
$V_{+}^\ell(r)$ & $E_{+}^0$ & $E_{+}^1$ & $E_{+}^2$ & $E_{+}^3$  \\
 
\hline
\hline
$\ell=0$ & $-6.32877$ & $-2.68596$ & $-$0.299762 & $\substack{\text{first unbound}\\ (0.281553)}$  \\
$\ell=1$ & 0 & $\substack{\text{first unbound}\\ (0.296584)}$ & \text{unbound} & \text{unbound} \\
$\ell=2$ & $\substack{\text{first unbound}\\ (0.314928)}$ & \text{unbound} & \text{unbound} & \text{unbound}  \\
\hline
\hline
 $V_{-}^\ell(r)$ & $E_{-}^0$ & $E_{-}^1$ & $E_{-}^2$ & $E_{-}^3$ \\
\hline
\hline
$\ell=0$ & $-1.93887$ & 0 & $\substack{\text{first unbound}\\ (0.283307)}$ & \text{unbound} \\
$\ell=1$ & $\substack{\text{first unbound}\\ (0.296226)}$ & \text{unbound} & \text{unbound} & \text{unbound} \\
\hline
\hline
\end{tabular}
\caption{
Eigenspectra for $\omega=0.85$ and potentials $V_{-1}^\ell(r)$, $V_{+}^\ell(r)$, and $V_-^\ell(r)$, for relevant values $\ell$.
}
\label{tab:omega0-85}
$$$$
\end{table*}
\begin{table*}[h!]
\begin{tabular}{c||cccc}
\hline
\hline
 $V_{-1}^\ell(r)$ & $E_{-1}^0$ & $E_{-1}^1$ & $E_{-1}^2$ & $E_{-1}^3$ \\
\hline
\hline
$\ell=0$ & $-3.31383$ & $-0.910297$ & $\substack{\text{first unbound}\\ (0.145222)}$ & \text{unbound} \\
$\ell=1$ & $\substack{\text{first unbound}\\ (0.208235)}$ & \text{unbound} & \text{unbound} & \text{unbound} \\
\hline
\hline
$V_{+}^\ell(r)$ & $E_{+}^0$ & $E_{+}^1$ & $E_{+}^2$ & $E_{+}^3$ \\
 
\hline
\hline
$\ell=0$ & $-5.22681$ & $-2.12991$ & $-$0.273849 & $\substack{\text{first unbound}\\ (0.193224)}$ \\
$\ell=1$ & 0 & $\substack{\text{first unbound}\\ (0.209239)}$ & \text{unbound} & \text{unbound} \\
$\ell=2$ & $\substack{\text{first unbound}\\ (0.227407)}$ & \text{unbound} & \text{unbound} & \text{unbound} \\
\hline
\hline
 $V_{-}^\ell(r)$ & $E_{-}^0$ & $E_{-}^1$ & $E_{-}^2$ & $E_{-}^3$ \\
\hline
\hline
$\ell=0$ & $-1.48428$ & 0 & $\substack{\text{first unbound}\\ (0.195745)}$ & \text{unbound} \\
$\ell=1$ & $\substack{\text{first unbound}\\ (0.208676)}$ & \text{unbound} & \text{unbound} & \text{unbound} \\
\hline
\hline
\end{tabular}
\caption{Eigenspectra for $\omega=0.9$ and potentials $V_{-1}^\ell(r)$, $V_{+}^\ell(r)$, and $V_-^\ell(r)$, for relevant values $\ell$.
}
\label{tab:omega0-9}
$$$$
\end{table*}
\begin{table*}[t!]
\begin{tabular}{c||cccc}
\hline
\hline
 $V_{-1}^\ell(r)$ & $E_{-1}^0$ & $E_{-1}^1$ & $E_{-1}^2$ & $E_{-1}^3$ \\
\hline
\hline
$\ell=0$ & $-2.01353$ & $-0.533926$ & $\substack{\text{first unbound}\\ (0.0698135)}$ & \text{unbound} \\
$\ell=1$ & $\substack{\text{first unbound}\\ (0.115038)}$ & \text{unbound} & \text{unbound} & \text{unbound} \\
\hline
\hline
$V_{+}^\ell(r)$ & $E_{+}^0$ & $E_{+}^1$ & $E_{+}^2$ & $E_{+}^3$ \\
 
\hline
\hline
$\ell=0$ & $-3.23551$ & $-1.27336$ & $-$0.179464 & $\substack{\text{first unbound}\\ (0.0996173)}$ \\
$\ell=1$ & 0 & $\substack{\text{first unbound}\\ (0.117283)}$ & \text{unbound} & \text{unbound} \\
$\ell=2$ & $\substack{\text{first unbound}\\ (0.134789)}$ & \text{unbound} & \text{unbound} & \text{unbound}  \\
\hline
\hline
 $V_{-}^\ell(r)$ & $E_{-}^0$ & $E_{-}^1$ & $E_{-}^2$ & $E_{-}^3$ \\
\hline
\hline
$\ell=0$ & $-0.855891$ & 0 & $\substack{\text{first unbound}\\ (0.103121)}$ & \text{unbound} \\
$\ell=1$ & $\substack{\text{first unbound}\\ (0.115994)}$ & \text{unbound} & \text{unbound} & \text{unbound} \\
\hline
\hline
\end{tabular}
\caption{Eigenspectra for $\omega=0.95$ and potentials $V_{-1}^\ell(r)$, $V_{+}^\ell(r)$, and $V_-^\ell(r)$, for relevant values $\ell$.
}
\label{tab:omega0-95}
$$$$
\end{table*}


\end{document}